\documentclass[aps,nofootinbib,preprintnumbers]{revtex4}
\usepackage{eurosym}
\usepackage{lipsum}
\usepackage{CJK}
\usepackage{amsfonts}
\usepackage{amsmath}
\usepackage{amssymb}
\usepackage[english]{babel}
\usepackage{graphicx}
\usepackage{epsfig}
\usepackage{bm}
\usepackage{verbatim}
\usepackage[utf8]{inputenc}
\usepackage{booktabs}
\usepackage{multirow}
\usepackage{subfig}
\usepackage{slashed}
\usepackage{xcolor}
\usepackage[colorlinks=true,urlcolor=red,citecolor=red]{hyperref}
\usepackage[font=small]{caption}
\usepackage{float}
\usepackage{blindtext}
\usepackage{placeins}
\usepackage[utf8]{inputenc}
\usepackage{bbm}
\usepackage[colorinlistoftodos]{todonotes}

\newenvironment{Eqnarray}{\arraycolsep 0.14em\begin{eqnarray}}{\end{eqnarray}}
\newcommand{\ba}{\begin{Eqnarray}}
\newcommand{\ea}{\end{Eqnarray}}
\newcommand{\be}{\begin{equation}}
\newcommand{\ee}{\end{equation}}
\newcommand{\bal}{\begin{aligned}}
\newcommand{\eal}{\end{aligned}}
\newcommand{\bea}{\begin{eqnarray}}
\newcommand{\eea}{\end{eqnarray}}
\newcommand{\ben}{\begin{enumerate}}
\newcommand{\een}{\end{enumerate}}
\newcommand{\bit}{\begin{itemize}}
\newcommand{\eit}{\end{itemize}}
\newcommand{\bde}{\begin{widetext}}
\newcommand{\ede}{\end{widetext}}
\newcommand{\nn}{\nonumber}

\def\nn{\nonumber}
\def\lsim{\mathrel{\rlap{\lower4pt\hbox{\hskip1pt$\sim$}}
    \raise1pt\hbox{$<$}}}
\def\gsim{\mathrel{\rlap{\lower4pt\hbox{\hskip1pt$\sim$}}
    \raise1pt\hbox{$>$}}}

\def\3211{$\mathrm{SU(3) \otimes SU(2)_L \otimes U(1)_R \otimes U(1)_{B-L}}$ }
\def\321{$\mathrm{SU(3) \otimes SU(2) \otimes U(1)}$ }
\def\422{$\mathrm{SU(4) \otimes SU(2) \otimes SU(2)_R}$ }

\newcommand{\mathsym}[1]{{}}

\definecolor{bostonuniversityred}{rgb}{0.8, 0.0, 0.0}

\newcommand{\Catalina}[1]{{#1}}

\DeclareUnicodeCharacter{2212}{-}
\pdfoutput=1
\input{tcilatex}
\begin{document}

\title{Fermion masses and mixings, dark matter, leptogenesis and $g-2$ muon
anomaly in an extended 2HDM with inverse seesaw.}
\author{A. E. C\'{a}rcamo Hern\'{a}ndez${}^{a,b,c}$}
\email{antonio.carcamo@usm.cl}
\author{Catalina Espinoza${}^{d}$}
\email{m.catalina@fisica.unam.mx}
\author{Juan Carlos G\'{o}mez-Izquierdo${}^{e}$}
\email{jcgizquierdo1979@gmail.com}
\author{Myriam Mondrag\'on${}^{f}$}
\email{myriam@fisica.unam.mx}
\affiliation{$^{{a}}$Universidad T\'ecnica Federico Santa Mar\'{\i}a, Casilla 110-V,
Valpara\'{\i}so, Chile\\
$^{{b}}$Centro Cient\'{\i}fico-Tecnol\'ogico de Valpara\'{\i}so, Casilla
110-V, Valpara\'{\i}so, Chile\\
$^{{c}}$Millennium Institute for Subatomic physics at high energy frontier -
SAPHIR, Fernandez Concha 700, Santiago, Chile\\
$^{{d}}$C\'atedras Conacyt, Departamento de F\'{\i}sica Te\'{o}rica,
Instituto de F\'{\i}sica,\\
Universidad Nacional Aut\'onoma de M\'{e}xico\\
A.P. 20-364 01000, M\'{e}xico D.F,\\
$^{{e}}$Centro de Estudios Cient\'ificos y Tecnol\'ogicos No 16, \\
Instituto Polit\'ecnico Nacional, Pachuca: Ciudad del Conocimiento y la
Cultura,\\
Carretera Pachuca Actopan km 1+500, San Agust\'in Tlaxiaca, Hidalgo, M\'exico%
\\
$^{{f}}$Departamento de F\'{\i}sica Te\'{o}rica, Instituto de F\'{\i}sica,\\
Universidad Nacional Aut\'onoma de M\'{e}xico\\
A.P. 20-364 01000, CDMX, México.}
\date{\today }

\begin{abstract}
We propose a predictive $Q_4$ flavored 2HDM model, where the scalar sector
is enlarged by the inclusion of several gauge singlet scalars and the
fermion sector by the inclusion of right handed Majorana neutrinos. In our
model, the $Q_4$ family symmetry is supplemented by several auxiliary cyclic
symmetries, whose spontaneous breaking produces the observed pattern of SM
charged fermion masses and quark mixing angles. The light active neutrino
masses are generated from an inverse seesaw mechanism at one loop level
thanks to a remnant preserved $Z_2$ symmetry. Our model succesfully
reproduces the measured dark matter relic abundance and is consistent with
direct detection constraints for masses of the DM candidate around $\sim$ 6.3
TeV. Furthermore, our model is also consistent with the lepton and baryon
asymmetries of the Universe as well as with the muon anomalous magnetic
moment.
\end{abstract}

\maketitle


\section{Introduction}

Despite of being a highly successful theory in describing the
electromagnetic, strong and weak interactions, whose predictions have been
verified with the greatest degree of accuracy by the experiments at the
Large Hadron Collider (LHC), the SM model has several drawbacks. It fails in
providing a natural explanation for the very large hierarchy in the fermion
sector, which spans over a range of 13 orders of magnitude from the light
active neutrino mass scale up to the top quark mass. Furthermore the
observed pattern of fermion mixings characterized by small quark mixing
angles and sizeable leptonic mixing ones does not find an explanation within
the context of the SM. Besides that, the observed amount of dark matter
relic density of the universe and lepton asymmetry are not addressed by the
SM. These unaddressed issues motivate to consider extensions of the SM model
with augmented particle spectrum and extended symmetries. Discrete flavor
symmetries have been shown to be very useful in successfully describing the
observed pattern of SM fermion masses and mixings. Some reviews of discrete
flavor groups are provided in \cite%
{King:2013eh,Altarelli:2010gt,Ishimori:2010au,King:2015aea}. In particular,
the discrete flavor groups having small amount of doublets and singlets in
their irreducible representations, such as, for example $Q_{4}$ \cite%
{Lovrekovic:2012bz,Vien:2019lso} and $D_{4}$ \cite%
{Frampton:1994rk,Grimus:2003kq,Grimus:2004rj,Frigerio:2004jg,Blum:2007jz,Adulpravitchai:2008yp,Ishimori:2008gp,Hagedorn:2010mq,Meloni:2011cc,Vien:2013zra,Vien:2014ica,Vien:2014soa,CarcamoHernandez:2020ney,Vien:2020uzf,Bonilla:2020hct}
have been implemented in extensions of the SM since they allow to provide an
economical and simple way for obtaining viable fermion mass matrix textures,
then allowing to successfully explain and accommodate the current pattern of
SM fermion masses and mixings. Furthermore, several theories with extended
symmetries and particle spectrum have also been proposed to find an
explanation for the muon anomalous magnetic moment, see \cite{Athron:2021iuf}%
~for a very recent review. This muon anomaly was recently confirmed by the Muon $g-2$ experiment at
FERMILAB \cite{Abi:2021gix} and is one of the motivation for considering
extensions of the SM.

In the present paper, we propose an extended 2HDM with enlarged particle
spectrum where the SM gauge symmetry is supplemented by the $Q_{4}$ family
symmetry together with other auxiliary symmetries, thus allowing to get
predictive textures for the SM fermion sector consistent with the low energy
SM fermion flavor data. In the proposed model, the SM charged fermion mass
and quark mixing pattern is generated by the spontaneous breaking of the
discrete symmetries and the light active neutrino masses are produced by a
radiative inverse seesaw mechanism at one loop level, thanks to a remnant
preserved $Z_{2}$ symmetry. To the best of our knowledge our proposed model is the
first $Q_4$ flavoured theory with radiative inverse seesaw mechanism where a
cobimaximal mixing pattern governs the lepton mixings and the discrete
symmetries yield extended Gatto-Sartori-Tonin relations between the quark
masses and mixing angles. Furthermore, unlike other works about models with discrete flavor symmetry mostly focused in the implications of fermion masses and mixings, mainly in the lepton sector, in our current work we analyze in detail the consequences of our model in fermion masses and mixings, muon electric dipole and anomalous magnetic moments, dark matter and leptogenesis. In this way, under certain assumptions we attempt to solve several problems in one single flavor model.
While these assumptions are meant to lessen the complexity of the model, they are well motivated and allow for a thorough analysis of several aspects of it.
Concretely, in the matter sector, the only assumption is made in the neutrino sector, where we assume the equality of a pair of Yukawa couplings in order to get a light active neutrino mass matrix featuring the cobimaximal mixing pattern of lepton mixings, 
which is consistent with the neutrino oscillation experimental data. Regarding the quark sector no assumption is made and the extended Gatto-Sartori-Tonin relations are a direct consequence of the symmetries of the model and the particle assignments under the discrete and SM gauge groups. Likewise, in the treatment of the effective low energy scalar potential, while we give approximate analytical equations for the CP-even physical scalars we employ exact numerical  algorithms during the scan of parameter space. For the phenomenology involving collider limits for scalars, we neglect the masses of the first and second family of fermions and also off-diagonal terms in the Yukawa matrices. Deviations of the matter sector with respect to the SM are expected to be of negligible influence when analyzing present collider limits on scalars.

The layout of the remainder of the paper is as follows. In section \ref%
{model} we describe our extended 2HDM. Its implications on SM fermion masses
and mixings are analyzed in section \ref{fermionmixings}. The consequences
of our proposed theory in Dark matter, muon anomalous magnetic moment and
leptogenesis are discussed in sections \ref{DMsec}, \ref{gminus2} and \ref%
{lepto}, respectively. We conclude in section \ref{conclusions}. Some
technical details are given in the appendices. Appendix \ref{Q4} provides a
concise description of the $Q_{4}$ discrete group. The scalar potential for
two $Q_{4}$ doublets is analyzed in Appendix \ref{Q4scalarpotential}.

\newpage

\section{The model}

\label{model} We propose an extended 2HDM where the scalar sector is
augmented by the inclusion of several gauge singlet scalars and the fermion
sector is extended by the inclusion of six right handed Majorana neutrinos.
The SM gauge symmetry is extended by the inclusion of the $Q_{4}\times
Z_{3}^{\left( 1\right) }\times Z_{3}^{\left( 2\right) }\times Z_{4}\times
Z_{8}$ discrete group, whose spontaneous breaking generates predictive
fermion mass matrices consistent with the SM fermion masses and mixing
parameters. The role of the aforementioned cyclic symmetries is explained in
the following. The $Q_{4}$ symmetry shapes the textures of the SM fermion mass matrices thus reducing the model
parameters, especially in the SM lepton sector. We choose the $Q_{4}$
symmetry since it is the smallest non-Abelian discrete symmetry group having
five irreducible representations (irreps), explicitly, four singlets and one
doublet irreps. Besides that, the $Q_4$ flavour symmetry allows more freedom in assigning the fermionic and scalar fields in different
representations and having more suppressed Yukawa interactions when
compared with $S_3$. Moreover, the $D_4$ discrete group has very
similar tensor product rules as $Q_4$ and thus using $D_4$ instead of $Q_4$
will not yield significant changes in the model and the resulting physical
results would be very similar to the ones corresponding to the $Q_4$ flavoured theory. Replacing $Q_4$ by the $D_4$ flavor group will only yield important modifications in the neutrino Yukawa terms, due to the fact that the right handed Majorana neutrinos are the only fermionic fields of the model assigned as $Q_4$ doublets. This will affect the annihilation channels of fermionic dark matter candidates. Thus, the annihilation channels of the fermionic dark matter candidate $\Psi$, such as for instance $\Psi\Psi\rightarrow N^{\pm}_{a}N^{\pm}_{a}$ ($a=1,2,3$) can be an experimental test to distinguish our $Q_4$ flavored model from an alternative model based on the $D_4$ family symmetry.

 The $Z_{3}^{\left( 1\right) }$ separates the two $SU(2)_{L}$
scalar doublets $H_{1}$ and $H_{2}$, thus allowing to get viable and
predictive quark mass matrix textures where the Cabbibo mixing arises from
the down quark sector whereas the remaining quark mixing angles are
generated from the up quark sector. On the other hand, the $Z_{3}^{\left( 2\right) }$ and $%
Z_{8} $ discrete symmetries shape the hierarchical structure of the SM
charged fermion mass matrices crucial to yield the observed pattern of SM
charged fermion masses and mixing angles. The $Z_{4}$ discrete symmetry is
spontaneously broken to a preserved $Z_{2}$ symmetry, which allows the
implementation of one loop level inverse seesaw mechanism that produces the
tiny masses for the light active neutrinos. The $Q_{4}\times Z_{3}^{\left(
1\right) }\times Z_{3}^{\left( 2\right) }\times Z_{4}\times Z_{8}$
assignments for scalars, quarks and leptons are shown in Tables \ref%
{tab:scalars}, \ref{tab:quarks} and \ref{tab:leptons}, respectively. Here
the different $Z_{N}$ charges are given in additive notation. Let us note
that a field $\psi $ transforms under the $Z_{N}$ symmetry as: $\psi
\rightarrow e^{\frac{2\pi iq_{n}}{N}}\psi $, $n=0,1,2,3\cdots N-1$, where $%
q_{n}$ is its corresponding charge in additive notation. As shown in Tables %
\ref{tab:scalars} and \ref{tab:leptons}, the gauge singlet scalars $\eta
_{k} $ ($k=1,2$)\ and the right handed Majorana neutrino $\Psi _{R}$ are the
only particles having a complex $Z_{4}$ charge, corresponding to a
nontrivial charge under the preserved $Z_{2}$ symmetry. Due to the preserved 
$Z_{2}$ symmetry our model has stable scalar and fermionic dark matter
candidates. The scalar dark matter candidate is the lightest among $\func{Re}%
\eta _{k}$ and $\func{Im}\eta _{k}$ ($k=1,2$), whereas the fermionic dark
matter candidate is the gauge singlet neutral lepton $\Psi _{R}$. 
\begin{table}[th]
\begin{tabular}{|c|c|c|c|c|c|c|c|c|c|}
\hline
& $H_{1}$ & $H_{2}$ & $\sigma $ & $\rho $ & $\xi $ & $\eta _{1}$ & $\eta
_{2} $ & $\varphi $ & $\Phi $ \\ \hline
$Q_{4}$ & $\mathbf{1}_{++}$ & $\mathbf{1}_{++}$ & $\mathbf{1}_{++}$ & $%
\mathbf{1}_{-+}$ & $\mathbf{2}$ & $\mathbf{1}_{++}$ & $\mathbf{1}_{+-}$ & $%
\mathbf{1}_{++}$ & $\mathbf{2}$ \\ \hline
$Z_{3}^{\left( 1\right) }$ & 0 & 1 & 0 & 0 & 0 & 0 & 0 & 1 & 1 \\ \hline
$Z_{3}^{\left( 2\right) }$ & 0 & 0 & 0 & -1 & 0 & 0 & 0 & 0 & 0 \\ \hline
$Z_{4}$ & 0 & 0 & 0 & 0 & 0 & -1 & -1 & -2 & -2 \\ \hline
$Z_{8}$ & 0 & 0 & -1 & 0 & 0 & -1 & -1 & 0 & 0 \\ \hline
\end{tabular}%
\caption{Scalar assignments under $Q_{4}\times Z_{3}^{\left( 1\right)
}\times Z_{3}^{\left( 2\right) }\times Z_{4}\times Z_{8}$.}
\label{tab:scalars}
\end{table}

\begin{table}[th]
\begin{tabular}{|c|c|c|c|c|c|c|c|c|c|}
\hline
& $q_{1L}$ & $q_{2L}$ & $q_{3L}$ & $u_{1R}$ & $u_{2R}$ & $u_{3R}$ & $d_{1R}$
& $d_{2R}$ & $d_{3R}$ \\ \hline
$Q_{4}$ & $\mathbf{1}_{++}$ & $\mathbf{1}_{-+}$ & $\mathbf{1}_{--}$ & $%
\mathbf{1}_{--}$ & $\mathbf{1}_{++}$ & $\mathbf{1}_{--}$ & $\mathbf{1}_{+-}$
& $\mathbf{1}_{-+}$ & $\mathbf{1}_{+-}$ \\ \hline
$Z_{3}^{\left( 1\right) }$ & 0 & 0 & 1 & 0 & 0 & 1 & 2 & 2 & 1 \\ \hline
$Z_{3}^{\left( 2\right) }$ & 0 & 0 & 0 & 0 & 0 & 0 & -1 & -1 & 1 \\ \hline
$Z_{4}$ & 0 & 0 & 0 & 0 & 0 & 0 & 0 & 0 & 0 \\ \hline
$Z_{8}$ & 0 & 0 & 0 & 4 & 2 & 0 & 4 & 2 & 2 \\ \hline
\end{tabular}%
\caption{Quark assignments under $Q_{4}\times Z_{3}^{\left( 1\right) }\times
Z_{3}^{\left( 2\right) }\times Z_{4}\times Z_{8}$.}
\label{tab:quarks}
\end{table}


\begin{table}[th]
\begin{tabular}{|c|c|c|c|c|c|c|c|c|c|c|c|}
\hline
& $l_{1L}$ & $l_{L}$ & $l_{1R}$ & $l_{2R}$ & $l_{3R}$ & $\nu _{1R}$ & $\nu
_{2R}$ & $\nu _{3R}$ & $N_{1R}$ & $N_{R}$ & $\Psi_{R}$ \\ \hline
$Q_{4}$ & $\mathbf{1}_{++}$ & $\mathbf{2}$ & $\mathbf{1}_{--}$ & $\mathbf{1}%
_{+-}$ & $\mathbf{1}_{--}$ & $\mathbf{1}_{++}$ & $\mathbf{1}_{+-}$ & $%
\mathbf{1}_{--}$ & $\mathbf{1}_{++}$ & $\mathbf{2}$ & $\mathbf{1}_{++}$ \\ 
\hline
$Z_{3}^{\left( 1\right) }$ & 0 & 0 & 0 & 0 & 0 & 1 & 1 & 1 & 2 & 2 & 1 \\ 
\hline
$Z_{3}^{\left( 2\right) }$ & 0 & 0 & 0 & 0 & 0 & 0 & 0 & 0 & 0 & 0 & 0 \\ 
\hline
$Z_{4}$ & 0 & 0 & 0 & 0 & 0 & 0 & 0 & 0 & 0 & 0 & 1 \\ \hline
$Z_{8}$ & 0 & 0 & 0 & 0 & 0 & 0 & 0 & 0 & 0 & 0 & 0 \\ \hline
\end{tabular}%
\caption{Lepton assignments under $Q_{4}\times Z_{3}^{\left( 1\right)
}\times Z_{3}^{\left( 2\right) }\times Z_{4}\times Z_{8}$.}
\label{tab:leptons}
\end{table}


In order to get a predictive and viable pattern of SM fermion masses and
mixings we consider the following vacuum expectation value (VEV)
configuration for the $Q_{4}$ doublets SM gauge singlet scalars $\xi $ and $%
\Phi $; 
\begin{equation}
\left\langle \xi \right\rangle =v_{\xi }\left( 1,0\right) ,\hspace{1.5cm}%
\left\langle \Phi \right\rangle =v_{\Phi }\left( -e^{i\theta },e^{-i\theta
}\right) ,  \label{Q4VEVpattern}
\end{equation}%
Such VEV configuration is consistent with the scalar potential minimization
equations for a large region of parameter space as shown in detail in
Appendix \ref{Q4scalarpotential}.

Given that the observed SM charged fermion mass and quark mixing pattern is
caused by the spontaneous breaking of the $Q_{4}\times Z_{3}^{\left(
1\right) }\times Z_{3}^{\left( 2\right) }\times Z_{8}$ discrete group, the
vacuum expectation values (VEVs) of the gauge singlet scalars are set to
fullfill 
the following hierarchy: 
\begin{equation}
v<<v_{\varphi }\sim v_{\Phi }\sim \mathcal{O}(10\text{TeV})<<v_{\sigma }\sim
v_{\rho }\sim v_{\xi }\sim \lambda \Lambda \sim \mathcal{O}(100\text{TeV}),
\label{VEVsinglets}
\end{equation}%
where $v=246$ GeV, $\lambda =0.225$ is the Wolfenstein parameter and $%
\Lambda $ is the model cutoff. Notice that the gauge singlet scalar field $%
\varphi $ is assumed to acquire a VEV at the TeV scale, in order to get TeV
scale sterile neutrinos in the leptonic spectrum, thus allowing to have
sterile neutrino signatures testable at colliders.

In what follows we provide a discussion about the different scales (\ref{VEVsinglets}) of the vacuum expectation values of the singlet scalar fields. Notice that there is no symmetry that protects this pattern from large radiative corrections. Thus, in order to stabilize it, we need to apply certain tuning of the model parameters. The corresponding vacuum stability conditions arise from the Coleman-Weinberg type 1-loop effective potential. This analysis is left beyond the scope of the present paper. However, since in our model the VEV hierarchy (\ref{VEVsinglets}) is rather moderate, not exceeding three orders of magnitude, we expect that the quadratic divergences dangerous for a strong hierarchy can be tamed here by a moderate tuning of the model parameters. At the same time, for the scales larger than $v_ {\Phi} $ in (\ref{VEVsinglets}), where this is not possible, we proceed to assume that our model is embedded into a more fundamental theory with additional symmetries that protect the hierarchy up to the Planck scale. Some well-known and motivated examples of such theories are supersymmetry and warped five-dimensions.

The relevant Yukawa terms are: 
\begin{eqnarray}
\mathcal{L}^{\left( u\right) } &=&x_{13}^{\left( u\right) }\overline{q}_{1L}%
\widetilde{H}_{2}u_{3R}\frac{\left( \xi \xi \right) _{\mathbf{1}_{+-}}\left(
\xi \xi \right) _{\mathbf{1}_{-+}}}{\Lambda ^{4}}+x_{23}^{\left( u\right) }%
\overline{q}_{2L}\widetilde{H}_{2}u_{3R}\frac{\left( \xi \xi \right) _{%
\mathbf{1}_{+-}}}{\Lambda ^{2}}+x_{33}^{\left( u\right) }\overline{q}_{3L}%
\widetilde{H}_{1}u_{3R}  \notag \\
&&+x_{22}^{\left( u\right) }\overline{q}_{2L}\widetilde{H}_{1}u_{2R}\frac{%
\left( \xi \xi \right) _{\mathbf{1}_{-+}}\sigma ^{2}}{\Lambda ^{4}}%
+x_{11}^{\left( u\right) }\overline{q}_{1L}\widetilde{H}_{1}u_{1R}\frac{%
\left( \xi \xi \right) _{\mathbf{1}_{+-}}\left( \xi \xi \right) _{\mathbf{1}%
_{-+}}\sigma ^{4}}{\Lambda ^{8}}+h.c  \label{Lyu}
\end{eqnarray}%
\begin{eqnarray}
\mathcal{L}^{\left( d\right) } &=&x_{11}^{\left( d\right) }\overline{q}%
_{1L}H_{2}d_{1R}\frac{\left( \xi \xi \right) _{\mathbf{1}_{+-}}\sigma
^{4}\rho ^{2}}{\Lambda ^{8}}+x_{12}^{\left( d\right) }\overline{q}%
_{1L}H_{2}d_{2R}\frac{\left( \xi \xi \right) _{\mathbf{1}_{-+}}\sigma
^{2}\rho ^{2}}{\Lambda ^{6}}  \notag \\
&&+x_{22}^{\left( d\right) }\overline{q}_{2L}H_{2}d_{2R}\frac{\left( \xi \xi
\right) _{\mathbf{1}_{-+}}\sigma ^{2}\rho ^{\ast }}{\Lambda ^{5}}%
+x_{33}^{\left( d\right) }\overline{q}_{3L}H_{2}d_{3R}\frac{\sigma ^{2}\rho 
}{\Lambda ^{3}}+h.c  \label{Lyd}
\end{eqnarray}%
\begin{eqnarray}
\mathcal{L}^{\left( l\right) } &=&x_{11}^{\left( l\right) }\overline{l}%
_{1L}H_{1}l_{1R}\frac{\left( \xi \xi \right) _{\mathbf{1}_{+-}}\left( \xi
\xi \right) _{\mathbf{1}_{-+}}\sigma ^{4}}{\Lambda ^{8}}+x_{12}^{\left(
l\right) }\overline{l}_{1L}H_{1}l_{2R}\frac{\left( \xi \xi \right) _{\mathbf{%
1}_{+-}}\sigma ^{4}}{\Lambda ^{6}}  \notag \\
&&+x_{22}^{\left( l\right) }\overline{l}_{L}H_{1}l_{2R}\frac{\xi \sigma ^{4}%
}{\Lambda ^{5}}+x_{33}^{\left( l\right) }\overline{l}_{L}H_{1}l_{3R}\frac{%
\xi \sigma ^{2}}{\Lambda ^{3}}+h.c  \label{Lyl}
\end{eqnarray}%
\begin{eqnarray}
\mathcal{L}^{\left( \nu \right) } &=&y_{1}^{\left( \nu \right) }\overline{l}%
_{1L}\widetilde{H}_{2}\nu _{1R}+y_{2}^{\left( \nu \right) }\overline{l}_{L}%
\widetilde{H}_{2}\nu _{2R}\frac{\xi }{\Lambda }+y_{3}^{\left( \nu \right) }%
\overline{l}_{L}\widetilde{H}_{2}\nu _{3R}\frac{\xi }{\Lambda }+M_{1}%
\overline{\nu }_{1R}N_{1R}^{C}+y_{2}\overline{\nu }_{2R}N_{R}^{C}\xi +y_{3}%
\overline{\nu }_{3R}N_{R}^{C}\xi  \label{Lynu} \\
&&+y_{1N}N_{1R}\overline{\Psi _{R}^{C}}\varphi \frac{\eta _{1}}{\Lambda }%
+y_{2N}\left( N_{R}\Phi \right) _{\mathbf{1}_{++}}\overline{\Psi _{R}^{C}}%
\frac{\eta _{1}}{\Lambda }+y_{3N}\left( N_{R}\Phi \right) _{\mathbf{1}_{+-}}%
\overline{\Psi _{R}^{C}}\frac{\eta _{2}}{\Lambda }+y_{\Psi }\Psi _{R}%
\overline{\Psi _{R}^{C}}\varphi +M_{sb}\left( \overline{\nu }_{1R}\nu
_{2R}^{C}+\overline{\nu }_{2R}\nu _{1R}^{C}\right) +h.c  \notag
\end{eqnarray}%
where we have introduced the soft-breaking Majorana mass term $M_{sb}\left( 
\overline{\nu }_{1R}\nu _{2R}^{C}+\overline{\nu }_{2R}\nu _{1R}^{C}\right) $
in order to get the correct sign and magnitude of the muon anomalous
magnetic moment. Other possible soft-breaking mass terms of the form $M^{(1)}_{sb}\left(\overline{\nu }_{1R}\nu _{3R}^{C}+\overline{\nu }_{3R}\nu _{1R}^{C}\right) $ and $M^{(1)}_{sb}\left(\overline{\nu }_{2R}\nu _{3R}^{C}+\overline{\nu }_3R\nu _{2R}^{C}\right)$ in the lepton sector will generate corrections to the submatrix $\epsilon$ of the $(2,2)$ block of the full neutrino mass matrix. These corrections will add extra contributions to the mass matrices of the light active and sterile neutrinos neutrinos, along the same lines of \cite{Law:2013gma}. However such contributions are very subleading.

It is worth stressing that the $Q_{4}$ flavor symmetry is more
relevant in the lepton sector since some of the leptonic fields are assigned
as $Q_{4}$ doublets as seen in Table \ref{tab:leptons} and the considered
setup allows to get a predictive light active neutrino mass matrix featuring
the cobimaximal mixing pattern, as it will be shown in the next section. In the
concerning to the quark sector, despite there are no $Q_{4}$ doublets (as
follows from Table \ref{tab:leptons}), the importance of the $Q_{4}$ flavor
symmetry is that it allows to get, for example a twelve dimensional Yukawa
operator crucial for a naturally explanation of the smallness of the up
quark mass without relying on the inclusion of large cyclic symmetries like
for instance $Z_{16}$. This is due to the fact that there is no scalar field in
the particle spectrum assigned as $\mathbf{1}_{--}$ and thus the effective $%
\mathbf{1}_{--}$ scalar required to build the twelve dimensional Yukawa
operator (last term of Eq. (3)) that generates the up quark mass term, is
built from the quartic combination $\left( \xi \xi \right) _{\mathbf{1}%
_{+-}}\left( \xi \xi \right) _{\mathbf{1}_{-+}}$ \ involving the $Q_{4}$
scalar doublet $\xi $. This trick is also used in the construction of the up
type quark Yukawa operator (first term of Eq. (3)) that yields the $\theta
_{13}^{\left( q\right) }$ quark mixing angle.

\begin{figure}[tbp]
\centering
\includegraphics[width=0.5\textwidth]{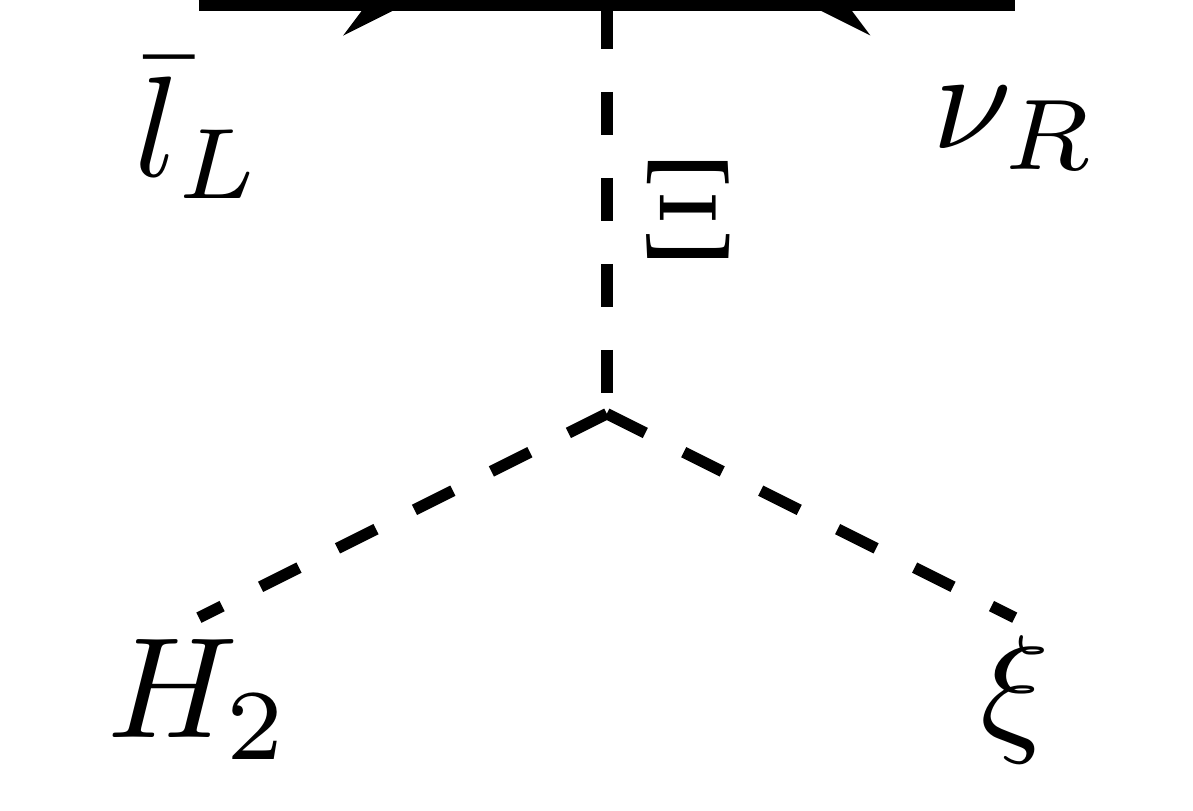}%
\includegraphics[width=0.5\textwidth]{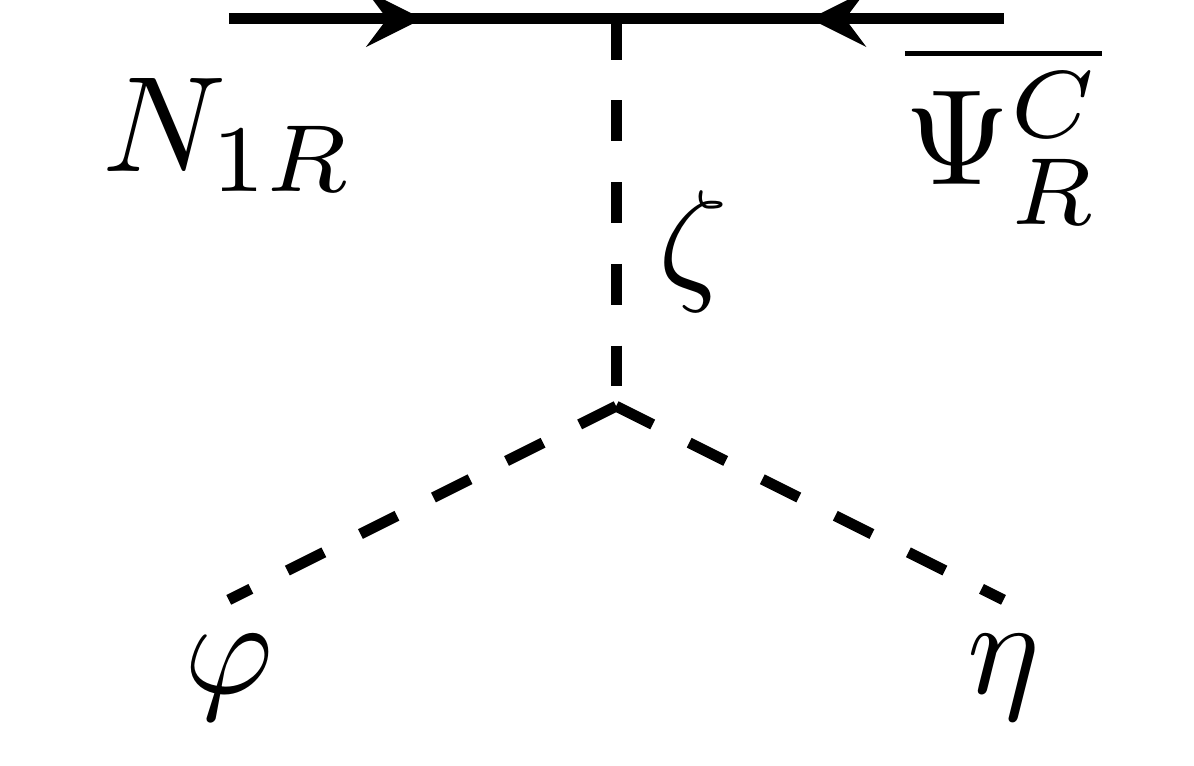}\newline
\includegraphics[width=0.5\textwidth]{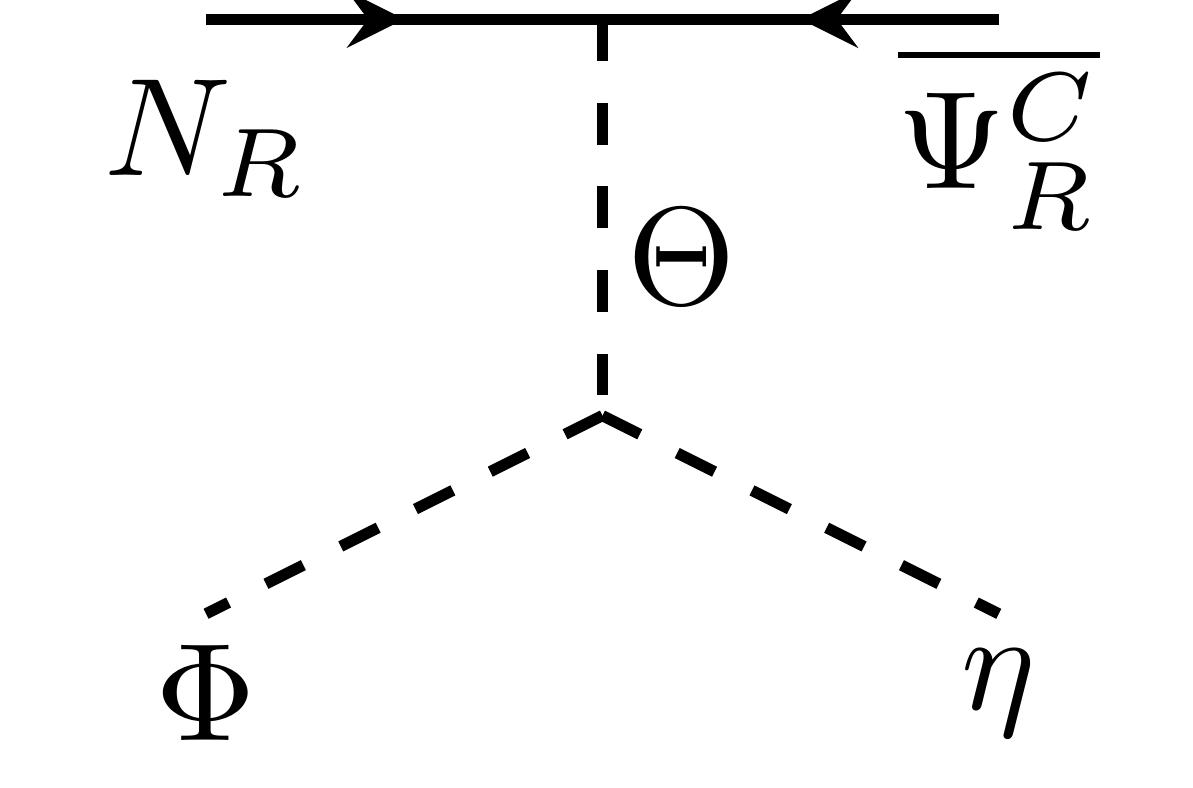}
\caption{Feynman diagrams that induce the non-renormalizable operators of
Eq. (\protect\ref{NeutrinoOP}).}
\label{NeutrinoUV}
\end{figure}

In what follows we will describe a plausible ultraviolet origin for these
non-renormalizable operators. As seen from Eqs. (\ref{Lyu}), (\ref{Lyd}), (%
\ref{Lyl}) and (\ref{Lynu}), we introduced several non-renormalizable Yukawa
operators. These allow us to explain the observed hierarchies in the SM
fermion mass spectrum and the fermion mixing parameters while keeping all
the Yukawa couplings of order unity. 
Notice that all of them have the following form: 
\begin{equation}
\overline{f}_{L}S_{1}F_{R}\left( \frac{\Sigma _{1}}{\Lambda }\right) ^{n_{1}}%
\hspace{1cm}\overline{F}_{L}S_{2}f_{R}\left( \frac{\Sigma _{2}}{\Lambda }%
\right) ^{n_{1}}  \label{NRoperators}
\end{equation}%
where $f$ and $F$ stand for light and heavy fermions, respectively, $n_{1}$, 
$n_{2}$ are integers and $S_{1}$, $S_{2}$, $\Sigma _{1}$ and $\Sigma _{2}$
are scalars. Here, for simplicity, we have omitted family and fermionic type
indices. One sees that these non-renormalizable operators in Eq. (\ref%
{NRoperators}) can all arise from the following renormalizable operators: 
\begin{eqnarray}
&&\overline{f}_{L}S_{3}\tilde{F}_{R}\hspace{1cm}\overline{\tilde{F}}%
_{L}S_{4}F_{R}\hspace{1cm}\overline{\tilde{F}}_{L}S_{5}\tilde{F}_{R}  \notag
\\
&&\overline{f}_{L}S_{6}\tilde{F}_{R}\hspace{1cm}\overline{\tilde{F}}%
_{L}S_{7}F_{R}\hspace{1cm}\overline{\tilde{F}}_{L}S_{8}\tilde{F}_{R}
\label{Roperators}
\end{eqnarray}%
where $S_{k}$ ($k=3,4,\cdots 8$) are extra scalars and $\tilde{F}$ extra
very heavy fermions. Assuming that the $S_{5}$ and $S_{8}$ scalars acquire
vacuum expectation values much larger than the remaining scalars, the
fermions $\tilde{F}$ will get very large masses. As a result, they can be
integrated out, yielding effective non-renormalizable operators as in Eq.~(%
\ref{NRoperators}). Now in order to make our discussion more explicit and we
are going to specify a possible ultraviolet origin of the following non
renormalizable neutrino Yukawa operators: 
\begin{equation}
\overline{l}_{L}\widetilde{H}_{2}\nu _{R}\frac{\xi }{\Lambda },\hspace{1cm}%
N_{1R}\overline{\Psi _{R}^{C}}\varphi \frac{\eta }{\Lambda },\hspace{1cm}%
N_{R}\overline{\Psi _{R}^{C}}\Phi \frac{\eta }{\Lambda }  \label{NeutrinoOP}
\end{equation}%
where we have suppressed the subscript $k$ of the scalar field $\eta _{k}$,
unessential for our discussion. These three non-renormalizable Yukawa terms
of Eq. (\ref{NeutrinoOP}) can be generated at low energies by the Feynman
diagrams shown in Figure \ref{NeutrinoUV} after integrating out the heavy
scalar fields $\zeta $ and $\Theta $ with characteristic masses of the order
of our model cutoff scale $\Lambda $. Their assignment under the symmetries
of the model is dictated by the requirement that the renormalizable
interactions in the vertices of these diagrams be invariant under these
symmetries. 
Thus, it follows that $\Xi $, $\xi $ and $\Theta $ are $Q_{4}$ doublets,
whereas $\zeta $ is a $Q_{4}$ singlet. Furthermore, $\Xi $ is a $SU(2)_{L}$
scalar doublet with hypercharge $\frac{1}{2}$ (as the usual SM Higgs
doublet), whereas $\xi $, $\zeta $ and $\Theta $ are electrically neutral scalars transforming as singlets under the SM gauge symmetry.
\begin{figure}[tbp]
\centering
\includegraphics[width=0.5\textwidth]{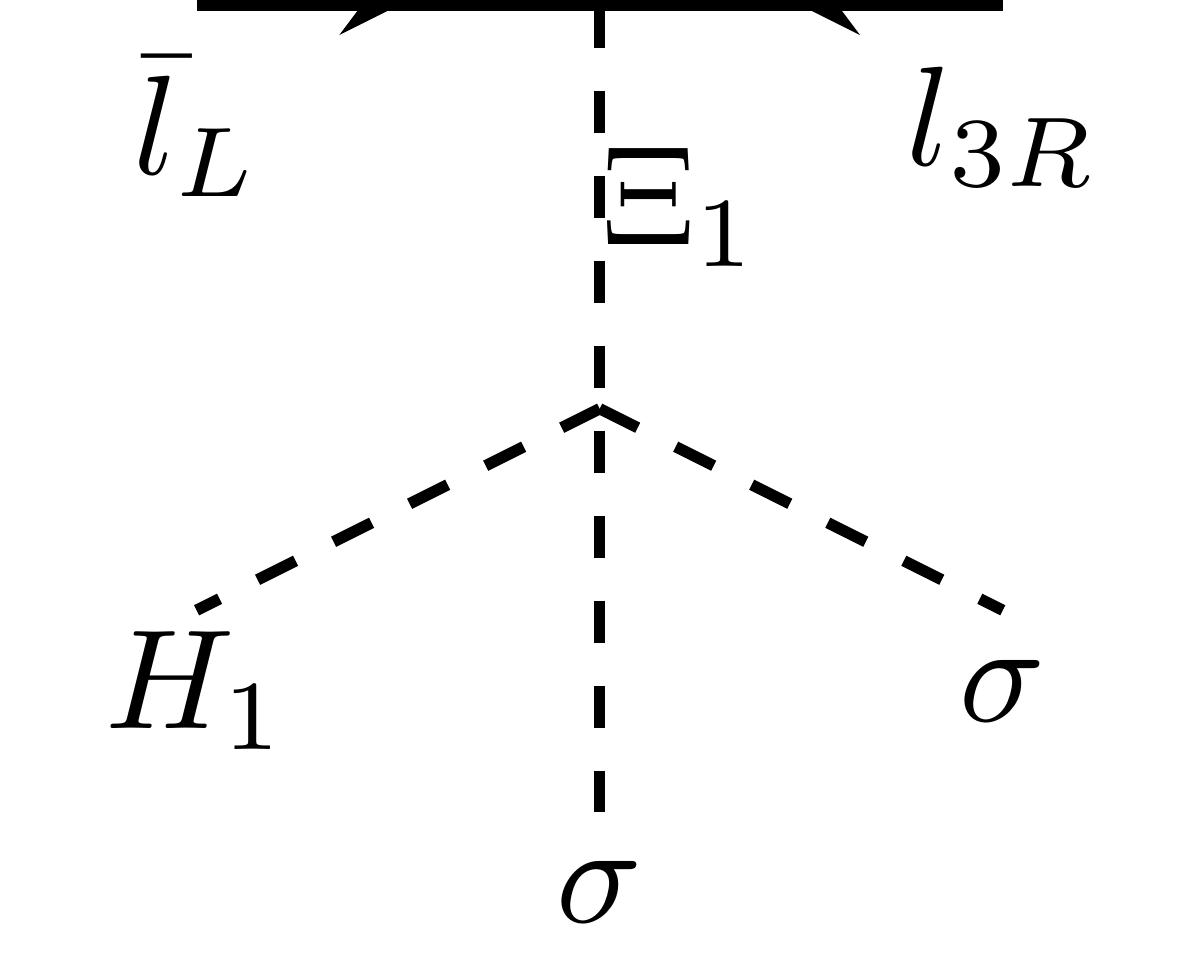}%
\includegraphics[width=0.5\textwidth]{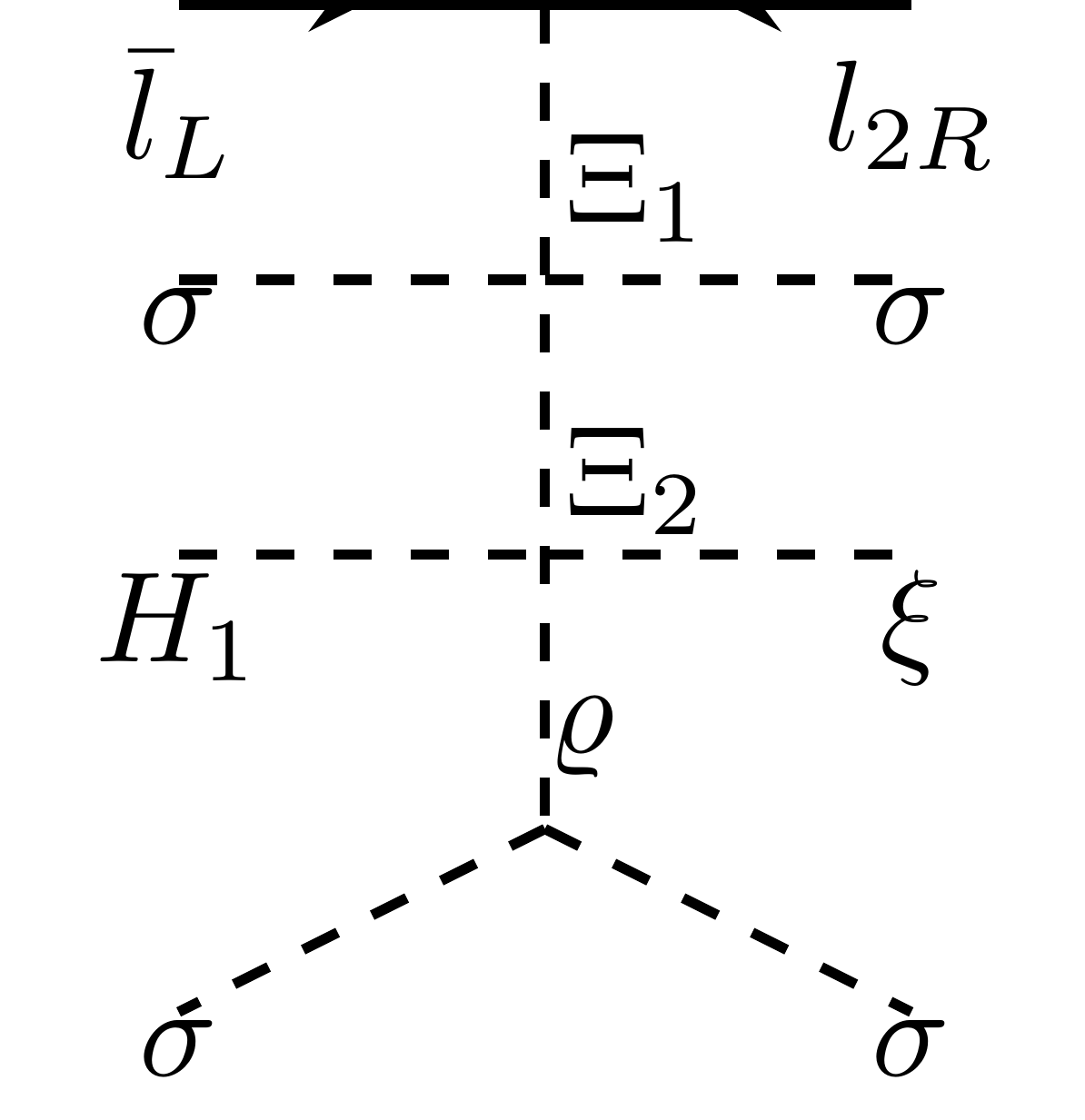}\newline
\vspace{5mm} \includegraphics[width=0.5\textwidth]{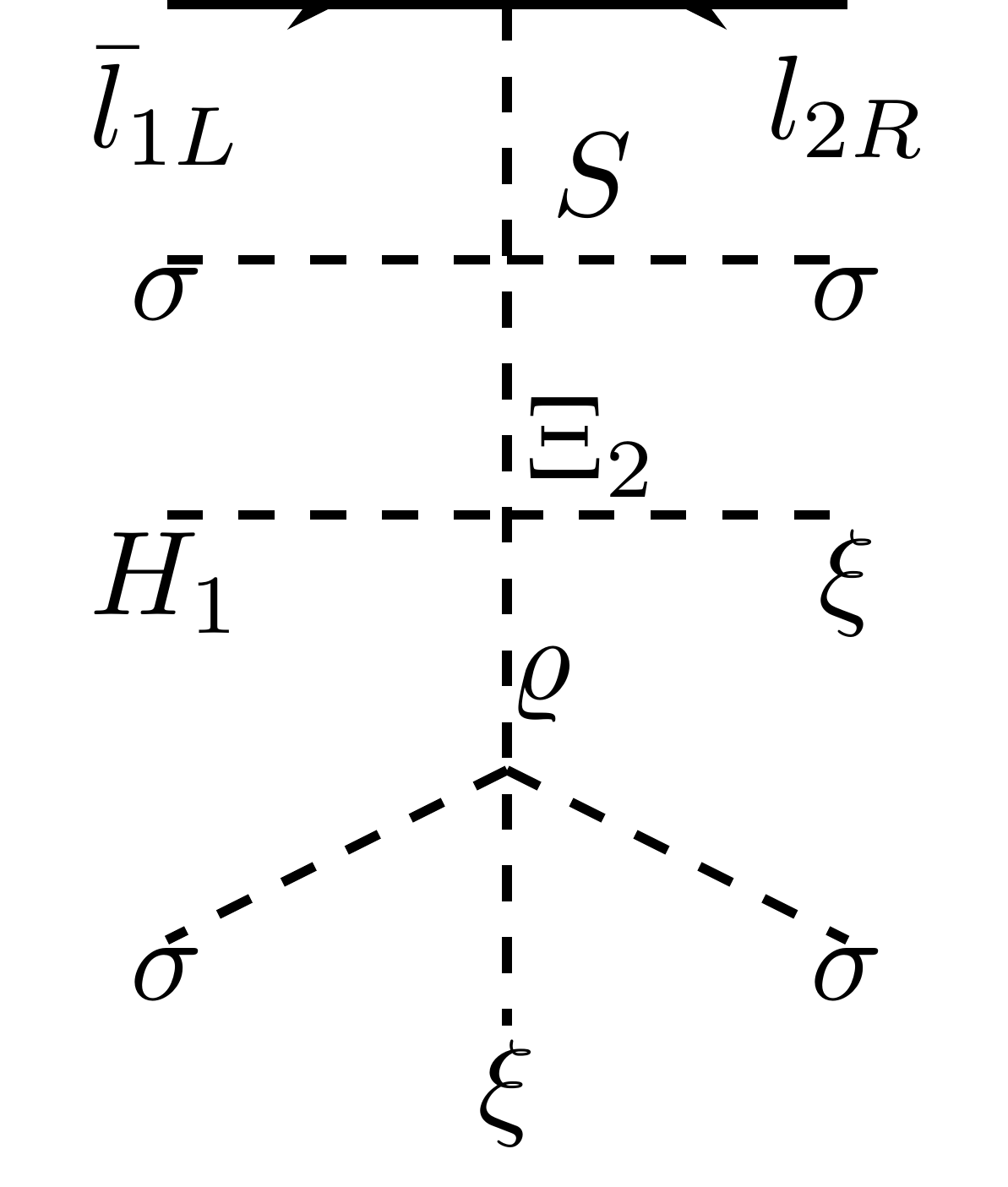}%
\includegraphics[width=0.5\textwidth]{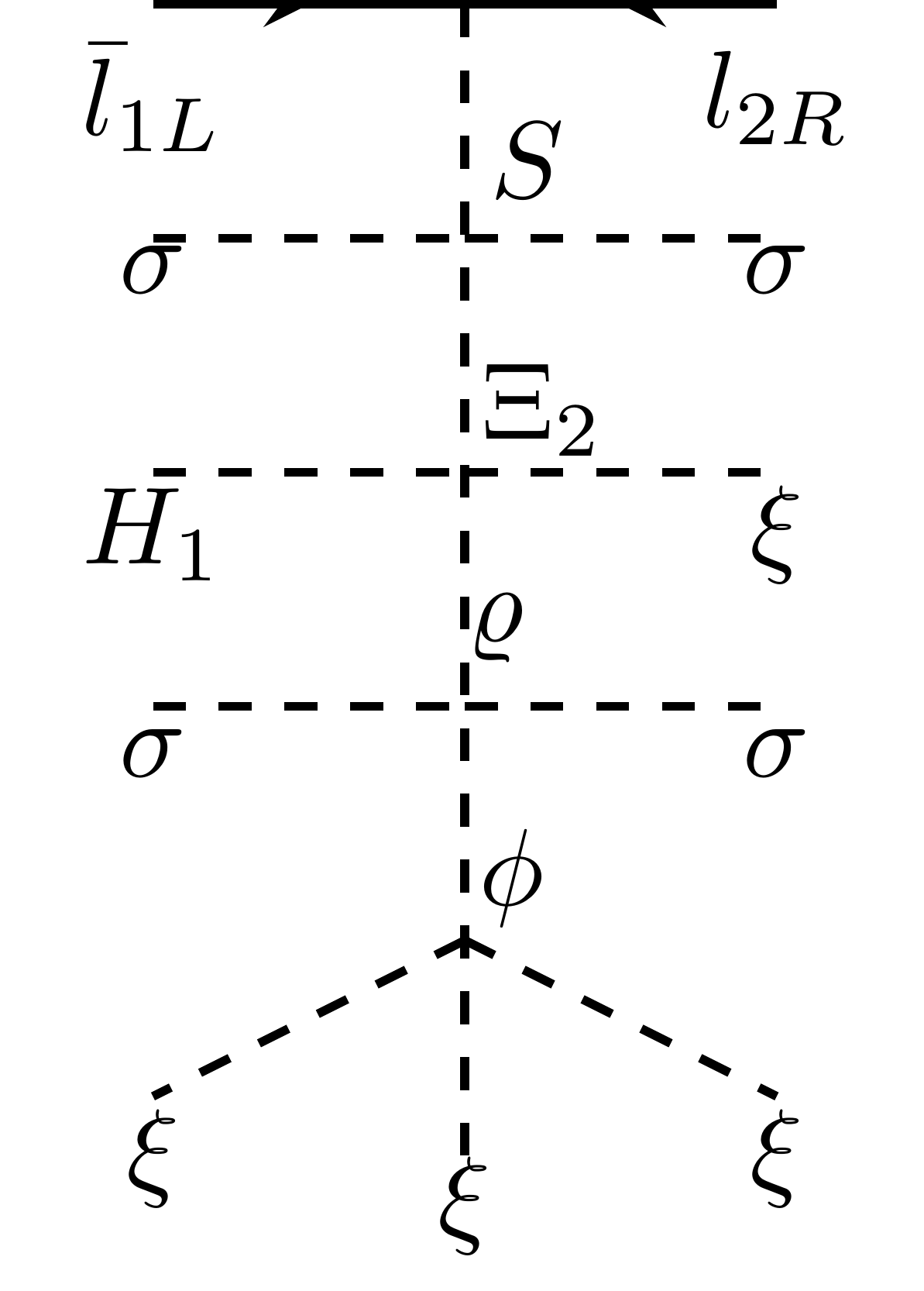}
\caption{Feynman diagrams that induce the non-renormalizable operators of
Eq. (\protect\ref{ChargedleptonOP}).}
\label{ChargedleptonUV}
\end{figure}
On the other hand, the non renormalizable charged lepton Yukawa operators: 
\begin{equation}
\overline{l}_{L}H_{1}l_{3R}\frac{\xi \sigma ^{2}}{\Lambda ^{3}},\hspace{1cm}%
\overline{l}_{L}H_{1}l_{2R}\frac{\xi \sigma ^{4}}{\Lambda ^{5}},\hspace{1cm}%
\overline{l}_{1L}H_{1}l_{2R}\frac{\left( \xi \xi \right) _{\mathbf{1}%
_{+-}}\sigma ^{4}}{\Lambda ^{6}},\hspace{1cm}\overline{l}_{1L}H_{1}l_{1R}%
\frac{\left( \xi \xi \right) _{\mathbf{1}_{+-}}\left( \xi \xi \right) _{%
\mathbf{1}_{-+}}\sigma ^{4}}{\Lambda ^{8}}  \label{ChargedleptonOP}
\end{equation}%
can be generated at low energies from the Feynman diagrams shown in Figure %
\ref{ChargedleptonUV} after integrating out the heavy scalar fields $\Xi _{1}$, $\Xi _{2}$, $\varrho $, $\phi $ and $S$ with masses of the order of the
model cutoff $\Lambda $. Here the invariance of the above given Yukawa interactions under the $Q_4$ flavor group requires that the fields $\Xi _{1}$ and $\Xi _{2}$ transform as $Q_{4}$ doublets, whereas $\phi $, $\varrho$ and $S$ as $Q_{4}$ singlets. Besides that, $\Xi _{1}$, $\Xi _{2}$ and $S$ are $SU(2)_{L}$
scalar doublets with hypercharge $\frac{1}{2}$ and $\varrho $, $\phi $ are gauge singlet scalars. Note that the symmetries of the model ensure that the only presented operators are the ones that are generated in the UV completion of the non renormalizable leptonic interactions. The analysis of the evolution of the couplings with energies requires careful and detailed studies beyond the scope of the present work.

\section{Fermion masses and mixings}

\label{fermionmixings}

\subsection{Quarks: masses and mixings}

In the standard basis, the quark mass term is given as 
\begin{equation}
\mathcal{L}=\bar{d}_{L}\mathbf{M}_{D}d_{R}+\bar{u}_{L}\mathbf{M}%
_{U}u_{R}+h.c.
\end{equation}

where the quark mass matrices can be written as 
\begin{equation}
M_{U}=\left( 
\begin{array}{ccc}
c_{1}\lambda ^{8} & 0 & a_{1}\lambda ^{4} \\ 
0 & b_{1}\lambda ^{4} & a_{2}\lambda ^{2} \\ 
0 & 0 & a_{3}%
\end{array}%
\right) \frac{v}{\sqrt{2}},\hspace{1cm}M_{D}=\left( 
\begin{array}{ccc}
e_{1}\lambda ^{8} & e_{4}\lambda ^{6} & 0 \\ 
0 & e_{2}\lambda ^{5} & 0 \\ 
0 & 0 & e_{3}\lambda ^{3}%
\end{array}%
\right) \frac{v}{\sqrt{2}}.  \label{Mq}
\end{equation}
where $a_i$ ($i=1,2,3$), $b_j$ ($j=1,2,3,4$), $c_1$ and $b_1$ are $\mathcal{O}(1)$ dimensionless parameters and $\lambda=0.225$ is the Wolfenstein parameter. The above given SM quark mass matrices can be rewritten in the form: 
\begin{equation}
M_{U}=\left( 
\begin{array}{ccc}
a_{u} & 0 & b_{u} \\ 
0 & c_{u} & d_{u} \\ 
0 & 0 & e_{u}%
\end{array}%
\right) ,\hspace{1cm} M_{D}=\left( 
\begin{array}{ccc}
a_{d} & b_{d} & 0 \\ 
0 & c_{d} & 0 \\ 
0 & 0 & e_{d}%
\end{array}%
\right).
\end{equation}
In here, the coefficients may be read of Eq. (\ref{Mq}). Then, the quark
mass matrices are diagonalized by the mixing matrices $\mathbf{U}_{f(L, R)}$
where $f=u,d$. Explicitly, we have $\mathbf{U}^{\dagger}_{f L}\mathbf{M}_{f}%
\mathbf{U}_{f R}=\mathbf{\hat{M}}_{f}$ where $\mathbf{\hat{M}}_{f}=\text{%
Diag.}\left(m_{f_{1}}, m_{f_{2}}, m_{f_{3}} \right)$ contains the physical
quark masses.

As it is well known, the CKM mixing matrix is given by $\mathbf{V}_{CKM}=%
\mathbf{U}^{\dagger}_{u L} \mathbf{U}_{d L}$, then we will obtain the
left-handed mixing matrix that takes place in the CKM matrix. Therefore, we
have to build the bilineal forms $\mathbf{U}^{\dagger}_{f L}\mathbf{M}_{f}%
\mathbf{M}^{\dagger}_{f} \mathbf{U}_{f L}=\mathbf{\hat{M}}_{f}\mathbf{\hat{M}%
}^{\dagger}_{f}$ so that let us start with the down sector. First of all, we
factorize the CP violating phases that come from $\mathbf{M}_{d}\mathbf{M}%
^{\dagger}_{d}$, this is, $\mathbf{M}_{d}\mathbf{M}^{\dagger}_{d}=\mathbf{P}%
_{d}\mathbf{m}_{d}\mathbf{m}^{\dagger}_{d}\mathbf{P}^{\dagger}_{d}$ where $\mathbf{P}_{d}=\text{Diag.}\left(1, e^{i\eta_{d}}, 1 \right)$ with 
\begin{eqnarray}
\eta_{d}=\alpha_{c_{d}}-\alpha_{b_{d}}, \quad
\alpha_{b_{d}}=arg(b_{d}),\quad \alpha_{c_{d}}=arg(c_{d})
\end{eqnarray}

In addition, we have
\begin{equation}
\mathbf{m}_{d}\mathbf{m}_{d}^{\dagger }=%
\begin{pmatrix}
|a_{d}|^{2}+|b_{d}|^{2} & |b_{d}||c_{d}| & 0 \\ 
|b_{d}||c_{d}| & |c_{d}|^{2} & 0 \\ 
0 & 0 & |e_{d}|^{2}%
\end{pmatrix}%
.
\end{equation}%
Three free parameters can be fixed in terms of the physical masses and one
unfixed parameter, explicitly, these are given as 
\begin{equation}
|a_{d}|=\sqrt{\frac{|m_{s}|^{2}+|m_{d}|^{2}-|b_{d}|^{2}-R_{d}}{2}},\quad
|c_{d}|=\sqrt{\frac{|m_{s}|^{2}+|m_{d}|^{2}-|b_{d}|^{2}+R_{d}}{2}},\quad
|e_{d}|=|m_{b}|;
\end{equation}%
where $R_{d}=\sqrt{\left( |m_{s}|^{2}+|m_{d}|^{2}-|b_{d}|^{2}\right)
^{2}-4|m_{s}|^{2}|m_{d}|^{2}}$. According to the parametrization, there is a
hierarchy among the free parameters, this is, $%
|e_{d}|>|m_{s}|>|c_{d}|>|b_{d}|>|m_{d}|>|a_{d}|>0$

Having done that, one can choose appropriately the left-handed mixing matrix 
$\mathbf{U}_{dL}=\mathbf{P}_{d} \mathbf{O}_{dL}$. In here, $\mathbf{O}_{dL}$
is an orthogonal real matrix that diagonalizes $\mathbf{m}_{d}\mathbf{m}%
^{\dagger}_{d}$. 
\begin{equation}
\mathbf{O}_{dL}= 
\begin{pmatrix}
\cos{\theta_{d}} & \sin{\theta_{d}} & 0 \\ 
-\sin{\theta_{d}} & \cos{\theta_{d}} & 0 \\ 
0 & 0 & 1%
\end{pmatrix}%
\end{equation}
with $\cos{\theta_{d}}=\sqrt{\frac{\vert m_{s}\vert^{2}-\vert
m_{d}\vert^{2}-\vert b_{d}\vert^{2}+R_{d}}{2\left(\vert m_{s}\vert^{2}-\vert
m_{d}\vert^{2}\right)}}$ and $\sin{\theta_{d}}=\sqrt{\frac{\vert
m_{s}\vert^{2}-\vert m_{d}\vert^{2}+\vert b_{d}\vert^{2}-R_{d}}{2\left(\vert
m_{s}\vert^{2}-\vert m_{d}\vert^{2}\right)}}$.

In similar way, for the up sector, we have to factorize the CP violating
phases that come from $\mathbf{M}_{u}\mathbf{M}^{\dagger}_{u}$ so that $%
\mathbf{M}_{u}\mathbf{M}^{\dagger}_{u}=\mathbf{P}_{u}\mathbf{m}_{u}\mathbf{m}%
^{\dagger}_{u}\mathbf{P}^{\dagger}_{u}$ where $\mathbf{P}_{u}=\text{Diag.}%
\left(1, e^{i\eta_{c}}, e^{i\eta_{t}} \right)$. The phases are given as 
\begin{equation}
\eta_{c}=\alpha_{d_{u}}-\alpha_{b_{u}},\quad
\eta_{t}=\alpha_{e_{u}}-\alpha_{b_{u}}
\end{equation}
with $\alpha_{b_{u}}=arg(b_{u})$, $\alpha_{d_{u}}=arg(d_{u})$ and $%
\alpha_{e_{u}}=arg(e_{u})$. At the same time, we have the real symmetric
matrix 
\begin{equation}
\mathbf{m}_{u}\mathbf{m}^{\dagger}_{u}=%
\begin{pmatrix}
\vert a_{u}\vert^{2}+\vert b_{u}\vert^{2} & \vert b_{u}\vert \vert d_{u}\vert
& \vert b_{u}\vert \vert e_{u}\vert \\ 
\vert b_{u}\vert \vert d_{u}\vert & \vert c_{u}\vert^{2}+\vert d_{u}\vert^{2}
& \vert d_{u}\vert \vert e_{u}\vert \\ 
\vert b_{u}\vert \vert e_{u}\vert & \vert d_{u}\vert \vert e_{u}\vert & 
\vert e_{u}\vert^{2}%
\end{pmatrix}%
,
\end{equation}
which has five free parameters. Three of them can be fixed in terms of the
physical masses, $\vert e_{u}\vert$ and $\vert a_{u}\vert$. Then, the fixed
parameters are written as 
\begin{eqnarray}
\vert b_{u}\vert=\sqrt{\frac{\vert e_{u}\vert^{2} \mathcal{N}_{1}\mathcal{N}%
_{2}\mathcal{N}_{3}}{\mathcal{K}}},\quad \vert c_{u}\vert=\frac{\vert
m_{t}\vert \vert m_{c}\vert \vert m_{u}\vert}{\vert e_{u}\vert \vert
a_{u}\vert},\quad \vert d_{u}\vert= \sqrt{\frac{\mathcal{M}_{1} \mathcal{M}%
_{2} \mathcal{M}_{3} }{\vert e_{u}\vert^{2}\vert a_{u}\vert^{2}\mathcal{K}}},
\end{eqnarray}
where 
\begin{eqnarray}
\mathcal{N}_{1}&=&\vert a_{u}\vert^{2}-\vert m_{u}\vert^{2},\qquad \mathcal{N%
}_{2}=\vert m_{c}\vert^{2}-\vert a_{u}\vert^{2},\qquad \mathcal{N}_{3}=\vert
m_{t}\vert^{2}-\vert a_{u}\vert^{2},  \notag \\
\mathcal{M}_{1}&=&\vert m_{t}\vert^{2}\vert m_{u}\vert^{2}-\vert
e_{u}\vert^{2}\vert a_{u}\vert^{2},\quad \mathcal{M}_{2}=\vert
m_{t}\vert^{2}\vert m_{c}\vert^{2}-\vert e_{u}\vert^{2}\vert a_{u}\vert^{2},
\notag \\
\mathcal{M}_{3}&=&\vert e_{u}\vert^{2}\vert a_{u}\vert^{2}-\vert
m_{c}\vert^{2}\vert m_{u}\vert^{2},\quad \mathcal{K}=\vert
m_{t}\vert^{2}\vert m_{c}\vert^{2}\vert m_{u}\vert^{2}-\vert
e_{u}\vert^{2}\vert a_{u}\vert^{4}.
\end{eqnarray}

With this parametrization, the hierarchy among the free parameters is $\vert
m_{t}\vert>\vert e_{u}\vert>\vert d_{u}\vert>\vert c_{u}\vert>\vert
m_{c}\vert>\vert b_{u}\vert>\vert a_{u}\vert>\vert m_{u}\vert$. So that, the
left-handed matrix is well determined as $\mathbf{U}_{uL}=\mathbf{P}_{u}%
\mathbf{O}_{uL}$ where the latter matrix diagonalizes the real symmetric
matrix, $\mathbf{m}_{u}\mathbf{m}^{\dagger}_{u}$. Explicitly, this is given
by

\begin{equation}
\mathbf{O}_{uL}=%
\begin{pmatrix}
-\frac{\vert m_{u}\vert}{\vert a_{u}\vert}\sqrt{\frac{\vert m_{u}\vert^{2}%
\mathcal{N}_{2}\mathcal{N}_{3} \mathcal{M}_{2}}{\mathcal{D}_{1}}} & -\frac{%
\vert m_{c}\vert}{\vert a_{u}\vert}\sqrt{\frac{\vert m_{c}\vert^{2}\mathcal{N%
}_{1}\mathcal{N}_{3} \mathcal{M}_{1}}{\mathcal{D}_{2}}} & \frac{\vert
m_{t}\vert}{\vert a_{u}\vert}\sqrt{\frac{\vert m_{t}\vert^{2}\mathcal{N}_{1}%
\mathcal{N}_{2} \mathcal{M}_{3}}{\mathcal{D}_{3}}} \\ 
-\sqrt{\frac{\mathcal{N}_{1} \mathcal{M}_{1}\mathcal{M}_{3}}{\mathcal{D}_{1}}%
} & \sqrt{\frac{ \mathcal{N}_{2} \mathcal{M}_{2}\mathcal{M}_{3}}{\mathcal{D}%
_{2}}} & \sqrt{\frac{ \mathcal{N}_{3} \mathcal{M}_{1}\mathcal{M}_{2}}{%
\mathcal{D}_{3}}} \\ 
\frac{1}{\vert a_{u}\vert}\sqrt{\frac{\mathcal{N}_{1} \mathcal{M}_{2}%
\mathcal{K}}{\mathcal{D}_{1}}} & -\frac{1}{\vert a_{u}\vert}\sqrt{\frac{%
\mathcal{N}_{2}\mathcal{M}_{1}\mathcal{K}}{\mathcal{D}_{2}}} & \frac{1}{%
\vert a_{u}\vert}\sqrt{\frac{\mathcal{N}_{3}\mathcal{M}_{3}\mathcal{K}}{%
\mathcal{D}_{3}}}%
\end{pmatrix}%
,
\end{equation}
with 
\begin{eqnarray}
\mathcal{D}_{1}&=&\left(\vert m_{t}\vert^{2}-\vert
m_{u}\vert^{2}\right)\left(\vert m_{c}\vert^{2}-\vert m_{u}\vert^{2}\right)%
\mathcal{K},  \notag \\
\mathcal{D}_{2}&=&\left(\vert m_{t}\vert^{2}-\vert
m_{c}\vert^{2}\right)\left(\vert m_{c}\vert^{2}-\vert m_{u}\vert^{2}\right)%
\mathcal{K},  \notag \\
\mathcal{D}_{3}&=&\left(\vert m_{t}\vert^{2}-\vert
m_{c}\vert^{2}\right)\left(\vert m_{t}\vert^{2}-\vert m_{u}\vert^{2}\right)%
\mathcal{K}.
\end{eqnarray}


\begin{figure}[]
\centering
\includegraphics[width=0.5\textwidth]{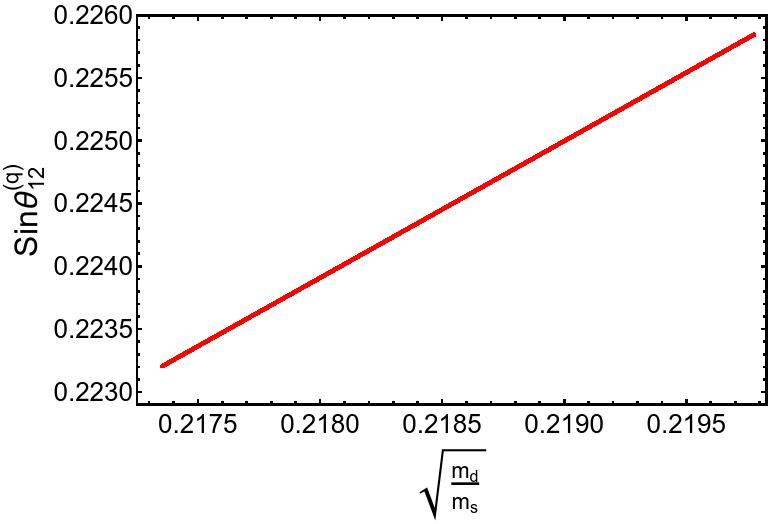}\includegraphics[width=0.5%
\textwidth]{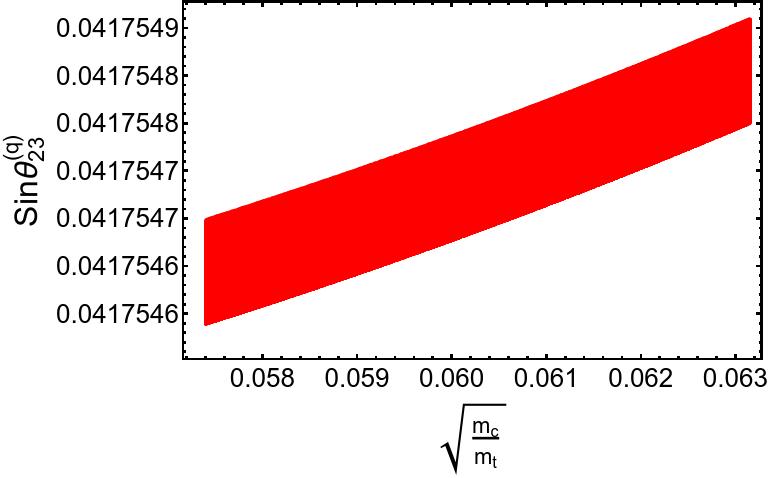}\newline
\includegraphics[width=0.5\textwidth]{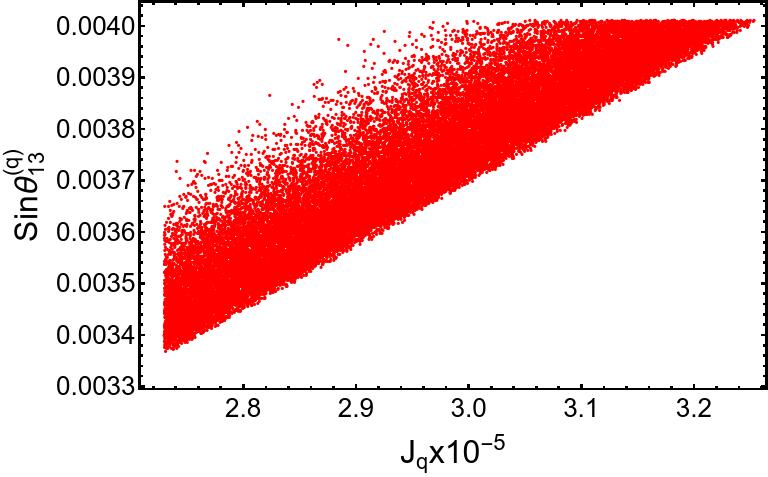}\includegraphics[width=0.5\textwidth]{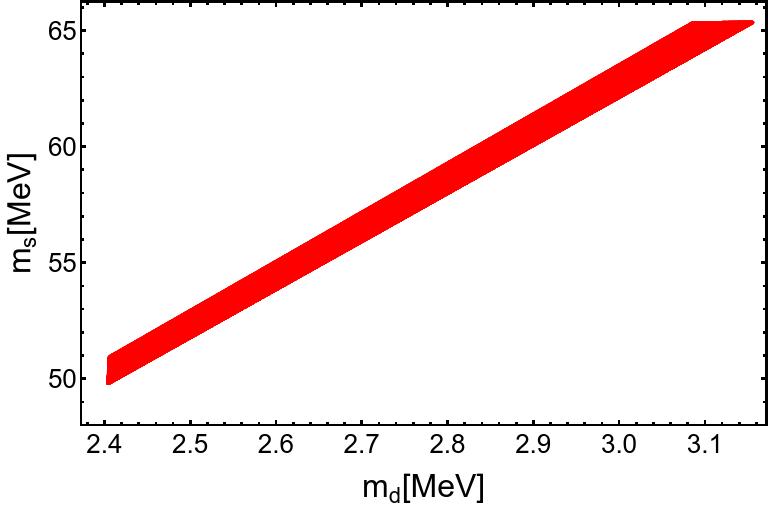}
\caption{Correlation between the different quark sector observables.}
\label{fig:correlationquarksector}
\end{figure}

Therefore, the CKM mixing matrix is written as
$\mathbf{V}_{CKM}=\mathbf{O}_{uL}^{T}\mathbf{\bar{P}}_{q}\mathbf{O}_{dL}$
where $\mathbf{\bar{P}}_{q}=\mathbf{P}_{u}^{\dagger
}\mathbf{P}_{d}=\text{Diag.}\left( 1,e^{-i\bar{\eta}_{c}},e^{-i\eta
_{t}}\right)$ with $\bar{\eta}_{c}=\eta _{c}-\eta _{d}$. In summary, the
theoretical CKM matrix depends of five parameters namely: $\vert
b_{d}\vert$, $\vert a_{u}\vert$, $\vert e_{u}\vert$ and two CP phases. From these five parameters, four of them have to be numerically determined with high precision in order to get values for the CKM parameters consistent with the experimental data, as follows from our numerical analysis. Only the phase $\eta_d$ can be varied in larger range when searching for the best point that reproduces the observed CKM mixing.

Remarkably, the $\eta _{t}$ phase is irrelevant for the magnitude of each entry of the third column. This facts will allow to reduce the free parameters so that four of them can be fitted. Before starting an $\chi^{2}$ analysis, we show that in this model, the extended Gatto-Sartori-Tonin
relations, are consequence of the hierarchical structure of the quark mass matrices, resulting from the symmetries of the model, which imply  
$|b_{d}|^{2}=|m_{s}||m_{d}|+|m_{d}|^{2}$. 
As result, one
gets 
\begin{equation}
\cos {\theta _{d}}\approx 1-\frac{1}{2}\frac{|m_{d}|}{|m_{s}|},\qquad \sin {\theta _{d}}\approx \sqrt{\frac{|m_{d}|}{|m_{s}|}}.
\end{equation}

Regarding, the up sector, the hierarchical structure of the up quark mass matrix implies that the free parameters must be $|a_{u}|\approx
|m_{u}|+|\delta _{a_{u}}|$ and $|e_{u}|^{2}\approx
|m_{t}|^{2}-|m_{t}||m_{c}| $ with $|\delta _{a_{u}}|\ll |m_{u}|$. In this
way, we respect the hierarchy among the free parameters. Therefore, after a
lengthy task, the following relations are obtained 
\begin{eqnarray}
(\mathbf{V}_{CKM})_{us}&\approx& -\sqrt{\frac{|m_{d}|}{|m_{s}|}},\nn\\ (\mathbf{V}_{CKM})_{cb}&\approx& -\sqrt{\frac{|m_{c}|}{|m_{t}|}}\left[ 1-\frac{|m_{t}|}{|m_{c}|}\frac{|\delta
_{a_{u}}|}{|m_{u}|}\right] e^{-i\eta _{t}},\nn\\ (\mathbf{V}_{CKM})_{td}&\approx& -\sqrt{\frac{|m_{c}|}{|m_{t}|}}\sqrt{\frac{|m_{d}|}{|m_{s}|}}\left[ 1-\frac{|m_{t}|}{|m_{c}|}\frac{|\delta _{a_{u}}|}{|m_{u}|}\right] e^{-i\bar{\eta}
_{c}}.
\end{eqnarray} After the above given analytical analysis of the quark
spectrum and CKM mixing matrix we carry out a numerical analysis. From this
analysis we find that the experimental values of the quark mass spectrum and
CKM parameters can be very well reproduced from the following benchmark
point: 
\begin{equation}
\begin{array}{c}
c_{1}\simeq 0.906\,,\hspace*{1cm}b_{1}\simeq 1.435\,,\hspace*{1cm}%
a_{3}\simeq 0.990\,, \\ 
\left\vert a_{1}\right\vert \simeq 1.359\,,\hspace*{1cm}\arg (a_{1})\simeq
105.14^{\circ }\,,\hspace*{1cm}a_{2}\simeq 0.824\,, \\ 
e_{1}\simeq 2.489\,,\hspace*{1cm}e_{2}\simeq 0.536\,,\hspace*{1cm}%
e_{3}\,\simeq 1.463,\hspace*{1cm}e_{4}\simeq 0.549\,.%
\end{array}
\label{eq:bm-values}
\end{equation}%
An important feature of the above result is that the absolute values of all
these parameters are of the order of unity. Thus, the symmetries of our
model allow us to naturally explain the hierarchy of quark mass spectrum and
quark mixing angles without appreciable tuning of these effective
parameters. 
\begin{table}[tbh]
\begin{center}
\begin{tabular}{c|l|l}
\hline\hline
Observable & Model value & Experimental value \\ \hline
$m_{u}(m_Z)$ & \quad $1.02$ & \quad $1.24\pm 0.22$ \\ \hline
$m_{c}(m_Z)$ & \quad $0.63$ & \quad $0.63\pm 0.02$ \\ \hline
$m_{t}(m_Z)$ & \quad $172.3$ & \quad $172.9\pm 0.4$ \\ \hline
$m_{d}(m_Z)$ & \quad $2.72$ & \quad $2.69\pm 0.19$ \\ \hline
$m_{s}(m_Z)$ & \quad $54.6$ & \quad $53.5\pm 4.6$ \\ \hline
$m_{b}(m_Z)$ & \quad $2.88$ & \quad $2.86\pm 0.03$ \\ \hline
$\sin \theta _{12}$ & \quad $0.2248$ & \quad $0.2245\pm 0.00044$ \\ \hline
$\sin \theta _{23}$ & \quad $0.0419$ & \quad $0.0421\pm 0.00076$ \\ \hline
$\sin \theta _{13}$ & \quad $0.00349$ & \quad $0.00365\pm 0.00012$ \\ \hline
$J_q$ & \quad $3.09\times 10^{-5}$ & \quad $\left(3.18\pm 0.15\right)\times
10^{-5}$ \\ \hline
\end{tabular}%
\end{center}
\caption{Model and experimental values of the quark masses and CKM
parameters.}
\label{QuarkObs}
\end{table}
The result given in Eq. \eqref{eq:bm-values} motivates to consider the
simplified benchmark scenario: 
\begin{eqnarray}
\begin{gathered} c_{1}=a_3=1,\hspace{0.6cm}b_1\simeq
1.451\hspace{0.6cm}\left\vert a_{1}\right\vert\simeq 1.434,
\hspace{0.6cm}\arg \left( a_{1}\right) \simeq -90^{\circ },\hspace{0.6cm}
a_{2}\simeq 0.830,\hspace{0.6cm} \\ e_1\simeq
2.867,\hspace{0.6cm}e_{2}=e_4\simeq 0.465,\hspace{0.6cm}e_{3}\simeq 1.451
\label{eq:Quark-benchmark-point2} \end{gathered}
\end{eqnarray}
As seen from Table \ref{QuarkObsbench}, the 10 quark observables are
reproduced with a good precision in the above given 7-parameter scenario. 
\begin{table}[tbh]
\begin{center}
\begin{tabular}{c|l|l}
\hline\hline
Observable & Model value & Experimental value \\ \hline
$m_{u}(m_Z)$ & \quad $1.12$ & \quad $1.24\pm 0.22$ \\ \hline
$m_{c}(m_Z)$ & \quad $0.64$ & \quad $0.63\pm 0.02$ \\ \hline
$m_{t}(m_Z)$ & \quad $174.1$ & \quad $172.9\pm 0.4$ \\ \hline
$m_{d}(m_Z)$ & \quad $3.14$ & \quad $2.69\pm 0.19$ \\ \hline
$m_{s}(m_Z)$ & \quad $47.3$ & \quad $53.5\pm 4.6$ \\ \hline
$m_{b}(m_Z)$ & \quad $2.86$ & \quad $2.86\pm 0.03$ \\ \hline
$\sin \theta _{12}$ & \quad $0.22$ & \quad $0.2245\pm 0.00044$ \\ \hline
$\sin \theta _{23}$ & \quad $0.0418$ & \quad $0.0421\pm 0.00076$ \\ \hline
$\sin \theta _{13}$ & \quad $0.00364$ & \quad $0.00365\pm 0.00012$ \\ \hline
$J_q$ & \quad $3.27\times 10^{-5}$ & \quad $\left(3.18\pm 0.15\right)\times
10^{-5}$ \\ \hline
\end{tabular}%
\end{center}
\caption{Model and experimental values of the quark masses and CKM
parameters for the simplified benchmark scenario given in Eq. (\protect\ref%
{eq:Quark-benchmark-point2}).}
\label{QuarkObsbench}
\end{table}

\begin{figure}[th]
\includegraphics[width=0.9\textwidth]{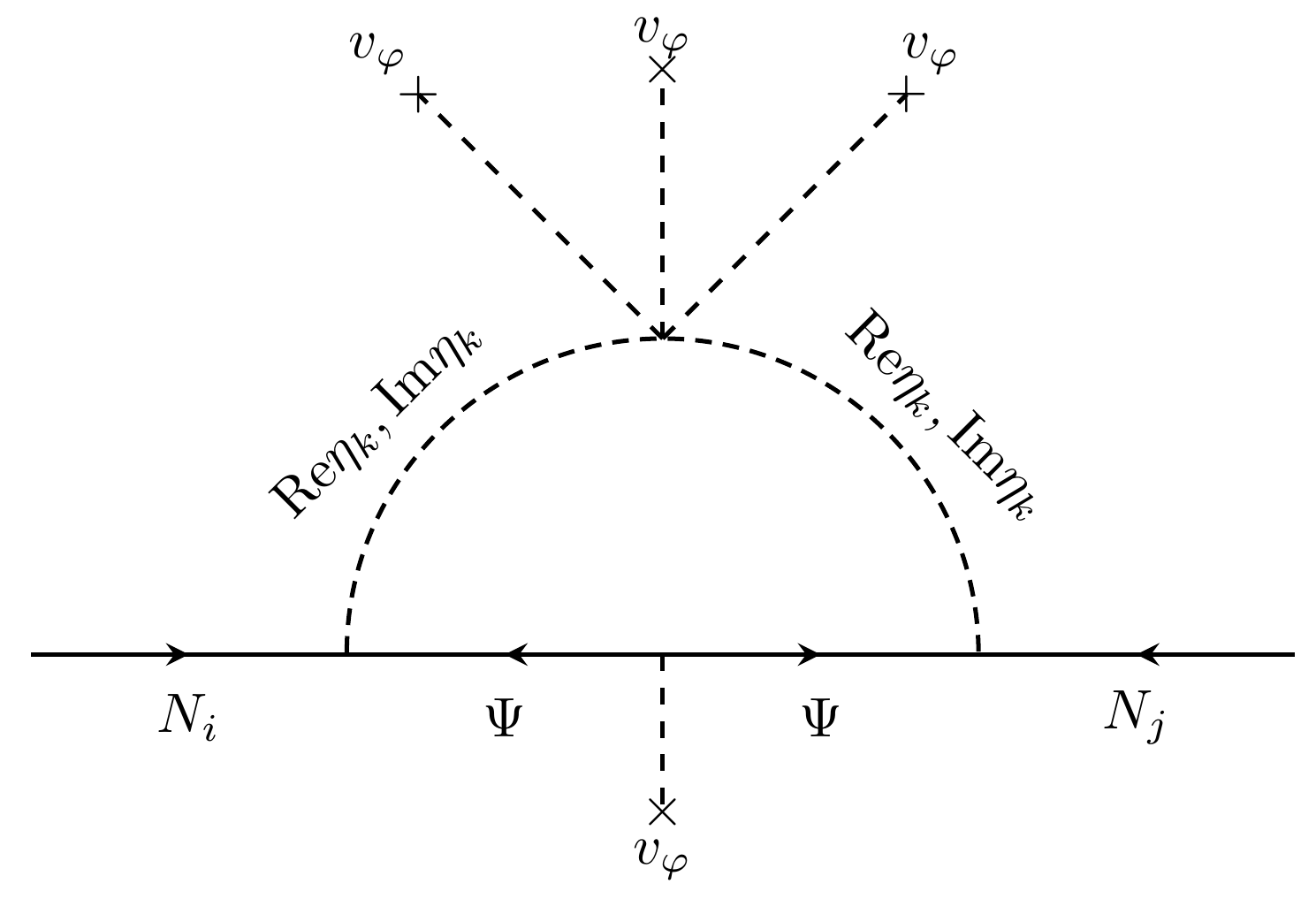} 
\caption{One-loop Feynman diagram contributing to the Majorana neutrino mass
submatrix $\protect\mu $. Here $i,j=1,2,3$ and $k=1,2$.}
\label{Loopdiagramsmu}
\end{figure}

To close this section, we briefly discuss the implications of our model in
Flavour Changing Neutral Currents (FCNC). As seen from the charged fermion
Yukawa terms of Eqs. (\ref{Lyu}), (\ref{Lyd}), (\ref{Lyl}), there is only
one Higgs doublet appearing in the charged lepton and down type quark Yukawa
interactions, thus implying the absence of FCNC at tree level, as follows
from the Weinberg-Glasgow-Pascos theorem. Consequently, we expect similar
predictions for the $K^{0}-\bar{K}^{0}$, $B_{d}^{0}-\bar{B}_{d}^{0}$ and $%
B_{s}^{0}-\bar{B}_{s}^{0}$ meson mixings as in the Standard Model. On the
other hand, there are two Higgs doublets in the up type quark Yukawa
interactions, thus implying the appearance of tree level FCNC in the up type
quark sector that will yield a tree level contribution mediated by neutral
scalars to the $D^{0}-\bar{D}^{0}$ meson oscillation. However, we expect
that the strong hierarchical structure in the Yukawa couplings of the
neutral scalars with up type quarks together with the very small mixing
between the first and second family of up type quarks, will provide a strong
suppression for the tree level contribution to the $D^{0}-\bar{D}^{0}$ meson
mixing.

\subsection{Lepton masses and mixings}

From the charged lepton Yukawa interactions, we find that the SM charged
lepton mass matrix reads: 
\begin{equation}
M_{l}=\left( 
\begin{array}{ccc}
f_{1}\lambda ^{8} & f_{4}\lambda ^{6} & 0 \\ 
0 & f_{2}\lambda ^{5} & 0 \\ 
0 & 0 & f_{3}\lambda ^{3}%
\end{array}%
\right) \frac{v}{\sqrt{2}}.
\end{equation}
The above given charged lepton mass matrix can be rewritten as:

\begin{equation}
M_{e}=\left( 
\begin{array}{ccc}
a_{e} & b_{e} & 0 \\ 
0 & c_{e} & 0 \\ 
0 & 0 & f_{e}%
\end{array}%
\right).
\end{equation}
Then, in similar way to the down quark sector, three free parameters may be
fixed in terms of the physical masses and the unfixed parameter, $\vert
b_{e}\vert$. This is 
\begin{eqnarray}
\vert a_{e}\vert=\sqrt{\frac{\vert m_{\mu}\vert^{2}+ \vert
m_{e}\vert^{2}-\vert b_{e}\vert^{2}-R_{e}}{2}}\quad \vert c_{e}\vert=\sqrt{%
\frac{\vert m_{\mu}\vert^{2}+ \vert m_{e}\vert^{2}-\vert b_{e}\vert^{2}+R_{e}%
}{2}},\quad \vert f_{e}\vert= \vert m_{\tau}\vert,
\end{eqnarray}
where $R_{e}=\sqrt{\left(\vert m_{\mu}\vert^{2}+ \vert m_{e}\vert^{2}-\vert
b_{e}\vert^{2}\right)^{2}-4\vert m_{\mu}\vert^{2}\vert m_{e}\vert^{2}}$. In
this case, the free parameters satisfy the following ordering $\vert
f_{e}\vert>\vert m_{\mu}\vert>\vert c_{e}\vert>\vert b_{e}\vert>\vert
m_{e}\vert>\vert a_{e}\vert>0$

Along with this, left-handed matrix that takes places in the PMNS mixing
matrix is given by $\mathbf{U}_{eL}=\mathbf{P}_{e}\mathbf{O}_{eL}$ with

\begin{equation}
\mathbf{O}_{eL}= 
\begin{pmatrix}
\cos{\theta_{e}} & \sin{\theta_{e}} & 0 \\ 
-\sin{\theta_{e}} & \cos{\theta_{e}} & 0 \\ 
0 & 0 & 1%
\end{pmatrix}%
,\quad \mathbf{P}_{e}=%
\begin{pmatrix}
e^{i\eta_{e}} & 0 & 0 \\ 
0 & 1 & 0 \\ 
0 & 0 & 1%
\end{pmatrix}%
\end{equation}

with $\cos {\theta _{e}}=\sqrt{\frac{|m_{\mu
}|^{2}-|m_{e}|^{2}-|b_{e}|^{2}+R_{e}}{2\left( |m_{\mu
}|^{2}-|m_{e}|^{2}\right) }}$ and $\sin {\theta _{e}}=\sqrt{\frac{|m_{\mu
}|^{2}-|m_{e}|^{2}+|b_{e}|^{2}-R_{e}}{2\left( |m_{\mu
}|^{2}-|m_{e}|^{2}\right) }}$; $\eta _{e}=\alpha _{b_{e}}-\alpha _{c_{e}}$
where $\alpha _{b_{e}}=arg(b_{e})$ and $\alpha _{c_{e}}=arg(c_{e})$.

Regarding the neutrino sector, from the Eq. (\ref{Lynu}), we find the
following neutrino mass terms: 
\begin{equation}
-\mathcal{L}_{mass}^{\left( \nu \right) }=\frac{1}{2}\left( 
\begin{array}{ccc}
\overline{\nu _{L}^{C}} & \overline{\nu _{R}} & \overline{N_{R}}%
\end{array}%
\right) M_{\nu }\left( 
\begin{array}{c}
\nu _{L} \\ 
\nu _{R}^{C} \\ 
N_{R}^{C}%
\end{array}%
\right) +H.c,  \label{Lnu}
\end{equation}%
where the neutrino mass matrix is given by: 
\begin{equation}
M_{\nu }=\left( 
\begin{array}{ccc}
0_{3\times 3} & m_{\nu D} & 0_{3\times 3} \\ 
m_{\nu D}^{T} & \varepsilon _{3\times 3} & M \\ 
0_{3\times 3} & M^{T} & \mu 
\end{array}%
\right) ,
\end{equation}%
and the submatrices are given by: 
\begin{eqnarray}
m_{\nu _{D}} &=&\left( 
\begin{array}{ccc}
y_{1}^{\left( \nu \right) }\frac{v_{H_{2}}}{\sqrt{2}} & 0 & 0 \\ 
0 & y_{2}^{\left( \nu \right) }\frac{v_{H_{2}}}{\sqrt{2}} & 0 \\ 
0 & 0 & y_{3}^{\left( \nu \right) }\frac{v_{H_{2}}}{\sqrt{2}}%
\end{array}%
\right) ,\hspace{1cm}M=\left( 
\begin{array}{ccc}
M_{1} & 0 & 0 \\ 
0 & y_{2}v_{\xi } & 0 \\ 
0 & 0 & y_{3}v_{\xi }%
\end{array}%
\right) ,\hspace{1cm}\varepsilon =\left( 
\begin{array}{ccc}
0 & M_{sb} & 0 \\ 
M_{sb} & 0 & 0 \\ 
0 & 0 & 0%
\end{array}%
\right) ,  \notag \\
\mu  &\simeq &\frac{y_{\Psi }\left( m_{R}^{2}-m_{I}^{2}\right) v_{\varphi
}^{3}}{8\pi ^{2}\left( m_{R}^{2}+m_{I}^{2}\right) \Lambda ^{2}}\left( 
\begin{array}{ccc}
y_{1N}^{2} & y_{1N}y_{2N}\frac{v_{\Phi }}{v_{\varphi }}e^{-i\theta } & 
y_{1N}y_{2N}\frac{v_{\Phi }}{v_{\varphi }}e^{i\theta } \\ 
y_{1N}y_{2N}\frac{v_{\Phi }}{v_{\varphi }}e^{-i\theta } & \left(
y_{2N}^{2}e^{-2i\theta }+y_{3N}^{2}e^{2i\theta }\right) \frac{v_{\Phi }^{2}}{%
v_{\varphi }^{2}} & \left( y_{2N}^{2}+y_{3N}^{2}\right) \frac{v_{\Phi }^{2}}{%
v_{\varphi }^{2}} \\ 
y_{1N}y_{2N}\frac{v_{\Phi }}{v_{\varphi }}e^{i\theta } & \left(
y_{2N}^{2}+y_{3N}^{2}\right) \frac{v_{\Phi }^{2}}{v_{\varphi }^{2}} & \left(
y_{2N}^{2}e^{2i\theta }+y_{3N}^{2}e^{-2i\theta }\right) \frac{v_{\Phi }^{2}}{%
v_{\varphi }^{2}}%
\end{array}%
\right) ,  \label{MR}
\end{eqnarray}%
where $m_{R}=m_{\func{Re}\eta _{k}}$, $m_{I}=m_{\func{Im}\eta _{k}}$ ($k=1,2$%
) and\ for the sake of simplicity, we have assumed that the singlet scalar
fields $\eta _{k}$\ are physical fields degenerate in mass and heavier than
the right-handed Majorana neutrino $\Psi $, thus allowing to consider the
scenario 
\begin{equation}
m_{R}^{2},m_{I}^{2}\gg m_{\Psi }^{2}.
\end{equation}%
Furthermore, we have assumed $m_{R}^{2}-m_{I}^{2}\ll m_{R}^{2}+m_{I}^{2}$ as
well as $M_{sb}<<246$ GeV. The $\mu $ block is generated at one loop level
due to the exchange of $\Psi $, $\func{Re}\eta $ and $\func{Im}\eta $ in the
internal lines, as shown in figure \ref{Loopdiagramsmu}. To close the
corresponding one loop diagram, the following non renormalizable scalar
interactions are needed: 
\begin{equation}
\frac{\kappa _{k}}{\Lambda ^{6}}\left( \eta _{k}^{\ast }\right) ^{2}\left(
\Phi \Phi \right) _{\mathbf{1}_{-+}}\varphi \left( \sigma ^{\ast }\right)
^{2}\rho ^{3},\hspace{1cm}\hspace{1cm}k=1,2
\label{NRI}
\end{equation}%
such interaction generates a small splitting between the masses $m_{R}$ and $%
m_{I}$, which is crucial to produce the tiny masses for the light active
neutrinos. Taking into account the VEV hierarchy given in Eq. (%
\ref{VEVsinglets}), this small splitting can be estimated as follows: 
\begin{equation}
\Delta m_{\eta _{k}}\sim \sqrt{\frac{\kappa _{k}}{\Lambda ^{6}}v_{\rho
}^{3}v_{\sigma }^{2}v_{\Phi }^{2}v_{\varphi }}\sim \sqrt{\kappa _{k}\lambda
^{5}\frac{v_{\varphi }}{\Lambda }}v_{\Phi }\sim \mathcal{O}(10-100)\text{GeV}%
,\hspace{1cm}{\kappa _{k}}\sim \mathcal{O}(1-10)\hspace{1cm}k=1,2
\end{equation}%
where the exact values depend on the specific magnitudes of the VEVs of the
scalar singlets as well as of the quartic scalar couplings $\kappa
_{k}$ ($k=1,2$). These quartic scalar couplings $\kappa _{k}$ can take
values up to their upper perturbativity bound of $4\pi $. 

The light active masses arise from an inverse seesaw mechanism and the
physical neutrino mass matrices are: 
\begin{eqnarray}
\widetilde{\mathbf{M}}_{\nu } &=&m_{\nu D}\left( M^{T}\right) ^{-1}\mu
M^{-1}m_{\nu D}^{T},\hspace{0.7cm}  \label{M1nu} \\
\mathbf{M}_{\nu }^{\left( 1\right) } &=&-\frac{1}{2}\left( M+M^{T}\right) +%
\frac{1}{2}\left( \mu +\varepsilon \right) ,\hspace{0.7cm} \\
\mathbf{M}_{\nu }^{\left( 2\right) } &=&\frac{1}{2}\left( M+M^{T}\right) +%
\frac{1}{2}\left( \mu +\varepsilon \right) .
\end{eqnarray}%
where $\widetilde{\mathbf{M}}_{\nu }$ corresponds to the mass matrix for
light active neutrinos ($\nu _{a}$), whereas $\mathbf{M}_{\nu }^{\left(
1\right) }$ and $\mathbf{M}_{\nu }^{\left( 2\right) }$ are the mass matrices
for sterile neutrinos ($N_{a}^{-},N_{a}^{+}$) which are superpositions of
mostly $\nu _{aR}$ and $N_{aR}$ as $N_{a}^{\pm }\sim \frac{1}{\sqrt{2}}%
\left( \nu _{aR}\mp N_{aR}\right) $. In the limit $\mu \rightarrow 0$, which
corresponds to unbroken lepton number, the light active neutrinos become
massless. The smallness of the $\mu $- parameter makes the mass splitting of
three pairs of sterile neutrinos to become small, thus implying that the
sterile neutrinos form pseudo-Dirac pairs.

From Eqs. (\ref{MR}) and (\ref{M1nu}), we find that the light active
neutrino mass matrix takes the form: 
\begin{eqnarray}
\widetilde{\mathbf{M}}_{\nu } &=&\frac{y_{\Psi }\left(
m_{R}^{2}-m_{I}^{2}\right) v_{\varphi }^{3}}{8\pi ^{2}\left(
m_{R}^{2}+m_{I}^{2}\right) \Lambda ^{2}}\left( 
\begin{array}{ccc}
\alpha ^{2}y_{1N}^{2} & \alpha \beta y_{1N}y_{2N}\frac{v_{\Phi }}{v_{\varphi
}}e^{-i\theta } & \alpha \gamma y_{1N}y_{2N}\frac{v_{\Phi }}{v_{\varphi }}%
e^{i\theta } \\ 
\alpha \beta y_{2N}y_{1N}\frac{v_{\Phi }}{v_{\varphi }}e^{-i\theta } & \beta
^{2}\left( y_{2N}^{2}e^{-2i\theta }+y_{3N}^{2}e^{2i\theta }\right) \frac{%
v_{\Phi }^{2}}{v_{\varphi }^{2}} & \beta \gamma \left(
y_{2N}^{2}+y_{3N}^{2}\right) \frac{v_{\Phi }^{2}}{v_{\varphi }^{2}} \\ 
\alpha \gamma y_{2N}y_{1N}\frac{v_{\Phi }}{v_{\varphi }}e^{i\theta } & \beta
\gamma \left( y_{2N}^{2}+y_{3N}^{2}\right) \frac{v_{\Phi }^{2}}{v_{\varphi
}^{2}} & \gamma ^{2}\left( y_{2N}^{2}e^{2i\theta }+y_{3N}^{2}e^{-2i\theta
}\right) \frac{v_{\Phi }^{2}}{v_{\varphi }^{2}}%
\end{array}%
\right) ,\allowbreak  \notag \\
\alpha &=&\frac{y_{1}^{\left( \nu \right) }v_{H_{2}}}{\sqrt{2}M_{1}},\hspace{%
1cm}\hspace{1cm}\beta =\frac{y_{2}^{\left( \nu \right) }v_{H_{2}}}{\sqrt{2}%
y_{2}v_{\xi }},\hspace{1cm}\hspace{1cm}\gamma =\frac{y_{3}^{\left( \nu
\right) }v_{H_{2}}}{\sqrt{2}y_{3}v_{\xi }}.  \label{Mnulow}
\end{eqnarray}%
The above given effective neutrino mass matrix, in the simplified benchmark
scenario $\beta =$ $\gamma $ can be written as: 
\begin{equation}
\mathbf{M}_{\nu }=\left( 
\begin{array}{ccc}
A_{\nu } & \tilde{B}_{\nu } & \tilde{B}_{\nu }^{\ast } \\ 
\tilde{B}_{\nu } & \tilde{C}_{\nu } & D_{\nu } \\ 
\tilde{B}_{\nu }^{\ast } & D_{\nu } & \tilde{C}_{\nu }^{\ast }%
\end{array}%
\right) ,
\end{equation}%
where $\tilde{B}_{\nu }=B_{\nu }e^{-i\theta }$ and $\tilde{C}_{\nu }=C_{\nu
}e^{2i\zeta }$. This matrix is diagonalized by the mixing matrix $\mathbf{U}%
_{\nu }$, this is, $\mathbf{U}_{\nu }^{\dagger }\mathbf{M}_{\nu }\mathbf{U}%
_{\nu }^{\ast }=\hat{\mathbf{M}}_{\nu }$ with $\hat{\mathbf{M}}_{\nu }=\text{%
Diag.}(|m_{1}|,|m_{2}|,|m_{3}|)$. As it has been shown, the neutrino mixing
matrix is parametrized by $\mathbf{U}_{\nu }=\mathbf{U}_{\alpha }\mathbf{O}%
_{23}\mathbf{O}_{13}\mathbf{O}_{12}\mathbf{U}_{\beta }$. Explicitly, we have 
\begin{eqnarray}
\mathbf{U}_{\alpha } &=&%
\begin{pmatrix}
e^{i\alpha _{1}} & 0 & 0 \\ 
0 & e^{i\alpha _{2}} & 0 \\ 
0 & 0 & e^{i\alpha _{3}}%
\end{pmatrix}%
,\qquad \mathbf{U}_{\beta }=%
\begin{pmatrix}
1 & 0 & 0 \\ 
0 & e^{i\beta _{1}} & 0 \\ 
0 & 0 & e^{i\beta _{2}}%
\end{pmatrix}
\notag \\
\mathbf{O}_{23} &=&%
\begin{pmatrix}
1 & 0 & 0 \\ 
0 & \cos {\gamma }_{23} & \sin {\gamma }_{23} \\ 
0 & -\sin {\gamma }_{23} & \cos {\gamma }_{23}%
\end{pmatrix}%
,\quad \mathbf{O}_{13}=%
\begin{pmatrix}
\cos {\gamma }_{13} & 0 & \sin {\gamma }_{13}e^{-i\delta } \\ 
0 & 1 & 0 \\ 
-\sin {\gamma }_{13}e^{i\delta } & 0 & \cos {\gamma }_{13}%
\end{pmatrix}%
,\quad \mathbf{O}_{12}=%
\begin{pmatrix}
\cos {\gamma }_{12} & \sin {\gamma }_{12} & 0 \\ 
-\sin {\gamma }_{12} & \cos {\gamma }_{12} & 0 \\ 
0 & 0 & 1%
\end{pmatrix}%
.
\end{eqnarray}%
In the above matrices, $\alpha _{i}$ ($i=1,2,3$) are unphysical phases; $%
\beta _{j}$ ($j=1,2$) stands for the Majorana phases. In addition, there are
three angles and one phase that parameterize the rotations.

As one can verify, the $\alpha _{i}$ and $\beta _{j}$ phases are not
arbitrary since they can be fixed by inverting the expression, $\mathbf{U}%
_{\nu }^{\dagger }\mathbf{M}_{\nu }\mathbf{U}_{\nu }^{\ast }=\hat{\mathbf{M}}%
_{\nu }$ to obtain the effective mass matrix. This means explicitly, $%
\mathbf{M}_{\nu }=\mathbf{U}_{\nu }\hat{\mathbf{M}}_{\nu }\mathbf{U}_{\nu
}^{T}$, then, we obtain 
\begin{eqnarray}
A_{\nu } &=&\cos ^{2}{\gamma _{13}}\left( |m_{1}|\cos ^{2}{\gamma _{12}}%
+|m_{2}|\sin ^{2}{\gamma _{12}}\right) +|m_{3}|\sin ^{2}{\gamma _{13}}; 
\notag \\
\tilde{B}_{\nu } &=&\frac{\cos {\gamma _{13}}}{\sqrt{2}}\left[ |m_{1}|\cos {%
\gamma _{12}}\left( \sin {\gamma _{12}}-i\cos {\gamma _{12}}\sin {\gamma
_{13}}\right) -|m_{2}|\sin {\gamma _{12}}\left( \cos {\gamma _{12}}+i\sin {%
\gamma _{12}}\sin {\gamma _{13}}\right) +i|m_{3}|\sin {\gamma _{13}}\right] ;
\notag \\
\tilde{C}_{\nu } &=&\frac{1}{2}\left[ |m_{1}|\left( \sin {\gamma _{12}}%
-i\cos {\gamma _{12}}\sin {\gamma _{13}}\right) ^{2}+|m_{2}|\left( \cos {%
\gamma _{12}}+i\sin {\gamma _{12}}\sin {\gamma _{13}}\right)
^{2}-|m_{3}|\cos ^{2}{\gamma _{13}}\right]  \notag \\
D_{\nu } &=&\frac{1}{2}\left[ |m_{1}|\left( \sin ^{2}{\gamma _{12}}+\cos ^{2}%
{\gamma _{12}}\sin ^{2}{\gamma _{13}}\right) +|m_{2}|\left( \cos ^{2}{\gamma
_{12}}+\sin ^{2}{\gamma _{12}}\sin ^{2}{\gamma _{13}}\right) +|m_{3}|\cos
^{2}{\gamma _{13}}\right] .
\end{eqnarray}%
These matrix elements are obtained with $\alpha _{1}=\alpha _{3}=0$ and $%
\alpha _{2}=\pi $; $\beta _{1}=0$ and $\beta _{2}=\pi /2$. Along with these, 
$\gamma _{23}=\pi /4$ and $\delta =-\pi /2$.

Having given the above conditions, let us write explicitly the neutrino
mixing matrix 
\begin{eqnarray}
\mathbf{U}_{\nu}=%
\begin{pmatrix}
\cos{\gamma_{12}}\cos{\gamma_{13}} & \sin{\gamma_{12}}\cos{\gamma_{13}} & 
-\sin{\gamma_{13}} \\ 
\frac{1}{\sqrt{2}}\left(\sin{\gamma_{12}}-i\cos{\gamma_{12}}\sin{\gamma_{13}}%
\right) & -\frac{1}{\sqrt{2}}\left(\cos{\gamma_{12}}+i\sin{\gamma_{12}}\sin{%
\gamma_{13}}\right) & -\frac{i\cos{\gamma_{13}}}{\sqrt{2}} \\ 
\frac{1}{\sqrt{2}}\left(\sin{\gamma_{12}}+i\cos{\gamma_{12}}\sin{\gamma_{13}}%
\right) & -\frac{1}{\sqrt{2}}\left(\cos{\gamma_{12}}-i\sin{\gamma_{12}}\sin{%
\gamma_{13}}\right) & \frac{i\cos{\gamma_{13}}}{\sqrt{2}}%
\end{pmatrix}%
\end{eqnarray}

Therefore, the PMNS mixing matrix is given by $\mathbf{U}^{j}=\mathbf{U}%
^{\dagger}_{\ell}\mathbf{U}^{j}_{\nu}$ where $j=n,i$ denotes the normal and
inverted hierarchy, respectively.

In here, we add a important comment on the $\mathbf{U}_{\nu}$ matrix. If the
charged lepton mass matrix was diagonal, then the $\mathbf{U}_{\nu}$ matrix
would be identified with the well known cobimaximal mixing matrix. In the
current model, the charged lepton is not diagonal so that we expect some
deviations to cobimaximal mixing matrix.

The expression for the mixing angles are obtained by comparing our PMNS
mixing matrix with the standard parametrization. Then, one obtains 
\begin{eqnarray}
\sin{\theta}_{13}&=&\left| \mathbf{U}_{13} \right|= \left| -\cos{\theta_{e}}%
\sin{\gamma_{13}}e^{-i\eta_{e}}+\frac{i}{\sqrt{2}}\sin{\theta_{e}}\cos{%
\gamma_{13}} \right|;  \notag \\
\sin{\theta}_{23}&=&\frac{\left| \mathbf{U}_{23} \right|}{\sqrt{1-\sin^{2}{%
\theta_{13}}}}=\frac{\left| \sin{\theta_{e}}\sin{\gamma_{13}}e^{-i\eta_{e}}+%
\frac{i}{\sqrt{2}}\cos{\theta_{e}}\cos{\gamma_{13}} \right|}{\sqrt{1-\sin^{2}%
{\theta_{13}}}};  \notag \\
\sin{\theta}_{12}&=&\frac{\left| \mathbf{U}_{12} \right|}{\sqrt{1-\sin^{2}{%
\theta_{13}}}}=\frac{\left| \cos{\theta_{e}}\sin{\gamma_{12}}\cos{\gamma_{13}%
}e^{-i\eta_{e}}+\frac{i}{\sqrt{2}}\sin{\theta_{e}}\left(\cos{\gamma_{12}}%
+i\sin{\gamma_{12}} \sin{\gamma_{13}}\right) \right|}{\sqrt{1-\sin^{2}{%
\theta_{13}}}}
\end{eqnarray}

We ought to comment that there are still free parameters in the PMNS mixing
matrix namely: $|b_{e}|$ (or $\theta _{e}$), $\gamma _{12}$, $\gamma _{13}$
and the phase $\eta _{e}$. As we notice, with $\theta _{e}\approx 0$, the
Cobimaximal predictions are recovered: $\theta _{23}=\gamma _{23}=\pi /4$
and the CP-violating phase, $\delta _{CP}=-\pi /2$. A numerical analysis has
to be done to constrain those parameters.

In addition, one gets for the Jarlskog invariant 
\begin{eqnarray}
\mathcal{J}_{CP} &=&Im\left[ U_{23}U_{13}^{\ast }U_{12}U_{22}^{\ast }\right]
\notag \\
&=&\frac{1}{8}\sin {2\theta _{12}}\sin {2\theta _{23}}\sin {2\theta _{13}}%
\cos {\theta _{13}}\sin {\delta _{CP}}.
\end{eqnarray}%
Here, we want to show explicitly that the CP-violating phase, $\delta _{CP}$%
, is deviated from $-\pi /2$ by the charged lepton sector. To do this, we
perform an approximation as follows: let us consider $|b_{e}|\approx |m_{e}|$
in the charged lepton mass matrix, which implies $\sin {\theta _{e}}\approx
|m_{e}|/|m_{\mu }|$ and $\cos {\theta _{e}}\approx 1$. Then, we obtain the
following PMNS matrix elements

\begin{eqnarray}
U_{12} &\approx &\sin {\gamma _{12}}\cos {\gamma _{13}}e^{-i\eta
_{e}},\qquad |U_{12}|\approx \sin {\gamma _{12}}\cos {\gamma _{13}},  \notag
\\
U_{13} &\approx &-\sin {\gamma _{13}}e^{-i\eta _{e}},\qquad |U_{13}|\approx
\sin {\gamma _{13}},  \notag \\
U_{23} &\approx &-i\frac{\cos {\gamma _{13}}}{\sqrt{2}},\qquad
|U_{23}|\approx \frac{\cos {\gamma _{13}}}{\sqrt{2}},  \notag \\
U_{22} &\approx &-\frac{1}{\sqrt{2}}\left( \cos {\gamma _{12}}+i\sin {\gamma
_{12}}\sin {\gamma _{13}}\right) .
\end{eqnarray}

\begin{figure}[tbp]
\centering
\includegraphics[width=0.5\textwidth]{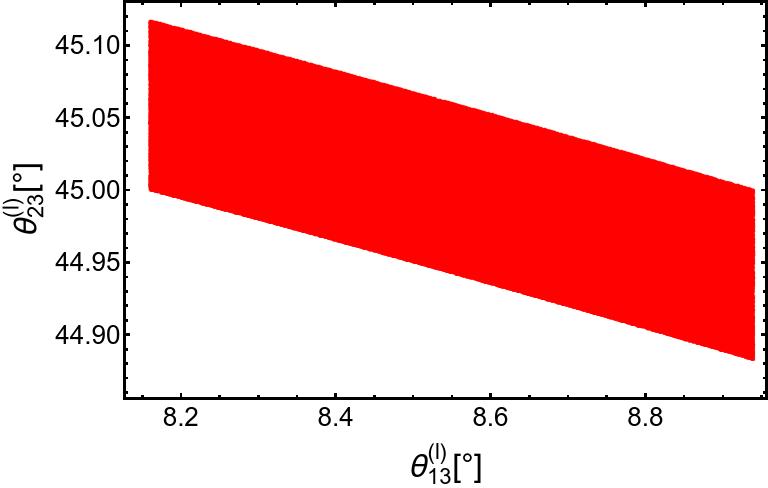}%
\includegraphics[width=0.5\textwidth]{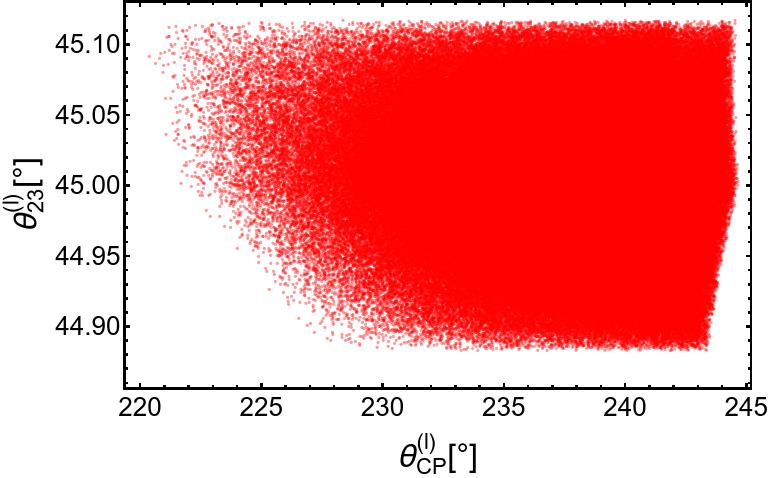}
\caption{Correlations of the atmospheric with the reactor mixing angle and
with the leptonic Dirac CP violating phase.}
\label{leptoncorrelations}
\end{figure}

Having done that, using the Jarlskog invariant one obtains 
\begin{equation}
\sin{\delta_{CP}}=\frac{Im\left[U_{23} U^{\ast}_{13}U_{12}U^{\ast}_{22}%
\right] \notag}{\cos{\theta_{12}}\cos{\theta_{23}}\vert U_{12}\vert \vert
U_{13}\vert \vert U_{23}\vert}
\end{equation}
where 
\begin{equation}
Im\left[U_{23} U^{\ast}_{13}U_{12}U^{\ast}_{22}\right] \approx-\frac{1}{2}%
\sin{\gamma_{12}}\cos{\gamma_{12}}\sin{\gamma_{13}}\cos^{2}{\gamma_{13}} 
\notag
\end{equation}
therefore

\begin{equation}
\sin{\delta_{CP}}\approx-\frac{\sqrt{2}}{2}\frac{\cos{\gamma_{12}} }{\cos{%
\theta_{12}}\cos{\theta_{23}}}
\end{equation}
In the above expression, we could consider $\cos{\theta_{23}}\approx 1/\sqrt{%
2}$ in good approximation. In addition, in the already mentioned
approximation $\cos{\theta_{12}}\approx\cos{\gamma_{12}}$ so that the Dirac
CP phase is near to $270^{\circ}$. The correlations of the atmospheric with
the reactor mixing angle and with the leptonic Dirac CP violating phase are
shown in figure \ref{leptoncorrelations}.%

\section{Muon anomalous magnetic moment}

\label{gminus2} 
\begin{figure}[tbp]
\centering
\includegraphics[width=12.0cm, height=9cm]{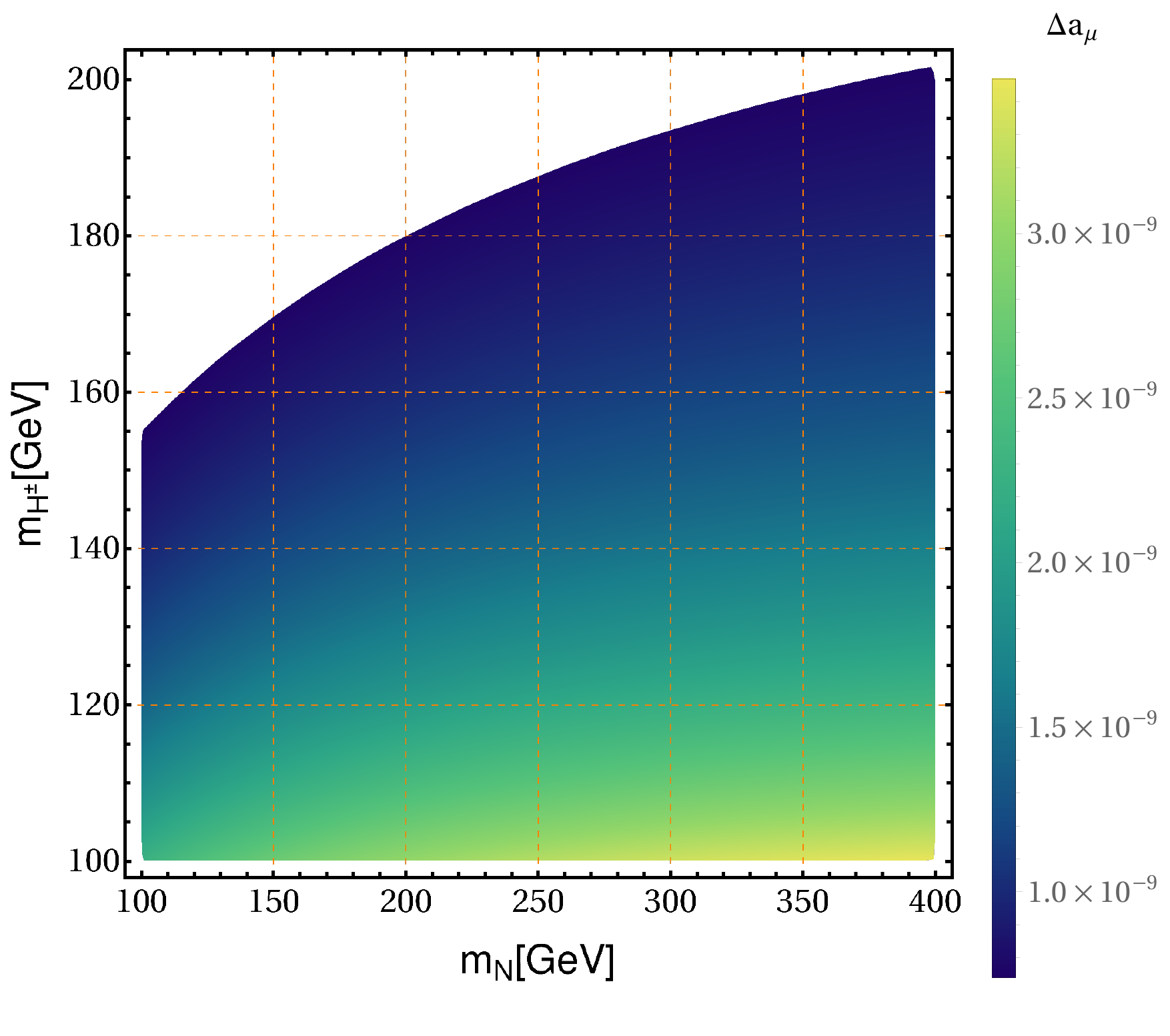} 
\caption{Allowed parameter space in the $m_{N}-m_{H^{\pm}}$ plane consistent
with the muon anomalous magnetic moment.}
\label{gminus2c}
\end{figure}

In this section we will discuss the consequences of our
model in the muon anomalous magnetic moment. The dominant contribution to
the muon anomalous magnetic moment arises from the one-loop diagram
involving the exchange of electrically charged scalars and nearly degenerate
sterile neutrinos running in the internal lines. Unlike the model of \cite%
{Mondragon:2007nk}, the muon anomalous magnetic moment does not receive
contributions involving electrically neutral virtual scalars since in our
model only one scalar doublet participates in the charged lepton Yukawa
interactions, which prevents the appearance of flavor changing neutral
scalar interactions in the lepton sector. Then, in our model the leading
contribution to the muon anomalous magnetic moment is given by:

\begin{eqnarray}
\Delta a_{\mu } &=&\frac{y_{1}^{\left( \nu \right) }z_{2}^{\left( \nu
\right) }m_{\mu }^{2}\cos \theta _{e}\sin \theta _{e}}{8\pi ^{2}m_{H^{\pm
}}^{2}}J\left( \frac{M_{sb}}{m_{\mu }},\frac{M_{sb}}{m_{H^{\pm }}}\right) +%
\frac{\left( z_{2}^{\left( \nu \right) }\right) ^{2}\cos ^{2}\theta
_{e}m_{\mu }^{2}}{8\pi ^{2}m_{H^{\pm }}^{2}}J\left( \frac{m_{N}}{m_{\mu }},%
\frac{m_{N}}{m_{H^{\pm }}}\right) ,\qquad  \notag \\
z_{2}^{\left( \nu \right) } &=&y_{2}^{\left( \nu \right) }\frac{v_{\xi }}{%
\Lambda }\cos ^{2}\beta ,\qquad \qquad \tan \beta =\frac{v_{H_{2}}}{v_{H_{1}}%
},
\end{eqnarray}%
where loop integral $J\left( \frac{m_{N}}{m_{\mu }},\frac{m_{N}}{m_{H^{\pm }}%
}\right) $ has the form \cite%
{Diaz:2002uk,Jegerlehner:2009ry,Kelso:2014qka,Lindner:2016bgg,Kowalska:2017iqv}
\begin{equation}
J\left( \frac{m_{N}}{m_{\mu }},\frac{m_{N}}{m_{H^{\pm }}}\right)
=\int_{0}^{1}dx\frac{P_{+}\left( x,\frac{m_{N}}{m_{\mu }}\right)
+P_{-}\left( x,\frac{m_{N}}{m_{\mu }}\right) }{\left( \frac{m_{N}}{m_{H^{\pm
}}}\right) ^{2}\left( 1-x\right) \left[ 1-\left( \frac{m_{\mu }}{m_{N}}%
\right) ^{2}x\right] +x},
\end{equation}%
where 
\begin{equation}
P_{\pm }(x,\epsilon )=-x\left( 1-x\right) \left( x\pm \epsilon \right) .
\end{equation}%
Considering that the muon anomalous magnetic moment is constrained to be in
the range \cite%
{Hagiwara:2011af,Davier:2017zfy,Blum:2018mom,Keshavarzi:2018mgv,Nomura:2018lsx,Nomura:2018vfz,Aoyama:2020ynm,Abi:2021gix}%
: 
\begin{equation}
\left( \Delta a_{\mu }\right) _{\exp }=\left( 2.51\pm 0.59\right) \times
10^{-9}\text{.}  \notag
\end{equation}%
We plot in figure \ref{gminus2c} the allowed parameter space in the $%
m_{N}-m_{H^{\pm }}$ plane consistent with the muon anomalous magnetic
moment. We have fixed $y_{1}^{\left( \nu \right) }=3.5$, $z_{2}^{\left( \nu
\right) }=-0.85$, $\theta _{e}=17.19^{\circ }$,$\ M_{sb}=1$ GeV. Notice that the complex phases in the couplings $y^{(\nu)}_1$ and $z^{(\nu)}_2$ can be rotated away by a phase redefinition of the right handed Majorana neutrino fields. Consequently, the couplings $y^{(\nu)}_1$ and $z^{(\nu)}_2$ can be taken real without a loss of generality. It is worth mentioning that the range of values for charged scalar masses is consistent with the collider constraints \cite{Sanyal:2019xcp,CMS:2020osd}.
We find that our model can successfully accommodate the experimental values
of the muon anomalous magnetic moment. On the other hand, there is an extra two loop level contribution to the muon anomalous magnetic moment arising from the Barr-Zee type mechanism \cite{Arhrib:2001xx}, however we have numerically checked that this contribution is of the order of $10^{-13}$ for electrically charged scalar masses of about $200$ GeV and quartic scalar couplings of order unity. It is worth mentioning that in the case where the muon anomaly does not get confirmed, the electrically charged scalars will have masses close to the TeV scale.
On the other hand, CP-violating interactions related to the Barr-Zee
two-loop mechanism can give rise values of the muon electric dipole moment,
several orders of magnitude larger than the SM prediction \cite{Barr:1990vd,Chang:1998uc,Heo:2008sr,Heo:2008dq}. In our model, the
CP violating interactions responsible for the generation of the muon
electric dipole moment only appear when one consider complex quartic scalar coupling, which corresponds to a CP violating scalar potential. In that case, these CP violating interactions are:
\begin{equation}
\mathcal{L}_{int}=\frac{\sqrt{2}m_{\mu }\tan \beta }{v}\overline{\mu }%
i\gamma ^{5}A^{0}\mu -\frac{\lambda _{A^{0}H^{+}H^{-}}v}{\sqrt{2}}%
A^{0}H^{+}H^{-},\qquad \qquad \tan \beta =\frac{v_{H_{2}}}{v_{H_{1}}}, 
\notag
\end{equation}
and the resulting two loop level induced muon electric dipole moment has the
form:
\begin{equation}
d_{\mu }=-\frac{\alpha _{em}m_{\mu }\lambda _{A^{0}H^{+}H^{-}}\tan \beta }{%
32\pi ^{3}}F\left( \frac{m_{H^{\pm }}^{2}}{m_{A^{0}}^{2}}\right) ,  \notag
\end{equation}
where the two loop integral $F\left( z\right) $ has the form:
\begin{equation}
F\left( z\right) =\int_{0}^{1}dz\frac{x\left( 1-x\right) }{z-x\left(
1-x\right) }\ln \left( \frac{x\left( 1-x\right) }{z}\right) ,  \label{Floop}
\end{equation}

\begin{figure}[tbp]
\centering
\includegraphics[width=12.0cm, height=9cm]{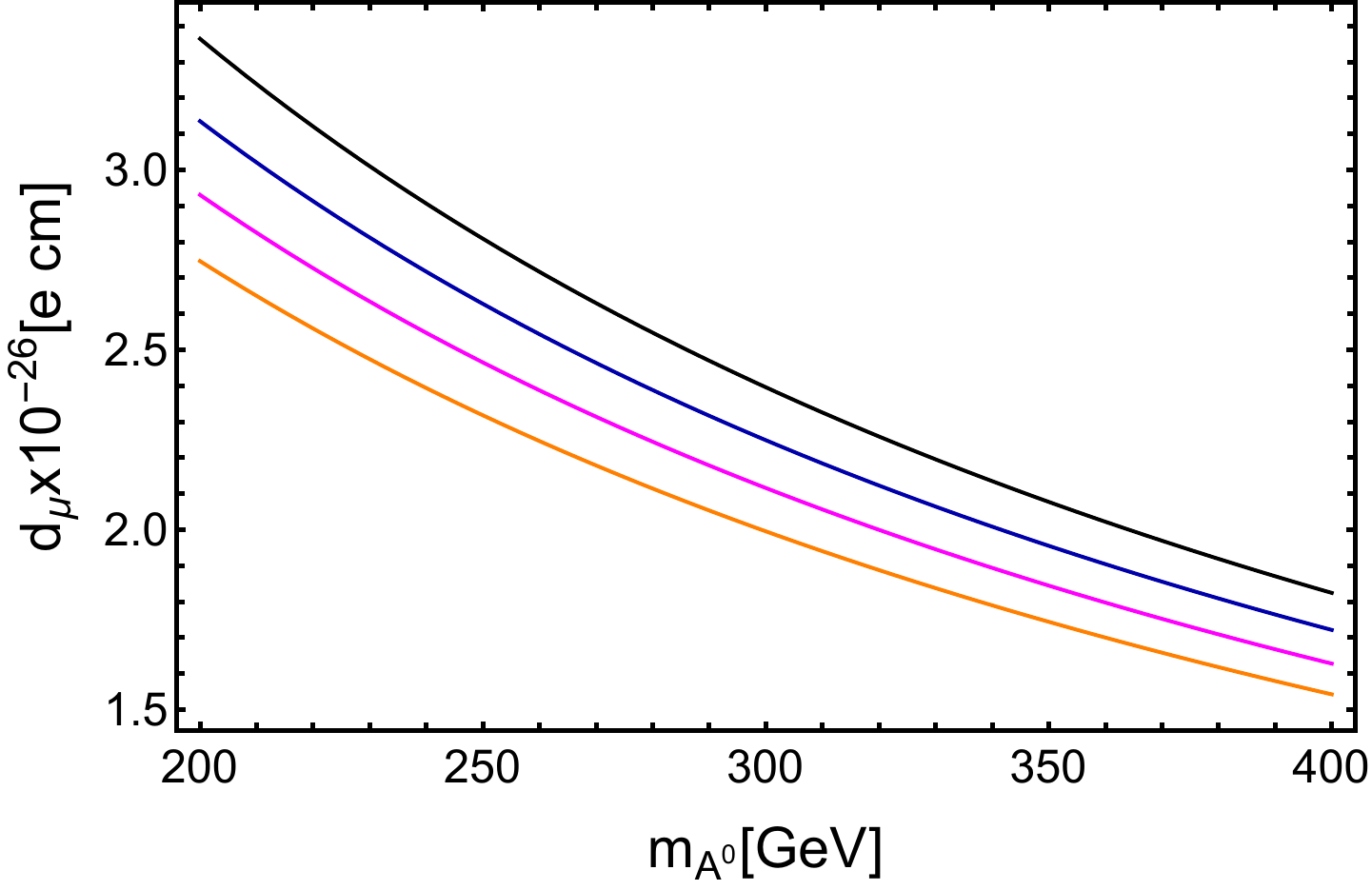} 
\caption{Muon electric dipole moment as a function of the CP odd scalar mass $m_{A^0}$. The black, blue, magenta and orange curves corresponds to charged scalar masses equal to $170$ GeV, $180$ GeV, $190$ GeV and $200$ GeV, respectively. Here we have set $\tan\beta=0.2$.}
\label{muonEDM}
\end{figure}
Figure \ref{muonEDM} displays the muon electric dipole moment as a function of the CP odd scalar mass $m_{A^0}$, for different values of the charged scalar masses, taken to be equal to $170$ GeV, $180$ GeV, $190$ GeV and $200$ GeV, for the black, blue, magenta and orange curves, respectively. As shown in figure \ref{muonEDM}, the muon electric dipole moment reach values around $10^{-26}$ e.cm, which is several orders of magnitude larger than the SM prediction $10^{-42}$ e.cm \cite{muonEDMinitiative:2022fmk}. Besides that, our obtained values of the muon electric dipole moment are lower than the experimental upper limit of $1.8\times 10^{-19}$ e.cm. Note that the electric dipole moment obtained in our model is larger than zero provided that the scalar coupling $\lambda_{A^{0}H^{+}H^{-}}$ is positive. We have numerically checked the two loop integral $F(z)$ of Eq. (\ref{Floop}) is always negative.

\section{Scalar and dark matter sectors}

\label{DMsec}

In this section we discuss the scalar and Dark Matter (DM) sectors of the
model with more detail. 
\Catalina{
We present several numerical results based on a scan of the parameter space of
the model where we construct likelihood profiles involving observables of
interest by comparing predictions with experimental measurements.
A complete composite likelihood global analysis is outside the scope of this
letter. We limit ourselves to include the information from the measured
values of the relic density $\Omega h^2_{\text{Planck}}$, Higgs mass $m_h$
and Baryon asymmetry of the Universe (BAU) $Y_B$
as basic Gaussian likelihoods $\mathcal{L}_\Omega$, $\mathcal{L}_{m_h}$ and
$\mathcal{L}_{Y_B}$
respectively. We also include a likelihood function $\mathcal{L}_{DD}$
based on recent results from the XENON1T Direct Detection Experiment,
we then maximize over the model's parameter space the
composite log-likelihood

\begin{equation}
	\log \mathcal{L} = \log \mathcal{L}_\text{DD} + \log \mathcal{L}_\Omega +
	\log \mathcal{L}_{m_h} + \log \mathcal{L}_{Y_B}
\end{equation}
Note that in the high statistic limit, twice the negative of the composite log-likelihood
approaches a $\chi$-square function so this procedure is equivalent to minimizing such function.
In the next subsections we detail the construction of these likelihood profiles.

Using such variety of physical observables to construct the total log-likelihood 
leads to a large number of free parameters, in our case we need 27\footnote{
In addition to the parameters of the mass and mixing matrices from the quark
and charged lepton sector,
which are kept fixed in the analysis of the scalar and DM sectors
(the neutrino sector parameters influence the baryon asymmetry observable).
}
to properly conduct the numerical analysis, however distinct observables depend mostly
on different subsets of the free parameters. From inspection of the analytic equations for the
predicted observables it is clear that only a few number of the free parameters have
influence in all the physical observables considered, and this leads to important
correlations between them.

\subsection{Scalar mass spectra}

}

For the purpose of this section, we will consider
that all scalars which have VEVs of order of the model cutoff $\Lambda$ are decoupled, since their masses will be around that of the cutoff scale. This
leaves us with an effective scalar potential $V$. For simplicity we will
assume that only the $\eta$ and $\varphi$ scalar singlets couple to the
Higgses and write the low energy scalar potential as $V=V_1+V_2$. For the
doublets $H_1$ and $H_2$ we'll take the simple CP-conserving potential given
by:

\begin{eqnarray}
V_{1} &=&m_{11}^{2}H_{1}^{\dagger }H_{1}+m_{22}^{2}H_{2}^{\dagger
}H_{2}-m_{12}^{2}\left( H_{1}^{\dagger }H_{2}+H_{2}^{\dagger }H_{1}\right) +%
\frac{\lambda _{1}}{2}\left( H_{1}^{\dagger }H_{1}\right) ^{2}+\frac{\lambda
_{2}}{2}\left( H_{2}^{\dagger }H_{2}\right) ^{2}  \notag \\
&&+\lambda _{3}H_{1}^{\dagger }H_{1}H_{2}^{\dagger }H_{2}+\lambda
_{4}H_{1}^{\dagger }H_{2}H_{2}^{\dagger }H_{1}+\frac{\lambda _{5}}{2}\left[
\left( H_{1}^{\dagger }H_{2}\right) ^{2}+\left( H_{2}^{\dagger }H_{1}\right)
^{2}\right] ,
\end{eqnarray}%
with all parameters real. In order to reduce the number of free parameters
for the numerical calculations, for the second part of the scalar potential
we will take simply: 
\begin{eqnarray}
V_{2} &=&\sum_{k=1}^{2}\left[ \mu _{\eta }^{2}\eta _{k}^{\ast }\eta _{k}+%
\frac{\lambda _{\eta }^{\left( k\right) }}{2}\left( \eta _{k}^{\ast }\eta
_{k}\right) ^{2}\right] +\mu _{\varphi }^{2}\varphi ^{2}+\frac{\lambda
_{\varphi }}{2}\varphi ^{4}  \notag \\
&&+\left( \sum_{k=1}^{2}\lambda _{6}^{\left( k\right) }\eta _{k}^{\ast }\eta
_{k}+\lambda _{7}\varphi ^{2}\right) H_{1}^{\dagger }H_{1}+\left(
\sum_{k=1}^{2}\lambda _{8}^{\left( k\right) }\eta _{k}^{\ast }\eta
_{k}+\lambda _{9}\varphi ^{2}\right) H_{2}^{\dagger
}H_{2}+\sum_{k=1}^{2}\lambda _{10}^{\left( k\right) }\eta _{k}^{\ast }\eta
_{k}\varphi ^{2}+h.c,
\end{eqnarray}%
%
%
%
%
%
%
%
%
%
%
%
%
%
%
%
%
%
%
%
%
%
%
%
%
%
%
%
%
%
Note that after Electroweak Symmetry Breaking (EWSB), the above scalar
potential induces a mixing between the neutral scalar components of $H_{1}$
and $H_{2}$ and the singlet $\varphi $. As a result the field content of the
model arises from the three field mass eigenstates from this mixing: $h$, $H$
and $H_{3}$, together with the pseudo scalar $A$ and the electrically
charged scalar $H^{+}$. 
\Catalina{
The minimization conditions for this potential take the form:

\begin{eqnarray}\label{tadpoleScalar}
	0 &=& m^2_{11} - m^2_{12} \tan\beta + \frac{1}{2} v^2 
	\left(   
	\lambda_1 \cos^2\beta + \lambda_{345} \sin^2\beta 
	\right)
	+ \frac{1}{2} \lambda_7 v_\phi^2 \notag \\
	0 &=& m^2_{22} - m^2_{12} \cot\beta + \frac{1}{2} v^2
	\left(
	\lambda_2 \sin^2\beta + \lambda_{345} \cos^2\beta 
	\right)
	+ \frac{1}{2} \lambda_9 v_\phi^2 \\
	0 &=& \mu_\phi^2 + \frac{1}{2} v^2 ( \lambda_7 \cos^2\beta 
	+ \lambda_9 \sin^2\beta ) + \frac{1}{2} \lambda_\phi v_\phi^2 \notag
\end{eqnarray}%
where $\lambda_{345}$ is short for $(\lambda_3 + \lambda_4 + \lambda_5)$
and as before $\tan\beta=v_{H_2}/v_{H_1}$.
From these, we eliminate $m^2_{11}$, $m^2_{22}$ and $\mu_\phi^2$
in terms of the remaining parameters, this however only means we would be 
sitting in an extremum of the potential. To ensure that the values of the
parameters correspond in fact to a minimum, we check numerically during the
scan of parameter space the stability of the potential at a given point
using the public tool \texttt{EVADE} \cite%
{Ferreira:2019iqb,Hollik:2018wrr},
which features the minimization of the scalar potential through 
polynomial homotopy continuation and an estimation of the decay rate of
a false vacuum. We apply a hard cut on the parameter points that do not
satisfy the stability criteria.

From the scalar potential we obtain the mass matrices for the different
scalar particles. The charged and pseudoscalar cases contain the 
two SM massless Goldstone states (the longitudinal modes of the SM massive
gauge bosons). The physical particles have masses given by:

\begin{equation}
	M_A^2 = m^2_{12} \csc{\beta} \sec{\beta} - v^2 \lambda_5
\end{equation}

\begin{equation}
	M_{H^\pm}^2 = m^2_{12} \csc{\beta} \sec{\beta} - \frac{1}{2} v^2 (
	\lambda_4 + \lambda_5)
\end{equation}
For the CP-even neutral scalars we can write the mass matrix as:

 \begin{equation}
	 M_{\textrm{scalar}}^2=\left( 
	 \begin{array}{ccc}
		 a & d & f \\ 
		 d & b & e \\ 
		 f & e & c%
		 \end{array}%
	 \right) ,  \label{mh2}
	 \end{equation}%
with

\begin{eqnarray}
	a &=& m^2_{12} \tan{\beta} + \lambda_1 \, v^2 \cos^2{\beta} \notag \\
	b &=& m^2_{12} \cot{\beta} + \lambda_2 \, v^2 \sin^2{\beta} \notag \\
	c &=& \lambda_\phi \, v^2_\phi  \notag \\
	d &=& - m^2_{12} + \lambda_{345} \, v^2 \cos{\beta} \sin{\beta} \\
	e &=& \lambda_9 \,  v \, v_\phi \sin{\beta} \notag \\
	f &=& \lambda_7 \, v \, v_\phi \cos{\beta} \notag \\
\end{eqnarray}
The neutral scalar mass matrix is diagonalized by the mixing matrix $Z^H$
such that

\begin{equation}\label{ZH}
	\textrm{Diag} (m_h^2, m_H^2, m_{H3}^2) = Z^H  M_{\textrm{scalar}}^2 Z^{H\textrm{T}}
\end{equation}

We find for the masses\footnote{These 
expressions are not general in the sense that they are not
valid for cases where there are degenerate eigenvalues or when one or
more of the matrix entries are zero, these {\it atypical} cases should be
treated separately. In particular, these equations are not expected to 
reduce to the correct results in the limit $\lambda_7 = \lambda_9 = 0$,
which is not contemplated since in this case four matrix entries reduce to zero. In the parameter scan we use standard numerical algorithms to
diagonalize the mass matrices.}
\cite%
{deledalle:hal-01501221}:

\begin{eqnarray}
	m_h^2 &=& \frac{1}{3} 
	\left(
	a + b + c - 2 \sqrt{x_1} \cos{[\Xi_s/3]}
	\right) \notag \\
	m_H^2 &=& \frac{1}{3} 
	\left(
	a + b + c + 2 \sqrt{x_1} \cos{[(\Xi_s-\pi)/3]}
	\right)  \\
	m_{H_3}^2 &=& \frac{1}{3} 
	\left(
	a + b + c + 2 \sqrt{x_1} \cos{[(\Xi_s+\pi)/3]}
	\right) \notag 
\end{eqnarray}
where

\begin{equation}
	x_1 = a^2 + b^2 + c^2 - a b - ac - bc + 3(d^2 + f^2 + e^2)
\end{equation}
and 

\begin{equation}
	\Xi_s = \left\{
	\begin{array}{lcc}
		\arctan \left(\frac{\sqrt{4x_1^3 - x_2^2}}{x_2} \right)
		& , & x_2 >0  \\
		\pi/2
		& , & x_2 = 0 \\
		\arctan \left(\frac{\sqrt{4x_1^3 - x_2^2}}{x_2} \right) + \pi
		& , & x_2 <0 
	\end{array}
	\right.
\end{equation}
with

\begin{eqnarray}\label{x2}
	x_2 &=& -(2a - b - c)(2b - a - c)(2c - a -  b) \notag \\
	& & + 9[(2c - a -b) d^2 + (2b -a - c) f^2 + (2a -b - c) e^2] - 54 d e f 
\end{eqnarray}
Note that $\Xi_s \in [-\pi/2,3\pi/2]$ so $m_H^2$ is always grater than $m_h^2$ 
but $m_{H3}^2$ can be smaller than $m_h$, this is an attractive feature of
the model since there are some potential excesses in searches for light
Higgs bosons reported by CMS\cite%
{CMS:2018cyk}, nevertheless a detailed study of this matter is outside
the scope of this work.  We do take into account experimental constraints
from scalar searches at colliders using the public tool 
\texttt{HiggsBounds}\cite%
{Bechtle:2020pkv}
and applying a hard cut on parameter space points not complying with
these limits\footnote{
For this part of the numerical scan we neglect the masses of the first and
second generation of fermions and neglect off-diagonal entries in the Yukawa
matrices. We expect deviations of the matter sector
relative to the SM to be of negligible influence in the phenomenology of the
scalar sector at present collider searches.
}
.

In figure (\ref{Scalars}) we present the low energy scalar mass spectra of the model, the regions
of parameter space that better match high values of the composite log-likelihood
are shown as bright zones, and the best fit point (BFP) is marked with a star.
For the best fit point we find that $x_2$ Eq. (\ref{x2}) is negative and
in turn $\Xi_s$ is very close to $\pi$. We thus find that the scalar $H$ is markedly heavier
than $H_3$ which is around twice as heavy as the SM-like higgs $h$.
Note that preferred values of the charged scalar $H^\pm$ mass are around 400 GeV, however
there are zones that also have high values of the likelihood function below
200 GeV.

\begin{figure}[tbp]
	\centering
	\includegraphics[width=9.0cm, height=6.5cm]{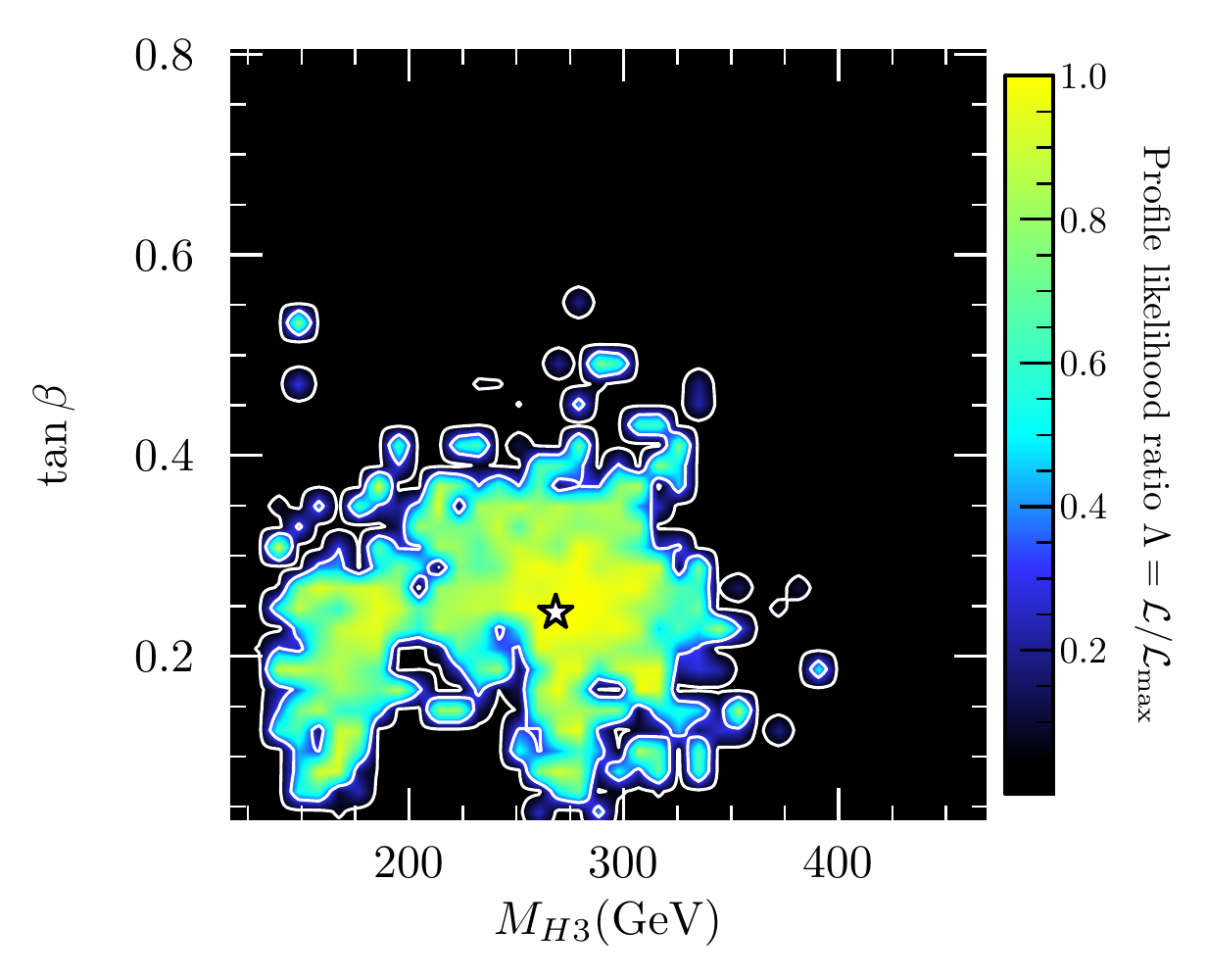}%
	\includegraphics[width=9.0cm, height=6.5cm]{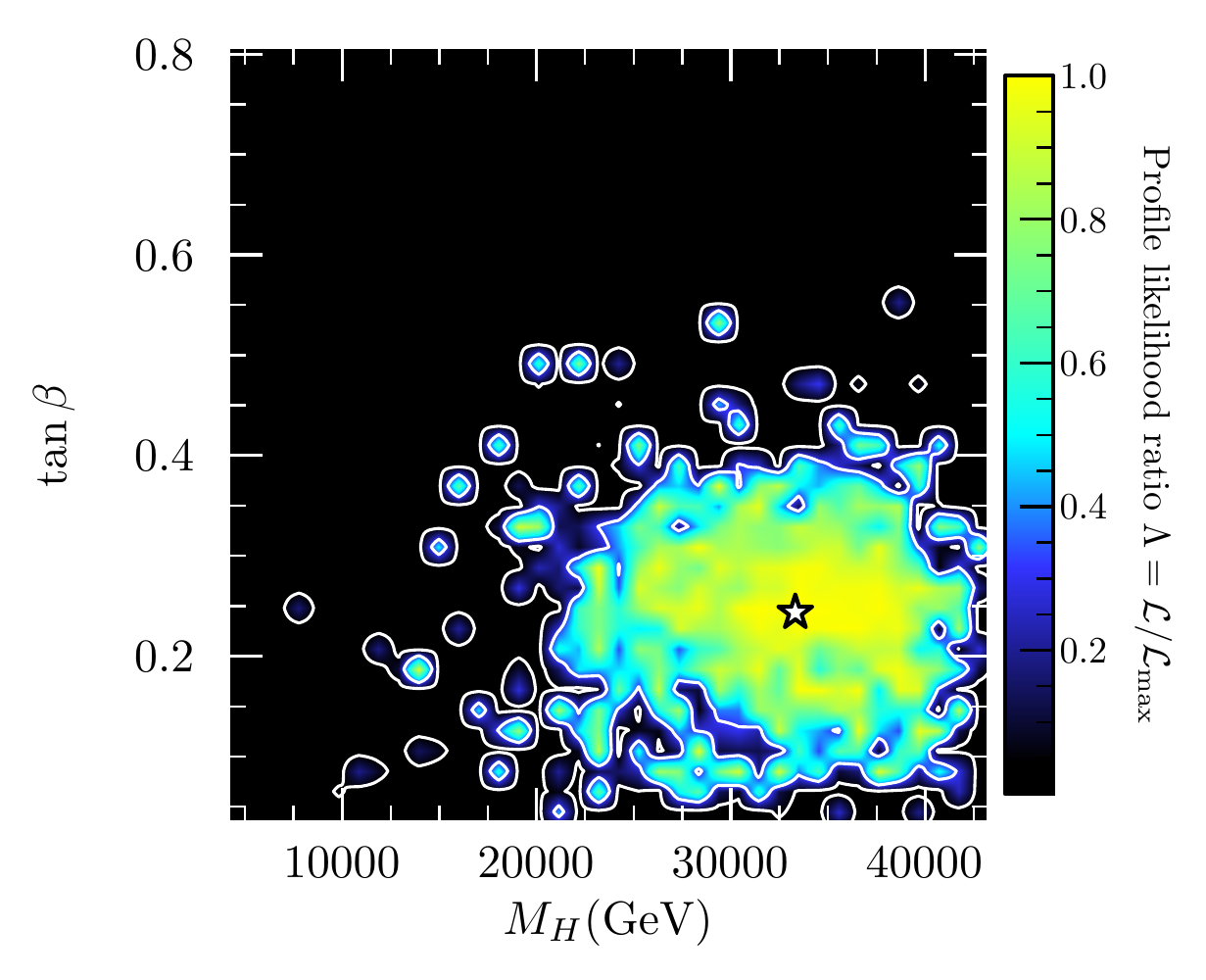}\newline
	\includegraphics[width=9.0cm, height=6.5cm]{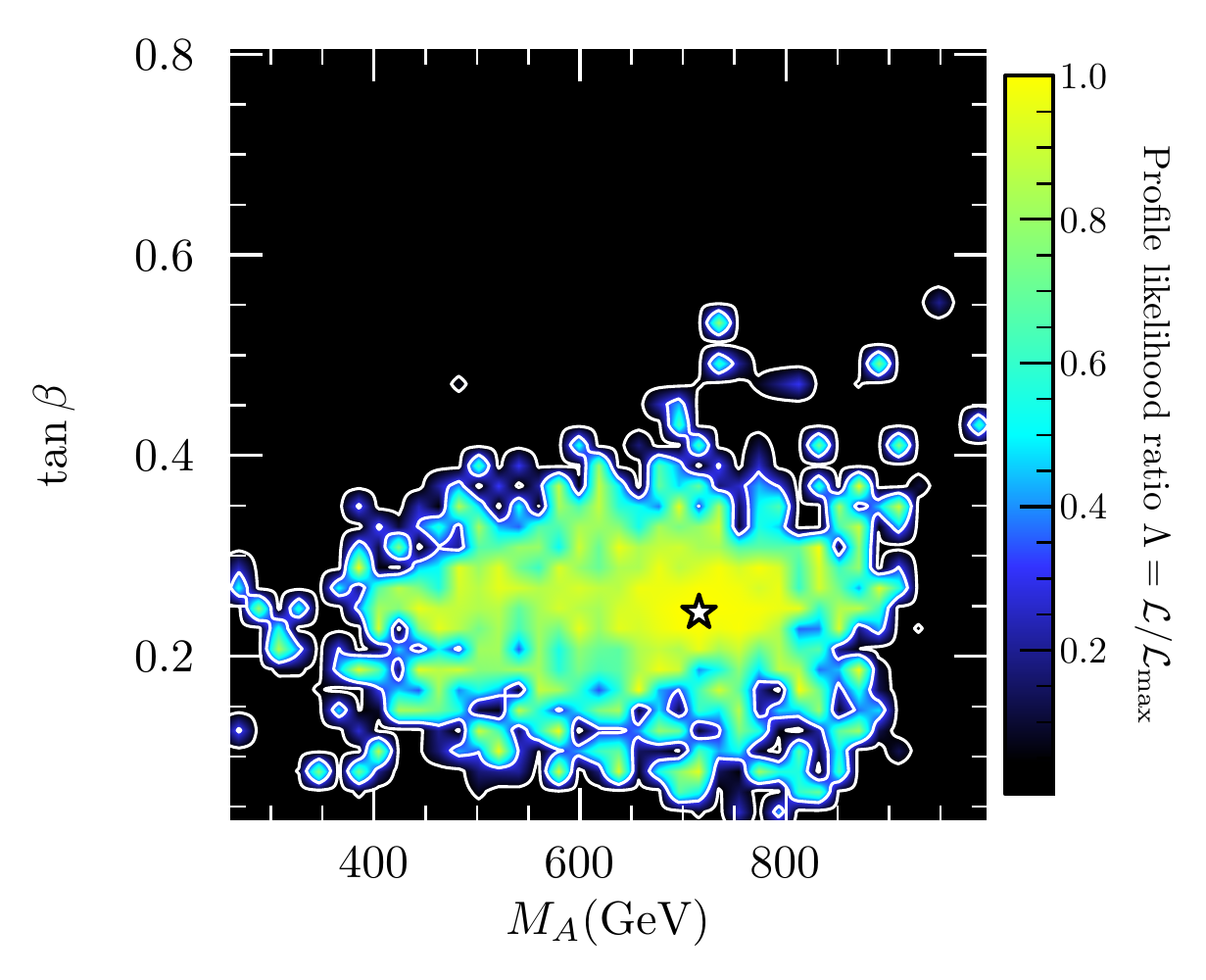}%
	\includegraphics[width=9.0cm, height=6.5cm]{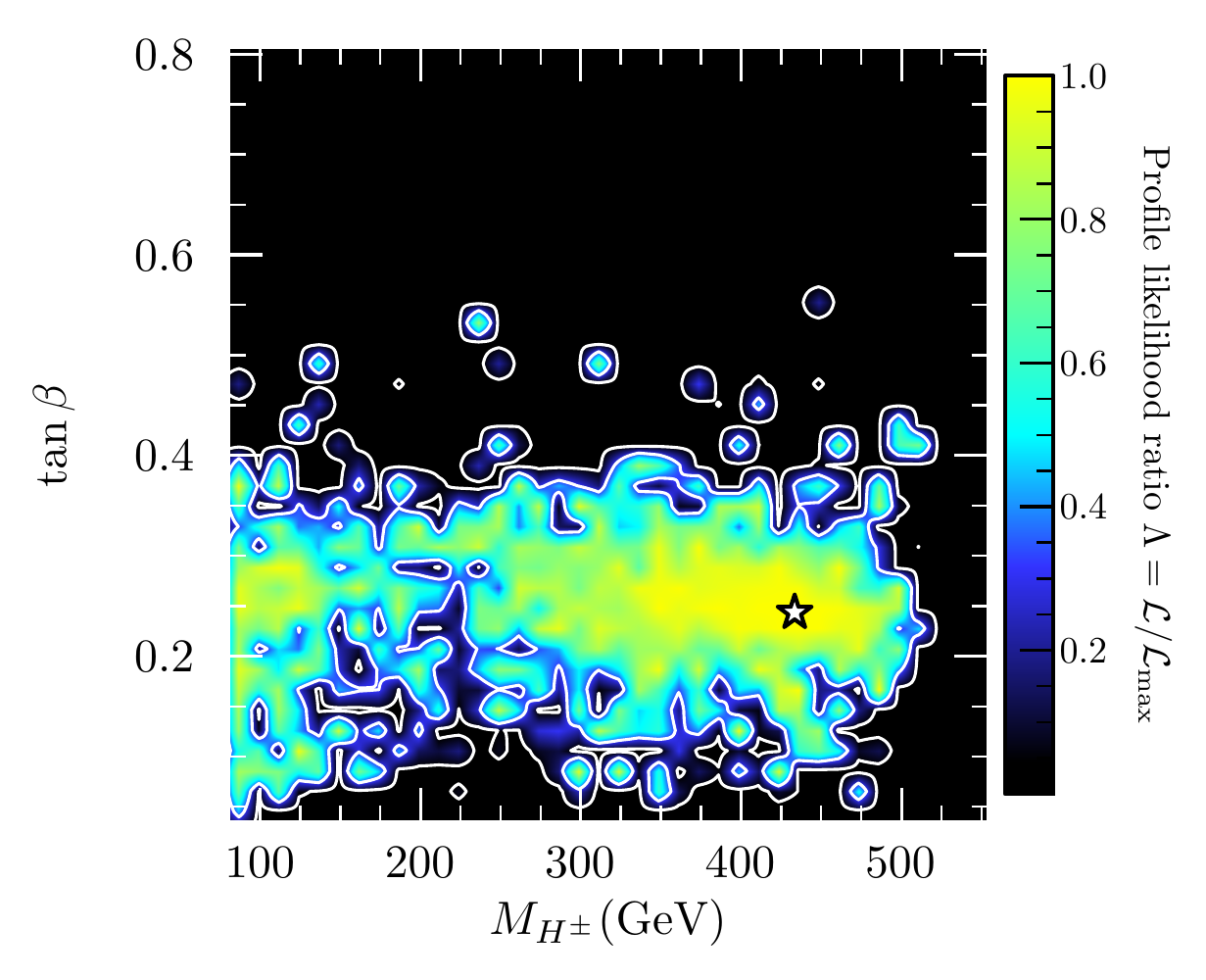}   
	\caption{
		Composite likelihoods as functions of
		the scalar masses and $\tan\beta$.
		Contours of 68\% and 95\% of CL are drawn and the best fit
		point is marked with a star.}
	\label{Scalars}
\end{figure}

\subsection{Relic density}
}

We will continue to assume that the components of the $Z_{2}$ odd fields $%
\eta _{k}$ ($k=1,2$)\ are heavier than the DM Majorana fermion $\Psi _{R}$
and thus consider the latter as our DM candidate\footnote{
\Catalina{
A second case, namely that one of the $\eta$ fields be the lightest of the
DM particles is of course also possible leading to a scalar DM candidate. 
In this letter we focus our attention on the fermion DM candidate
in part because of a matter of taste and in part because of the demanding
computational times required for the numerical analysis which make
unfeasible to present both cases in a single piece. We restrict our analysis to the scenario of fermionic Dark Matter only, because the case of scalar dark matter candidate is a bit generic and our expected results will be similar to those ones discussed in \cite{Abada:2021yot,Hernandez:2021zje,Espinoza:2018itz}, where the dark matter constraints set the mass of scalar dark matter candidates larger than about few TeVs or in a small window close to the half of the SM Higgs boson mass. Besides that, one can also consider the scenario of multicomponent dark matter candidates, however such scenario requires carefull analysis which are beyond the scope of the present work.
}
}
. 
Since the only interaction
of $\Psi _{R}$ that is not suppressed by the cutoff $\Lambda $ is the one
involving the Yukawa coupling $y_{\Psi }$, it follows that the DM
observables will mostly depend on its mass, the coupling $y_{\Psi }$ and the
mass of the mediators. In the region of the parameter space where the
couplings of $\varphi $ to the scalars are small, $\varphi $ will be
\textquotedblleft mostly\textquotedblright\ $H_{3}$, but in general the DM
candidate will communicate with the visible sector through all the above
scalar mass eigenstates.

For the numerical calculation of the relic density, 
the $\Lambda $ cutoff is taken as $\sim 10^{3}$ TeV and 
we keep the masses of the components of $\eta _{k}$ ($k=1,2$)\
large ($\sim $ 50 TeV) but with a small mass splitting between them to
ensure that the $\mu $ parameter influencing the masses of the active
neutrinos is of order $\sim 10^{-1}$ eV so that not much fine tunning of the
neutrino Yukawa couplings would be
required. It is worth mentioning that a naturally small mass splitting can arise from the higher-dimensional operators of Eq. (\ref{NRI}), as discussed in the previous section.

Finally, we implement the model in \texttt{SARAH} \cite%
{Staub:2008uz,Staub:2009bi,Staub:2010jh,Staub:2012pb,Staub:2013tta} from
which we obtain the \texttt{Micromegas} \cite%
{Belanger:2013oya,Belanger:2014vza,Barducci:2016pcb,Belanger:2018ccd} model
files to compute the value of the relic density and we perform a scan of the
parameter space using \texttt{Diver} \cite{Workgroup:2017htr} (in standalone
mode).

\begin{figure}[tbp]
\centering
\includegraphics[width=12.0cm, height=9.5cm]{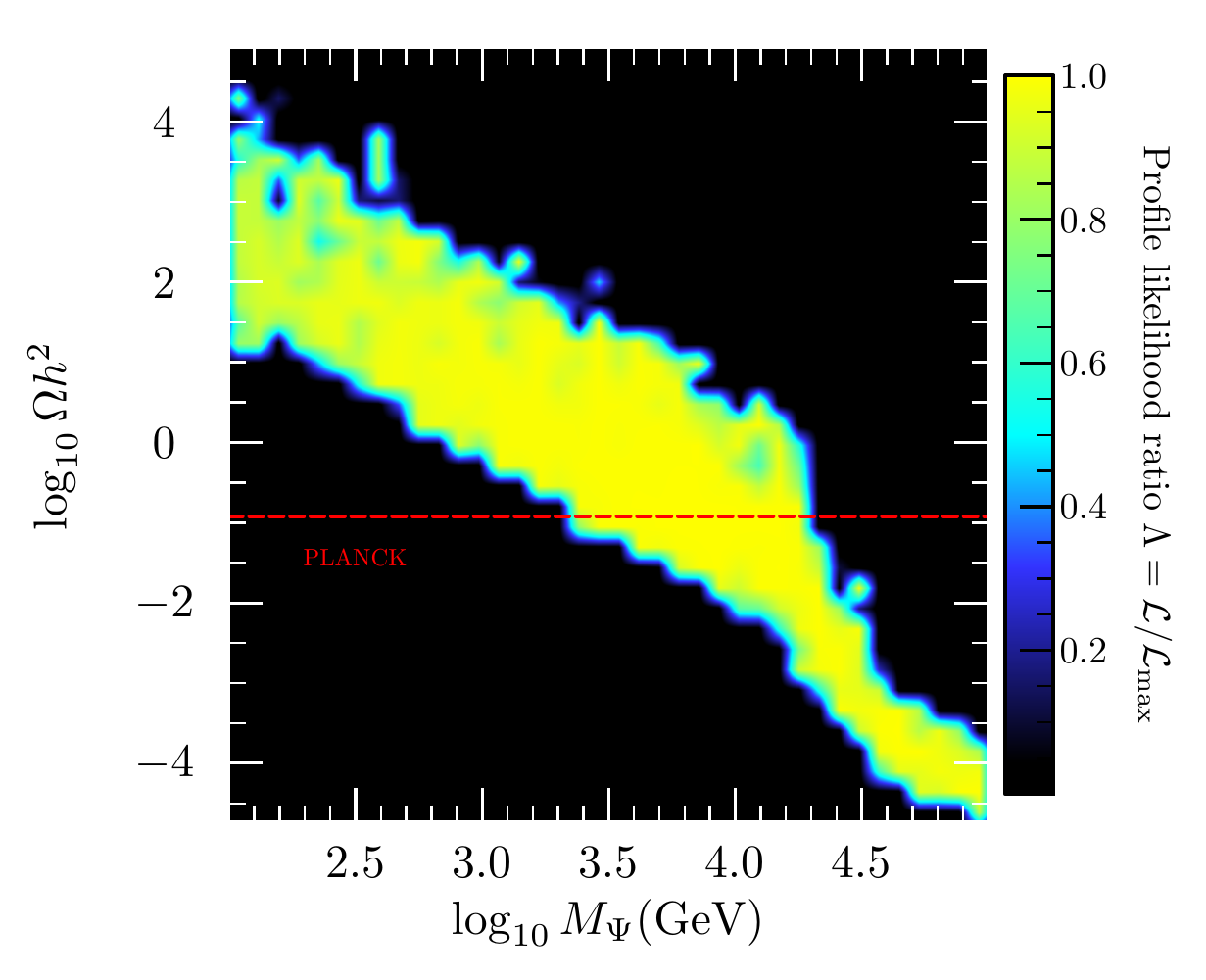}
\caption{
	Composite likelihood (not including the relic density likelihood) as a function of
	the DM candidate mass and its relic density. The Planck measured value is marked by
	the dashed horizontal line.
	}
\label{DM}
\end{figure}

In figure (\ref{DM}) we present the likelihood profile as a function of the 
mass of the DM candidate and its relic density (but not including the likelihood
from the relic density, the corresponding plot with the full log-likelihood
is just a slim horizontal bright band around the Planck measured value).
We infer from this figure that DM candidate masses below $\sim 2.5$ TeV,
though they can be compatible with e.g. direct detection limits, they
would be overproduced at the freeze out epoch. We observed also that, assuming
the DM candidate comprises 100\% of the dark matter of the universe, its
mass can only be around $\sim 2.5$ and $\sim 20$ TeV.

\subsection{Direct detection}

From the details brought up previously and inspecting the
model's Lagrangian, the DM candidate couples to fermions thanks to the
mixing between the scalars. For simplicity we will assume the DM Yukawa
coupling $y_\Psi$ to be real, then the only parity conserving effective
DM-quark interactions mediated by the physical scalars take the general form:

\begin{equation}
L_{\text{eff}} = \sum_k \overline{\Psi _{R}^{C}} c_\Psi^k \Psi _{R} \,\, h_k
+ \sum_{k,q} \overline{q} c_q^k q \,\, h_k
\end{equation}
where the sums are over the quark fields $q$ and the physical scalars $%
h_k=h, H, H_3$. The effective couplings $c_\Psi^k$ and $c_q^k$ are functions
of the free parameters and can be obtained explicitly from the Feynman rules
of the model,
\Catalina{
we find ($k,q=1,2,3$ and no summation over repeated indices):

\begin{equation}
	c_\Psi^k = Z^H_{k3} \, y_\Psi
\end{equation}
and for $d,s$ and $b$ type quarks:

\begin{equation}
	c_q^k = \frac{1}{2} Z^H_{k2} \, 
	\left[
	\lambda^8 \, x^{(d)}_{11} \, U^{dR}_{q1} \, U^{dL}_{q1}
	+ \lambda^3 \, x^{(d)}_{33} \, U^{dR}_{q3} \, U^{dL}_{q3} 
	+ 
	U^{dR}_{q2} \, 
	\left(
	\lambda^5 \, x^{(d)}_{22} \, U^{dL}_{q2}
	+ \lambda^6 \, x^{(d)}_{12} \, U^{dL}_{q1}  
	\right)
	\right] + \textrm{c.c.}
\end{equation}
while for $u,c$ and $t$ quarks we have:

\begin{eqnarray}
	c_q^k &=& \frac{1}{2}  \, 
	\left[
	\lambda^8 \, x^{(u)}_{11} \, U^{uR*}_{q1} \, U^{uL*}_{q1} \, Z^H_{k1} 
    + \lambda^4 \, x^{(u)}_{22} \, U^{uR*}_{q2} \, U^{uL*}_{q2} \, Z^H_{k1}
	\right. \notag \\ 
	& & + U^{uR*}_{q3} \,
	\left. 
	\left[
	\left(
	\lambda^2 \, x^{(u)}_{23} \, U^{uL*}_{q2} + \lambda^4 \, x^{(u)*}_{13} \, U^{uL*}_{q1}
	\right) \, Z^H_{k2}
	+ x^{(u)}_{33} \, U^{uL*}_{q3} \, Z^H_{k1}
	\right]
	\right]   + \textrm{c.c.}
\end{eqnarray}
where we have denoted the quark mixing matrices by $U^{f(L,R)}$ to avoid index
cluttering. The quark Yukawa couplings are obtained from the benchmark point
(\ref{eq:bm-values}) using the relations:

\begin{equation}
	x^{(u)}_{11} = - \frac{c_1}{\cos\beta} , 
	\quad x^{(u)}_{22} = - \frac{b_1}{\cos\beta} , 
	\quad x^{(u)}_{13} = - \frac{a_1}{\sin\beta} , 
	\quad x^{(u)}_{23} =  \frac{a_2}{\sin\beta} , 
	\quad x^{(u)}_{33} = - \frac{a_3}{\cos\beta} 
\end{equation}

\begin{equation}
	x^{(d)}_{11} =  \frac{e_1}{\sin\beta} , 
	\quad x^{(d)}_{12} =  \frac{e_4}{\sin\beta} , 
	\quad x^{(d)}_{22} =  \frac{e_2}{\sin\beta} , 
	\quad x^{(d)}_{33} =  \frac{e_3}{\sin\beta}  
\end{equation}

}

From these we obtain the DM-nucleon differential scattering
cross section (in the nonrelativistic limit):

\begin{equation}
\frac{d\sigma_N}{dE_R} = \frac{1}{32\pi M_\Psi m_N v^2} \,\, \left| 
\overline{\mathcal{M}} \right|^2
\end{equation}
here $E_R$ is the nucleon recoil energy, $m_N$ the nucleon mass and $v$ the
DM velocity. The scattering amplitude $\overline{\mathcal{M}}$ (averaged over
initial spins and summed over final spins) receives the contribution of
three diagrams (one for each scalar mediator) of the form:

\begin{equation}
\mathcal{M}_k = \frac{4 M_\Psi m_N}{q^2 + m_{h_k}^2} c_\Psi^k c_N^k \,\,
\delta_{ss^\prime}\delta_{rr^\prime}
\end{equation}
where $s,s^\prime$ and $r,r^\prime$ denote DM and nucleon spin indices
respectively, $q$ is the momentum transfer, $m_{h_k}$ the mass of the scalar
mediators and $c_N^k$ is defined as

\begin{equation}
c_N^k = \sum_q \frac{m_N}{m_q} c_q^k f_{T_q}^{N}
\end{equation}
with $m_q$ the quark valence masses and $f_{T_q}^{N}$ expresses the
quark-mass contributions to the nucleon mass. Numerical values for the
latter can be found e.g. in \cite{DelNobile:2021icc} and references therein.
The momentum transfer is related to the recoil energy through $q^2 = 2 m_N
E_R$, so that the total DM-nucleon spin independent cross section reads:

\begin{equation}  \label{dmNucleon}
\sigma^{\text{SI}}_N = \int_0^{E_R^{\text{max}}} \frac{d\sigma_N}{dE_R} dE_R
\end{equation}
with the maximum recoil energy given by

\begin{equation}
E_R^{\text{max}} = \frac{2 v^2 \mu^2}{m_N}
\end{equation}
$\mu$ being the DM-nucleon reduced mass.

We now present a likelihood analysis involving publicly available data from
the direct detection XENON1T experiment \cite{XENON:2018voc}. We make use
of the capabilities of the numerical tool \texttt{DDCalc}
to compute the Poisson likelihood given by

\begin{equation}
\mathcal{L}_\text{DD} = \frac{(b+s)^o e^{-(b+s)}}{o!}
\end{equation}
where $o$ is the number of observed events in the detector and $b$ is the
expected background count. From the model's predicted DM-nucleon cross
sections Eq. (\ref{dmNucleon}) as input, \texttt{DDCalc} computes the number
of expected signal events $s$ for given DM local halo and velocity
distribution models (we use the tool's default models, for specific details
on the implementation such as simulation of the detector efficiencies and
acceptance rates, possible binning etc. see \cite%
{GAMBITDarkMatterWorkgroup:2017fax,GAMBIT:2018eea}).

\begin{figure}[tbp]
\centering
\includegraphics[width=12.0cm, height=9.5cm]{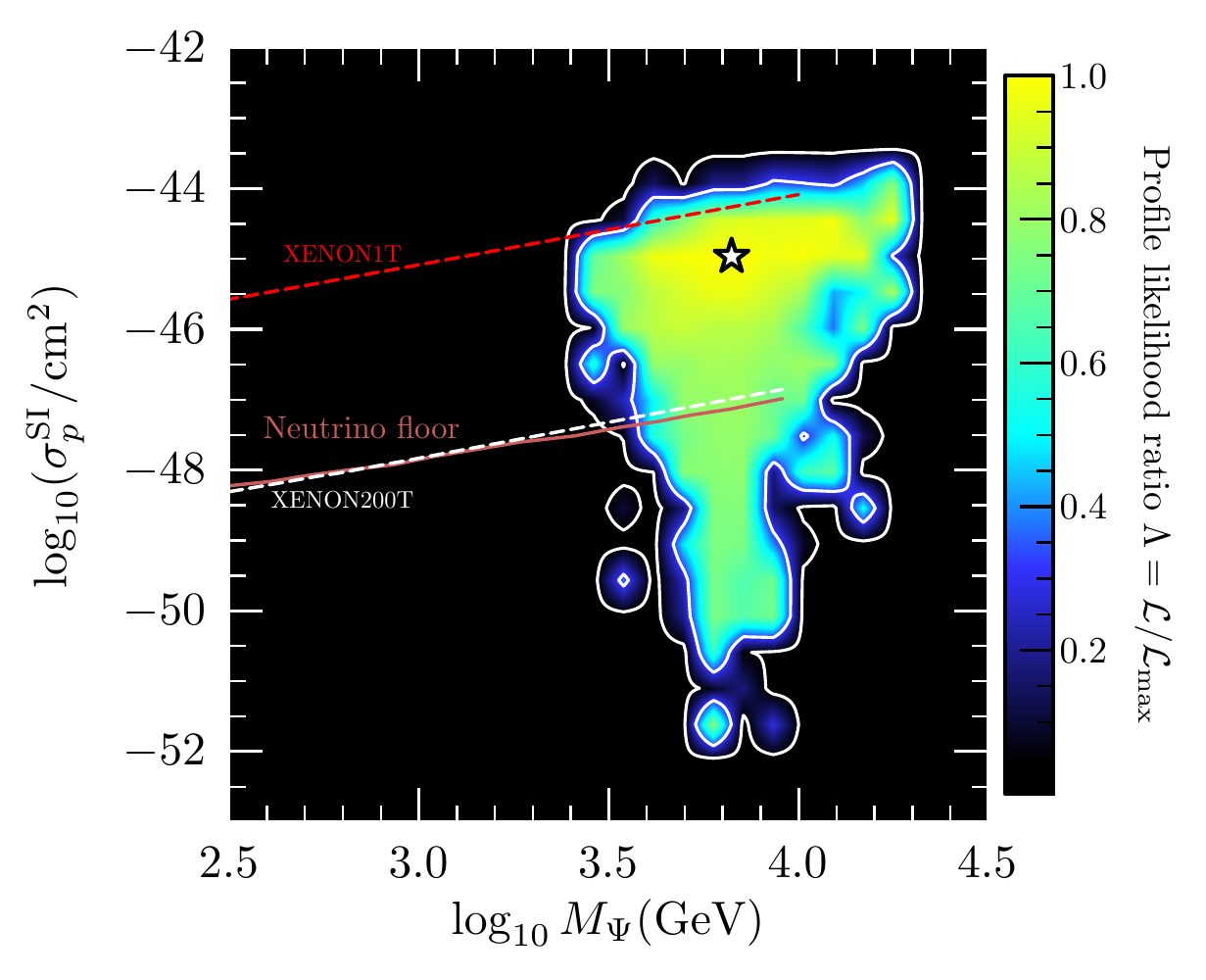}%
\caption{
	Composite likelihood as a function of
the DM candidate mass and SI DM-proton cross section for the case that the
candidate represents 100\% of the DM in the Universe. 
Contours of 68\% and 95\% of CL are drawn, and  also
shown are the 90\% CL upper limit from the 1t$\times$1yr XENON1T experiment, the
multi ton projection to 200t$\times$1yr and the neutrino floor. The best fit
point is marked with a star.}
\label{DD}
\end{figure}

\Catalina{
	
In figure (\ref{DD}), we present the profile likelihood
normalized to the value of $\mathcal{L}$ at the best fit point 
(signaled by a star) assuming the DM candidate
constitutes 100\% of the DM in the Universe. The plot shows the dependence
of the likelihood on the DM mass and the DM-proton spin independent (SI)
cross section; contours of 68\% and 95\% of confidence level (CL) are drawn.
We also depict the 90\% CL upper limit on the SI cross section from the
XENON1T (1t $\times$ yr) experiment \cite{XENON:2018voc}, alongside with the
multi ton-scale time projection to 200 t $\times$ yr of reference\footnote{
For better comparison with the other curves we extrapolated linearly the
data available from this reference from 1 TeV up to 10 TeV. 	
}
 \cite%
{Schumann:2015cpa} and an estimation of the neutrino floor \cite%
{Billard:2013qya}.

We note that almost all the region consistent
with the constraints including
the BFP lies below
the zone currently excluded by the XENON1T experiment.
However the figure also makes it evident that the
multi ton projection to 200t$\times$1yr
will be capable of probing zones well below the BFP of the model.

}


\section{Leptogenesis}
\label{lepto}
In this section we will analyze the implications of our model in
leptogenesis. Here we consider the case where $\left\vert y_{1}^{\left( \nu
\right) }\right\vert \ll \left\vert y_{2}^{\left( \nu \right) }\right\vert
,\left\vert y_{3}^{\left( \nu \right) }\right\vert $ and $\left\vert \frac{%
M_{1}}{v_{\xi }}\right\vert \ll \left\vert y_{2}\right\vert ,\left\vert
y_{3}\right\vert $. Therefore only the first generation of sterile neutrinos 
$N_{i}^{\pm }$ ($i=1,2,3$) can contribute to the Baryon asymmetry of the
Universe. We further assume that the gauge singlet neutral lepton $%
\Psi_{R}$ is heavier than the lightest pseudo-Dirac fermions $N_{1}^{\pm
}=N^{\pm }$. Then, the lepton asymmetry parameter, which is induced by decay
process of $N^{\pm }$, is given by \cite{Gu:2010xc,Pilaftsis:1997jf}: 
\begin{eqnarray}
\varepsilon _{\pm } &=&\dsum\limits_{i=1}^{3}\frac{\left[ \Gamma \left(
N_{\pm }\rightarrow l_{i}H^{+}\right) -\Gamma \left( N_{\pm }\rightarrow 
\bar{l}_{i}H^{-}\right) \right] }{\left[ \Gamma \left( N_{\pm }\rightarrow
l_{i}H^{+}\right) +\Gamma \left( N_{\pm }\rightarrow \bar{l}_{i}H^{-}\right) %
\right] }+\dsum\limits_{i=1}^{3}\frac{\left[ \Gamma \left( N_{\pm
}\rightarrow \nu _{i}A_{1}^{0}\right) -\Gamma \left( N_{\pm }\rightarrow \nu
_{i}A_{1}^{0}\right) \right] }{\left[ \Gamma \left( N_{\pm }\rightarrow \nu
_{i}A_{1}^{0}\right) +\Gamma \left( N_{\pm }\rightarrow \nu
_{i}A_{1}^{0}\right) \right] }  \notag \\
&&+\dsum\limits_{i=1}^{3}\frac{\left[ \Gamma \left( N_{\pm }\rightarrow \nu
_{i}h\right) -\Gamma \left( N_{\pm }\rightarrow \overline{\nu }_{i}h\right) %
\right] }{\left[ \Gamma \left( N_{\pm }\rightarrow \nu _{i}h\right) +\Gamma
\left( N_{\pm }\rightarrow \overline{\nu }_{i}h\right) \right] } \\
&\simeq &\frac{\func{Im}\left\{ \left( \left[ \left( y_{N_{+}}\right)
^{\dagger }\left( y_{N_{-}}\right) \right] ^{2}\right) _{11}\right\} }{8\pi
A_{\pm }}\frac{r}{r^{2}+\frac{\Gamma _{\pm }^{2}}{m_{N_{\pm }}^{2}}},
\end{eqnarray}%
with: 
\begin{eqnarray}
r &=&\frac{m_{N_{+}}^{2}-m_{N_{-}}^{2}}{m_{N_{+}}m_{N_{-}}},\hspace{0.7cm}%
\hspace{0.7cm}A_{\pm }=\left[ \left( y_{N_{\pm }}\right) ^{\dagger
}y_{N_{\pm }}\right] _{11},\hspace{0.7cm}\hspace{0.7cm}\Gamma _{\pm }=\frac{%
A_{\pm }m_{N_{\pm }}}{8\pi }, \\
y_{N_{\pm }} &=&\frac{m_{\nu D}}{v_{H_{2}}}\left( 1\mp S\right) =\frac{%
m_{\nu D}}{v_{H_{2}}}\left[ 1\pm \frac{1}{4}M^{-1}\left( \mu +\varepsilon
\right) \right]
\end{eqnarray}

Neglecting the interference terms involving the two different sterile
neutrinos $N^{\pm }$, the washout parameter $K_{N^{+}}+K_{N^{-}}$ is huge as
mentioned in \cite{Dolan:2018qpy}. However, the small mass splitting between
the pseudo-Dirac neutrinos leads to a destructive interference in the
scattering process \cite{Blanchet:2009kk}. The washout parameter including
the interference term has the following form: 
\begin{equation}
K^{eff}\simeq \left( K_{N^{+}}\delta _{+}^{2}+K_{N^{-}}\delta
_{-}^{2}\right) ,
\end{equation}%
where: 
\begin{equation}
\delta _{\pm }=\frac{m_{N^{+}}-m_{N^{-}}}{\Gamma _{N^{\pm }}},\hspace{0.7cm}%
\hspace{0.7cm}K_{N^{\pm }}=\frac{\Gamma _{\pm }}{H\left( T\right) },\hspace{%
0.7cm}\hspace{0.7cm}H(T)=\sqrt{\frac{4\pi ^{3}g^{\ast }}{45}}\frac{T^{2}}{%
M_{P}}
\end{equation}%
where $g^{\ast }=118$ is the number of effective relativistic degrees of
freedom, $M_{Pl}=1.2\times 10^{9}$ GeV is the Planck constant and $%
T=m_{N_{\pm }}$.

\begin{figure}[tbp]
\centering
\includegraphics[width=0.5\textwidth]{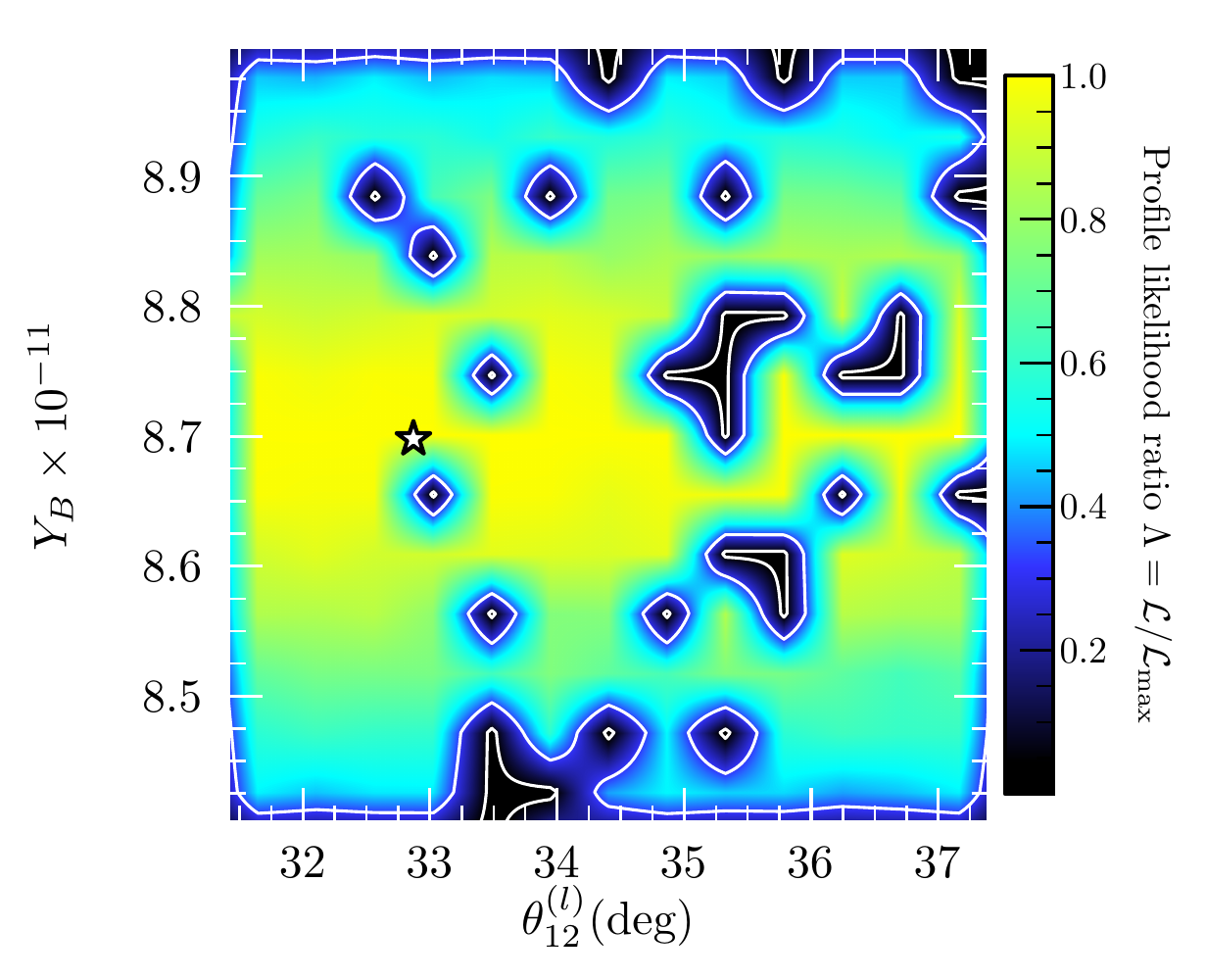}\includegraphics[width=0.5\textwidth]{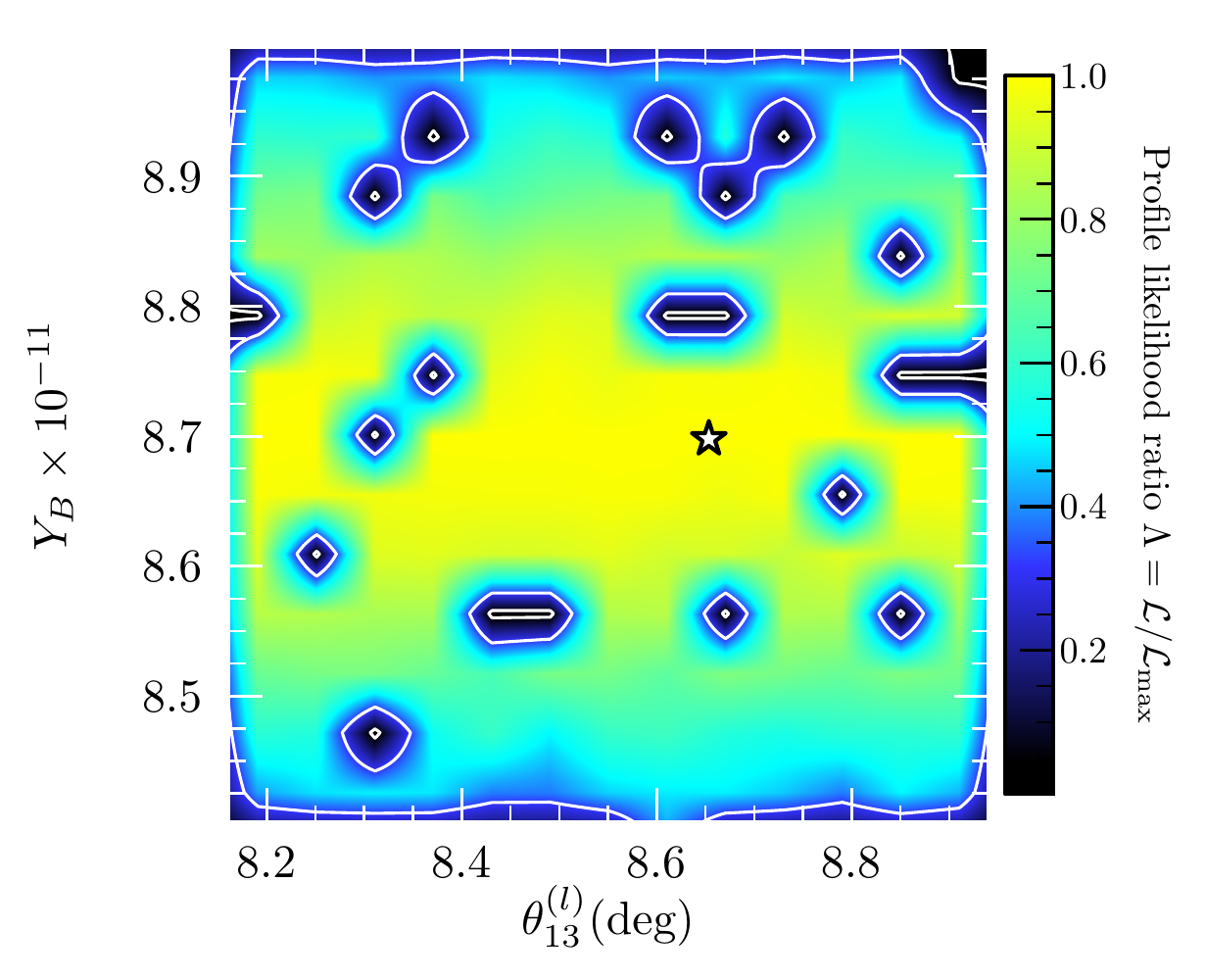}\newline
\includegraphics[width=0.5\textwidth]{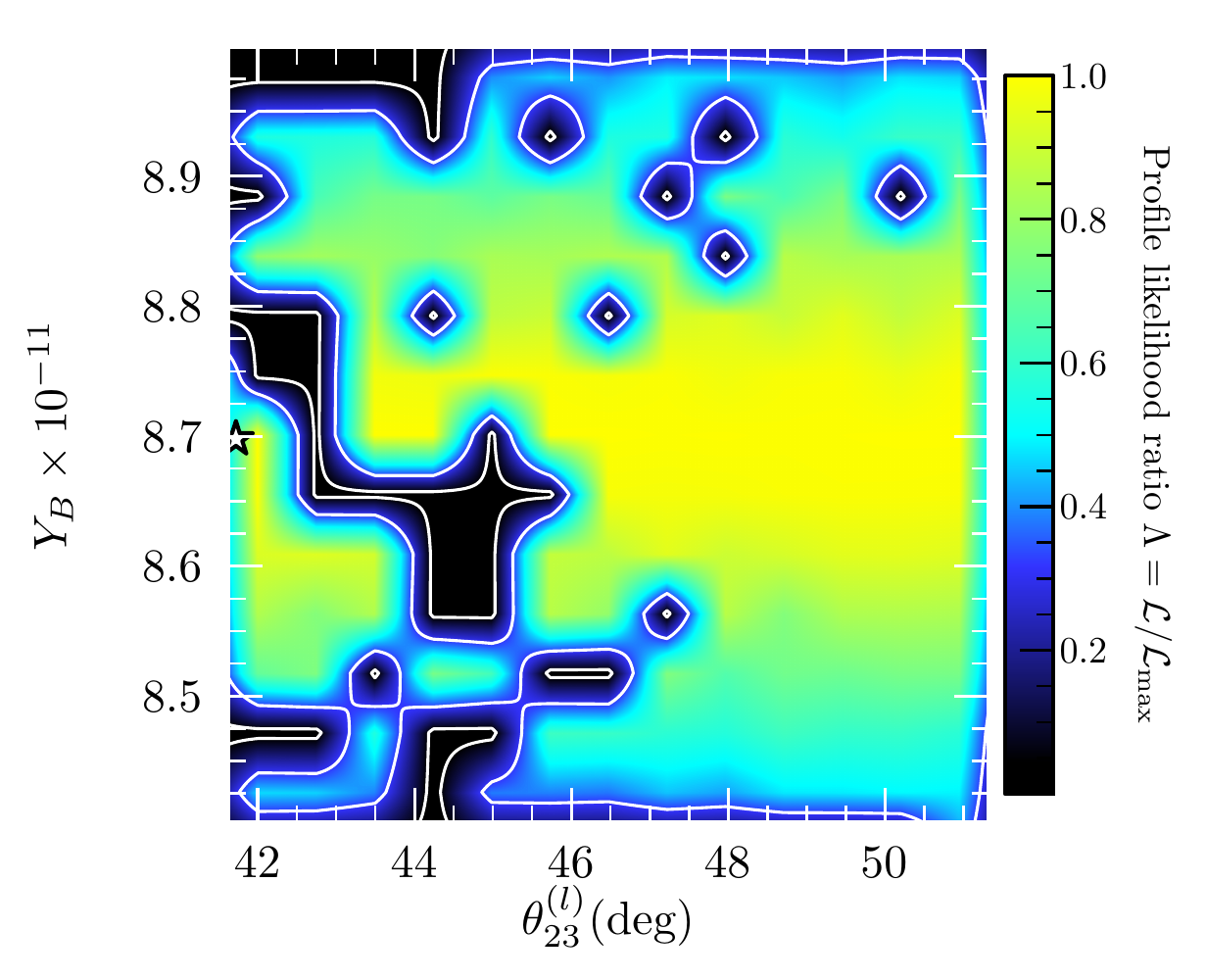}\includegraphics[width=0.5\textwidth]{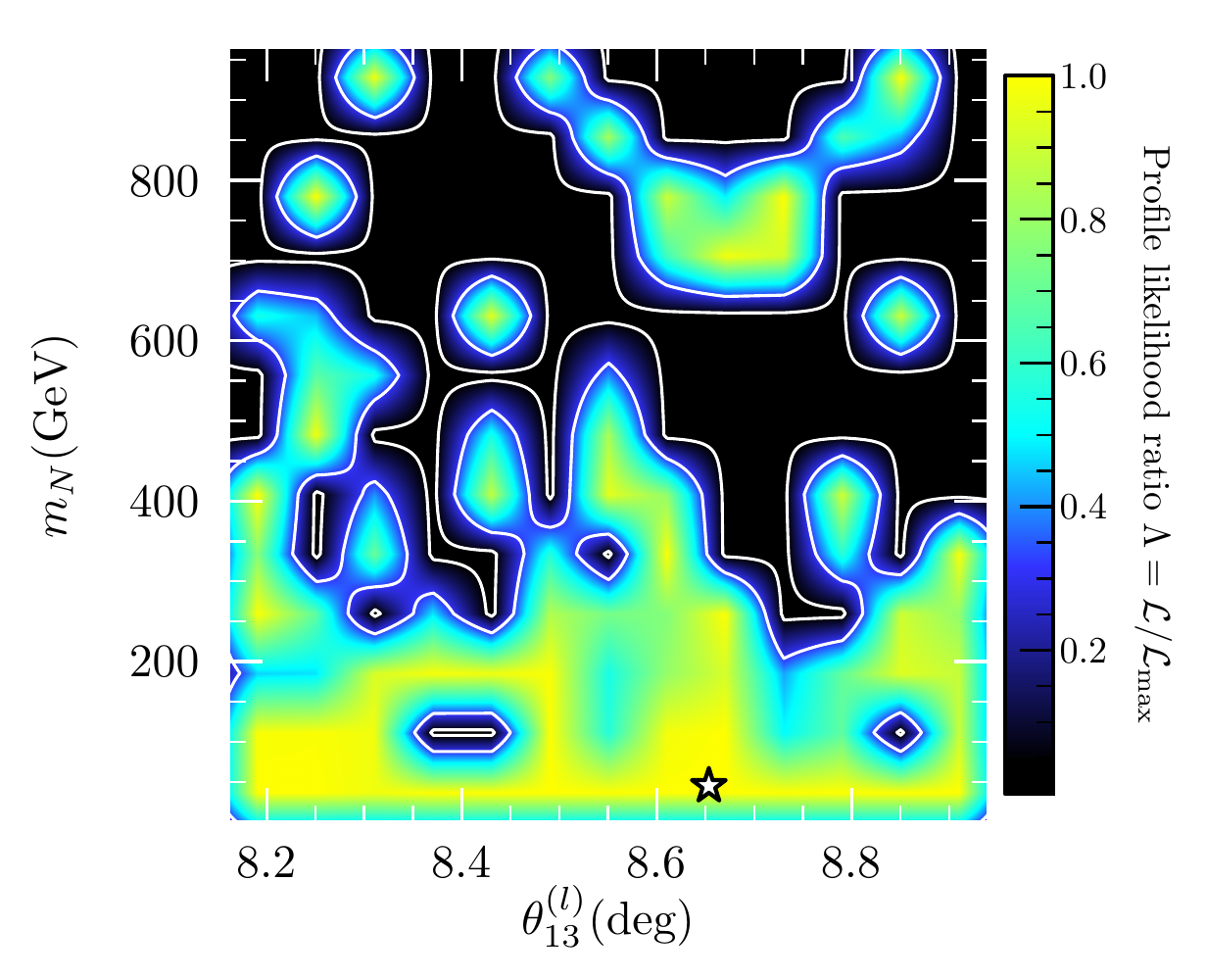}\newline
\includegraphics[width=0.5\textwidth]{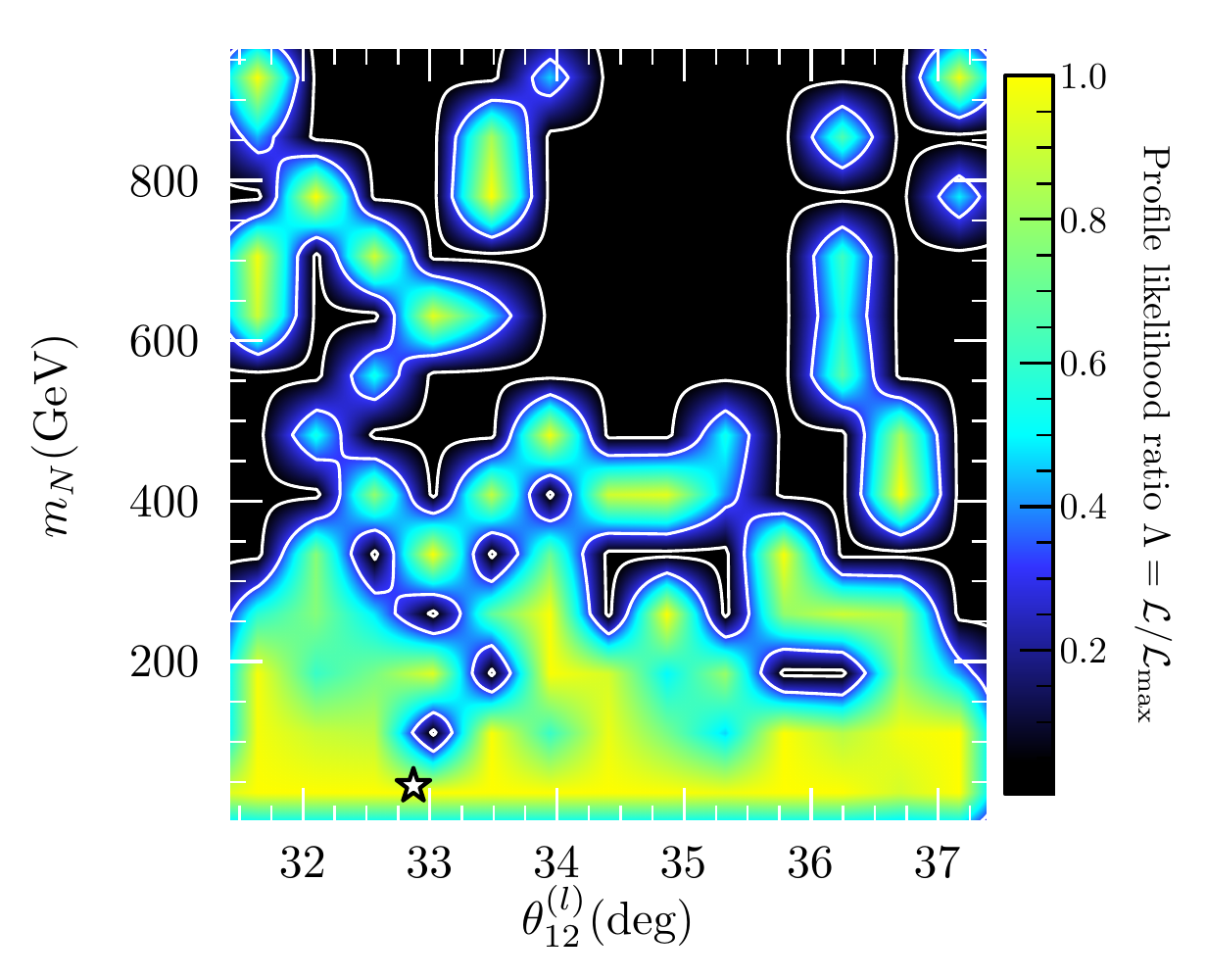}\includegraphics[width=0.5\textwidth]{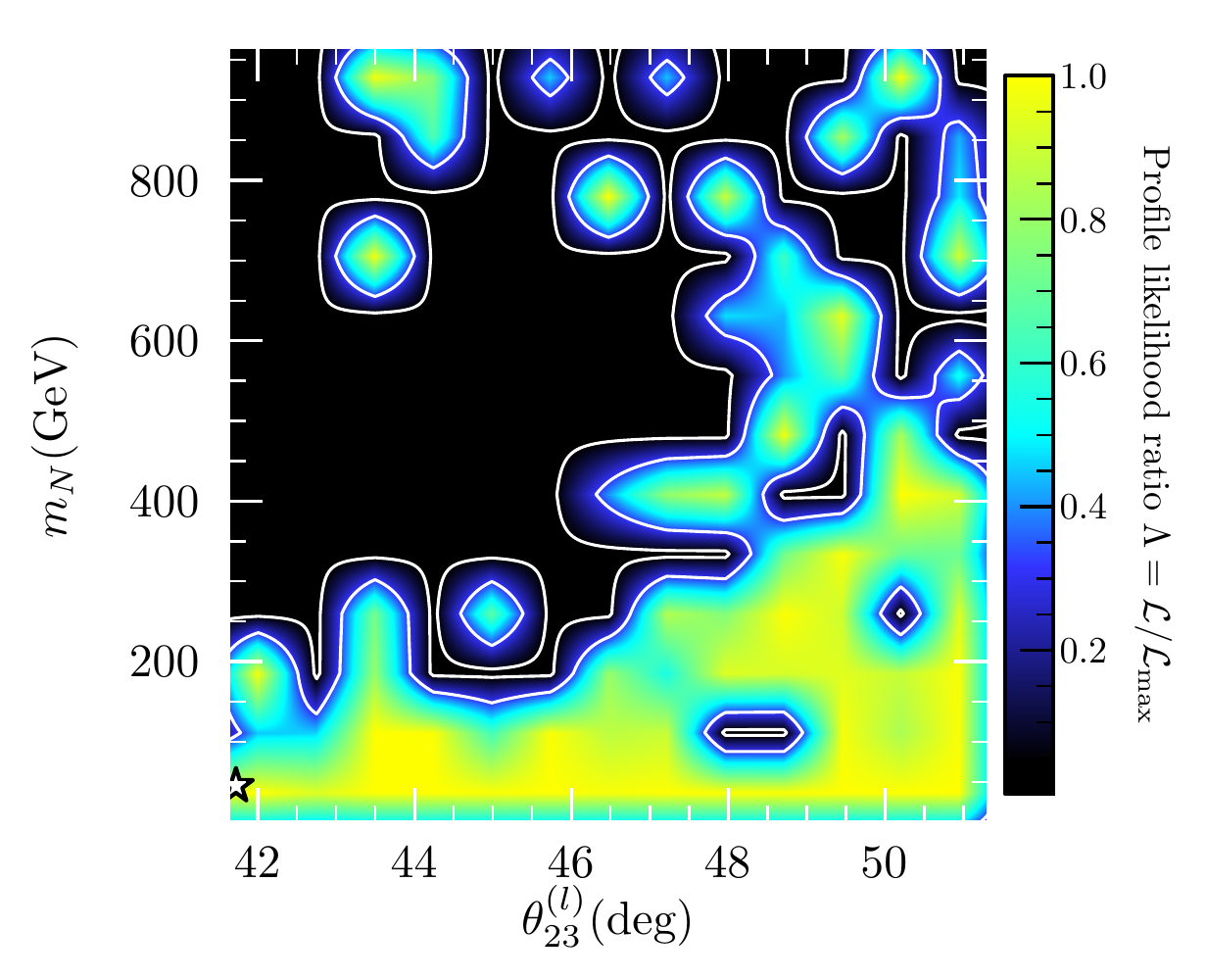}\newline
\caption{Allowed values of the baryon asymmetry parameter $Y_{B}$, leptonic mixing angles and the mass of the lightest pseudoDirac neutral lepton pair for the weak washout regime.}
\label{YBweak}
\end{figure}

In the weak and strong washout regimes, the baryon asymmetry is related to
the lepton asymmetry \cite{Pilaftsis:1997jf} as follows 
\begin{eqnarray}
Y_{\Delta B} &=&\frac{n_{B}-\overline{n}_{B}}{s}=-\frac{28}{79}\frac{%
\epsilon _{+}+\epsilon _{-}}{g^{\ast }},\hspace*{0.5cm}\text{for}\hspace*{%
0.5cm}K^{eff}\ll 1, \\
Y_{\Delta B} &=&\frac{n_{B}-\overline{n}_{B}}{s}=-\frac{28}{79}\frac{%
0.3\left( \epsilon _{+}+\epsilon _{-}\right) }{g^{\ast }K^{eff}\left( \ln
K^{eff}\right) ^{0.6}},\hspace*{0.5cm}\text{for}\hspace*{0.5cm}K^{eff}\gg 1,
\end{eqnarray}%
The correlation of the baryon asymmetry parameter $Y_{B}$ with the solar
mixing angle $\theta _{12}$ for the weak washout regime is shown in figure %
\ref{YBweak}. Our findings indicate that our model successfully accommodates
the experimental value of the baryon asymmetry parameter $Y_{B}$: 
\begin{equation}
Y_{\Delta B}=\left( 0.87\pm 0.01\right) \times 10^{-10}
\end{equation}%
Figure \ref{YBweak} shows the allowed values of the baryon asymmetry parameter $Y_{B}$, leptonic mixing angles and the mass of the lightest pseudoDirac neutral lepton pair for the weak washout regime. We find that the consistency with lepton masses and mixings, dark matter and baryon asymmetry constraints requires values for the leptonic mixing angles in the ranges $8.2^{\circ }\lesssim\theta^{(l)}_{13}\lesssim 8.9^{\circ }$, $31.5^{\circ }\lesssim\theta^{(l)}_{12}\lesssim 37.5^{\circ }$, $42^{\circ }\lesssim\theta^{(l)}_{23}\lesssim 51^{\circ }$ as well as a mass for the lightest heavy pseudo Dirac neutral lepton pair at the subTeV scale. 

\newpage

\section{Conclusions}

We have built a predictive and viable extended 2HDM, where the scalar and
fermion sectors are enlarged by the inclusion of gauge singlet scalars and
right handed Majorana neutrinos, respectively. The model incorporates the $%
Q_{4}$ family symmetry, which is supplemented by several auxiliary cyclic
symmetries, which allows to successfully describe the current pattern of SM
fermion masses and mixing angles, which is caused by the spontaneous
breaking of the discrete symmetries. The tiny masses of the light active
neutrinos are produced by an inverse seesaw mechanism at one loop level, due
to a remnant preserved $Z_{2}$ symmetry resulting from the spontaneous
breaking of the $Z_{4}$ discrete group. 
Under certain simplifying assumptions made in the scalar and neutrino sectors (equality of a pair of Yukawa couplings) and described in detail in the introduction and throughout the text, our 
model successfully accommodates
the experimental value of the dark matter relic density, the muon anomalous
magnetic moment as well as the lepton and baryon asymmetries of the
Universe. The consistency of our model with the constraints arising from 
collider searches for heavy scalars,
stability of the scalar potentials,
the dark matter relic density and current and future direct detection
experiments sets the mass of the scalar dark matter candidate to be
in between $2.5$ TeV and $20$ TeV. Finally our current flavor model intends to address and connect several problems such as the SM flavor puzzle, the current amount of dark matter and baryon asymmetries observed in the Universe, the muon anomalous magnetic moment. It predicts extended Gatto-Sartori-Tonin relations between the quark masses and mixing angles, a baryon asymmetry parameter between $8.4\times 10^{-11}$ and $9\times 10^{-11}$, leptonic mixing angles in the following ranges $8.2^{\circ }\lesssim\theta^{(l)}_{13}\lesssim 8.9^{\circ }$, $31.5^{\circ }\lesssim\theta^{(l)}_{12}\lesssim 37.5^{\circ }$, $42^{\circ }\lesssim\theta^{(l)}_{23}\lesssim 51^{\circ }$, the mass of the lightest heavy pseudo Dirac neutral lepton pair at the subTeV scale, the $\tan\beta$ parameter in the range $0.1\lesssim\tan\beta\lesssim 0.6$ and heavy non SM scalars at the subTeV scale with masses in the ranges $150$ GeV$\lesssim M_{H_3}\lesssim 400$ GeV, $300$ GeV$\lesssim M_{A}\lesssim 900$ GeV, $100$ GeV$\lesssim M_{H^{\pm}}\lesssim 500$ GeV with preferred values for charged scalar masses around $400$ GeV. It is worth mentioning that the extended Gatto-Sartori-Tonin relations predicted in the quark sector of the model are a direct consequence of the symmetries and the particle assignments under the discrete and SM gauge groups. The presence of heavy non SM scalar masses at the subTeV scale makes our model testable at colliders via the scalar production at the LHC by gluon fusion mechanism and Drell-Yan associated production with a SM gauge boson. Furthermore, our model has a heavy scalar above $20$ TeV, whose production can be relevant in a future $100$ TeV proton-proton collider. Besides that, in the simplified cobimaximal benchmark scenario considered in this work, we obtained values for the leptonic Dirac CP violating phase close to about $-90^{\circ }$.

\label{conclusions}

\section*{Acknowledgments}

A.E.C.H is supported by ANID-Chile FONDECYT 1210378, ANID PIA/APOYO
AFB180002 and ANID- Programa Milenio - code ICN2019\_044. C.E. acknowledges the support of
Conacyt (M\'{e}xico) C\'{a}tedra no. 341. This research is partially
supported by DGAPA PAPIIT IN109321. A.E.C.H is very grateful to the
Instituto de F\'{\i}sica, Universidad Nacional Aut\'{o}noma de M\'{e}xico
for hospitality and for financing his visit where part of this work was
done. JCGI is supported by SIP IPN Project 20211423.

\section*{Data Availability Statement}

This manuscript has no associated data or the data will not be deposited.
Authors comment: This article is based on research in theoretical physics.
Therefore, there are no associated data to be deposited.

\appendix

\section{The product rules for $Q_{4}$}

\label{Q4}

The irreducible representations of the $Q_{4}$ group are four singlets, $%
\mathbf{1}_{++}$, $\mathbf{1}_{+-}$, $\mathbf{1}_{-+}$ and $\mathbf{1}_{--}$%
, and one doublet $\mathbf{2}$. The tensor products of the $Q_{4}$
irreducible representation are given by \cite{Ishimori:2010au}:

\begin{eqnarray}
\left( 
\begin{array}{c}
z \\ 
\bar{z}%
\end{array}%
\right) _{\mathbf{2}}\otimes \left( 
\begin{array}{c}
z^{\prime } \\ 
\bar{z}^{\prime }%
\end{array}%
\right) _{\mathbf{2}} &=&\left( z\bar{z}^{\prime }-\bar{z}z^{\prime }\right)
_{\mathbf{1}_{++}}\oplus \left( z\bar{z}^{\prime }+\bar{z}z^{\prime }\right)
_{\mathbf{1}_{--}}  \notag \\
&&\oplus \left( zz^{\prime }-\bar{z}\bar{z}^{\prime }\right) _{\mathbf{1}%
_{+-}}\oplus \left( zz^{\prime }+\bar{z}\bar{z}^{\prime }\right) _{\mathbf{1}%
_{-+}},
\end{eqnarray}%
%
%
%
%
%
%
%
%
%
%
%
%
%
%
%
%
%
%
%
%
%
%
%
%
%
%
%
%
%
%
%
%
%
%
%
%
%
%
%
%
%
%
%
%
%
%
%
\begin{eqnarray}
&&\left( w\right) _{\mathbf{1}_{++}}\otimes \left( 
\begin{array}{c}
z \\ 
\bar{z}%
\end{array}%
\right) _{\mathbf{2}}=\left( 
\begin{array}{c}
wz \\ 
w\bar{z}%
\end{array}%
\right) _{\mathbf{2}},\quad \left( w\right) _{\mathbf{1}_{--}}\otimes \left( 
\begin{array}{c}
z \\ 
\bar{z}%
\end{array}%
\right) _{\mathbf{2}}=\left( 
\begin{array}{c}
wz \\ 
-w\bar{z}%
\end{array}%
\right) _{\mathbf{2}},  \notag \\
&&\left( w\right) _{\mathbf{1}_{+-}}\otimes \left( 
\begin{array}{c}
z \\ 
\bar{z}%
\end{array}%
\right) _{\mathbf{2}}=\left( 
\begin{array}{c}
w\bar{z} \\ 
wz%
\end{array}%
\right) _{\mathbf{2}},\quad \left( w\right) _{\mathbf{1}_{-+}}\otimes \left( 
\begin{array}{c}
z \\ 
\bar{z}%
\end{array}%
\right) _{\mathbf{2}}=\left( 
\begin{array}{c}
w\bar{z} \\ 
-wz%
\end{array}%
\right) _{\mathbf{2}},
\end{eqnarray}%
\begin{equation*}
\mathbf{1}_{s_{1}s_{2}}\otimes \mathbf{1}_{s_{1}^{\prime }s_{2}^{\prime }}=%
\mathbf{1}_{s_{1}^{\prime \prime }s_{2}^{\prime \prime }},
\end{equation*}%
where $s_{1}^{\prime \prime }=s_{1}s_{1}^{\prime }$ and $s_{2}^{\prime
\prime }=s_{2}s_{2}^{\prime }$.

\section{Scalar potential for two $Q_{4}$ doublets.}

\label{Q4scalarpotential}

The scalar potential for two $Q_{4}$ doublets $\xi $ and $\Phi $ (with $\xi $
real and $\Phi $ complex) has the form

\begin{eqnarray}
V &=&\mu _{\xi }^{2}\left( \xi \xi \right) _{\mathbf{1}_{++}}+\mu _{\Phi
}^{2}\left( \Phi \Phi ^{\dagger }\right) _{\mathbf{1}_{++}}+\lambda
_{1}\left( \xi \xi \right) _{\mathbf{1}_{++}}\left( \xi \xi \right) _{%
\mathbf{1}_{++}}+\lambda _{2}\left( \xi \xi \right) _{\mathbf{1}_{+-}}\left(
\xi \xi \right) _{\mathbf{1}_{+-}}+\lambda _{3}\left( \xi \xi \right) _{%
\mathbf{1}_{-+}}\left( \xi \xi \right) _{\mathbf{1}_{-+}}  \notag \\
&&+\lambda _{4}\left( \xi \xi \right) _{\mathbf{1}_{--}}\left( \xi \xi
\right) _{\mathbf{1}_{--}}+\lambda _{5}\left( \Phi \Phi ^{\dagger }\right) _{%
\mathbf{1}_{++}}\left( \Phi \Phi ^{\dagger }\right) _{\mathbf{1}%
_{++}}+\lambda _{6}\left( \Phi \Phi ^{\dagger }\right) _{\mathbf{1}%
_{+-}}\left( \Phi \Phi ^{\dagger }\right) _{\mathbf{1}_{+-}}+\lambda
_{7}\left( \Phi \Phi ^{\dagger }\right) _{\mathbf{1}_{-+}}\left( \Phi \Phi
^{\dagger }\right) _{\mathbf{1}_{-+}}  \notag \\
&&+\lambda _{8}\left( \Phi \Phi ^{\dagger }\right) _{\mathbf{1}_{--}}\left(
\Phi \Phi ^{\dagger }\right) _{\mathbf{1}_{--}}+\lambda _{9}\left( \xi \xi
\right) _{\mathbf{1}_{++}}\left( \Phi \Phi ^{\dagger }\right) _{\mathbf{1}%
_{++}}+\lambda _{10}\left( \xi \xi \right) _{\mathbf{1}_{+-}}\left( \Phi
\Phi ^{\dagger }\right) _{\mathbf{1}_{+-}}+\lambda _{11}\left( \xi \xi
\right) _{\mathbf{1}_{-+}}\left( \Phi \Phi ^{\dagger }\right) _{\mathbf{1}%
_{-+}}  \notag \\
&&+\lambda _{12}\left( \xi \xi \right) _{\mathbf{1}_{--}}\left( \Phi \Phi
^{\dagger }\right) _{\mathbf{1}_{--}}
\end{eqnarray}

The above given scalar potential can be rewritten as follows:

\begin{eqnarray}
V &=&\mu _{\Phi }^{2}\left( \Phi _{1}\Phi _{2}^{\dagger }-\Phi _{2}\Phi
_{1}^{\dagger }\right) +\lambda _{2}\left( \xi _{1}^{2}-\xi _{2}^{2}\right)
^{2}+\lambda _{3}\left( \xi _{1}^{2}+\xi _{2}^{2}\right) ^{2}+4\lambda
_{4}\xi _{1}^{2}\xi _{2}^{2}+\lambda _{5}\left( \Phi _{1}\Phi _{2}^{\dagger
}-\Phi _{2}\Phi _{1}^{\dagger }\right) ^{2}  \notag \\
&&+\lambda _{6}\left( \Phi _{1}\Phi _{1}^{\dagger }-\Phi _{2}\Phi
_{2}^{\dagger }\right) ^{2}+\lambda _{7}\left( \Phi _{1}\Phi _{1}^{\dagger
}+\Phi _{2}\Phi _{2}^{\dagger }\right) ^{2}+\lambda _{8}\left( \Phi _{1}\Phi
_{2}^{\dagger }+\Phi _{2}\Phi _{1}^{\dagger }\right) ^{2}  \notag \\
&&+\lambda _{10}\left( \xi _{1}^{2}-\xi _{2}^{2}\right) \left( \Phi _{1}\Phi
_{1}^{\dagger }-\Phi _{2}\Phi _{2}^{\dagger }\right) +\lambda _{11}\left(
\xi _{1}^{2}+\xi _{2}^{2}\right) \left( \Phi _{1}\Phi _{1}^{\dagger }+\Phi
_{2}\Phi _{2}^{\dagger }\right) +2\lambda _{12}\xi _{1}\xi _{2}\left( \Phi
_{1}\Phi _{2}^{\dagger }+\Phi _{2}\Phi _{1}^{\dagger }\right)
\end{eqnarray}

Due to hermiticity, the parameters are reals and the minimum conditions are
the following

\begin{eqnarray}
0&=&v_{\xi_{1}}\left[\lambda_{2}\left(v^{2}_{\xi_{1}}-v^{2}_{\xi_{2}}%
\right)+\lambda_{3}\left(v^{2}_{\xi_{1}}+v^{2}_{\xi_{2}}\right)+2%
\lambda_{4}v^{2}_{\xi_{2}}+v_{\Phi_{1}}v_{\Phi_{2}}\left\{\lambda_{10}\cos{%
(\alpha-\theta)}+i\lambda_{11}\sin{(\theta-\alpha)}\right\}+\lambda_{12}%
\frac{v_{\xi_{2}}}{2v_{\xi_{1}}}\left(v^{2}_{\Phi_{2}}-v^{2}_{\Phi_{1}}%
\right)\right],  \notag \\
0&=&v_{\xi_{2}}\left[-\lambda_{2}\left(v^{2}_{\xi_{1}}-v^{2}_{\xi_{2}}%
\right)+\lambda_{3}\left(v^{2}_{\xi_{1}}+v^{2}_{\xi_{2}}\right)+2%
\lambda_{4}v^{2}_{\xi_{1}}+v_{\Phi_{1}}v_{\Phi_{2}}\left\{-\lambda_{10}\cos{%
(\alpha-\theta)}+i\lambda_{11}\sin{(\theta-\alpha)}\right\}+\lambda_{12}%
\frac{v_{\xi_{1}}}{2v_{\xi_{2}}}\left(v^{2}_{\Phi_{2}}-v^{2}_{\Phi_{1}}%
\right)\right],  \notag \\
0&=&v_{\Phi_{1}}\left[\frac{\mu^{2}_{\Phi}}{2}+\lambda_{5}\left(v^{2}_{%
\Phi_{1}}+v^{2}_{\Phi_{2}}\right)+2v^{2}_{\Phi_{2}}\left\{\lambda_{6}\cos^{2}%
{(\alpha-\theta)}-\lambda_{7}\sin^{2}{(\theta-\alpha)}\right\}-\lambda_{8}%
\left(v^{2}_{\Phi_{2}}-v^{2}_{\Phi_{1}}\right)-\lambda_{12}v_{\xi_{1}}v_{%
\xi_{2}}\right.  \notag \\
&&\left.+\frac{v_{\Phi_{2}}}{2v_{\Phi_{1}}}\left\{\lambda_{10}\left(v^{2}_{%
\xi_{1}}-v^{2}_{\xi_{2}}\right)\cos{(\alpha-\theta)}+i\lambda_{11}%
\left(v^{2}_{\xi_{1}}+v^{2}_{\xi_{2}}\right)\sin{(\theta-\alpha)}\right\}%
\right],  \notag \\
0&=&v_{\Phi_{2}}\left[\frac{\mu^{2}_{\Phi}}{2}+\lambda_{5}\left(v^{2}_{%
\Phi_{1}}+v^{2}_{\Phi_{2}}\right)+2v^{2}_{\Phi_{1}}\left\{\lambda_{6}\cos^{2}%
{(\alpha-\theta)}-\lambda_{7}\sin^{2}{(\theta-\alpha)}\right\}+\lambda_{8}%
\left(v^{2}_{\Phi_{2}}-v^{2}_{\Phi_{1}}\right)+\lambda_{12}v_{\xi_{1}}v_{%
\xi_{2}}\right.  \notag \\
&&\left.+\frac{v_{\Phi_{1}}}{2v_{\Phi_{2}}}\left\{\lambda_{10}\left(v^{2}_{%
\xi_{1}}-v^{2}_{\xi_{2}}\right)\cos{(\alpha-\theta)}+i\lambda_{11}%
\left(v^{2}_{\xi_{1}}+v^{2}_{\xi_{2}}\right)\sin{(\theta-\alpha)}\right\}%
\right].  \label{ap2}
\end{eqnarray}

where we have considered in general

\begin{equation}
\left\langle \xi \right\rangle =\left(v_{\xi_{1} }, v_{\xi_{2} }\right) ,%
\hspace{1.5cm}\left\langle \Phi \right\rangle =\left(v_{\Phi_{1} }
e^{i\theta },v_{\Phi_{2} }e^{i\alpha }\right).
\end{equation}

According to our purpose, we need the alignment $\left\langle \xi
\right\rangle =v_{\xi }\left( 1,0\right) $ ($v_{\xi _{1}}\neq 0$ and $v_{\xi
_{2}}=0$), then we use the former two expressions in Eq. (\ref{ap2}) to
obtain 
\begin{eqnarray}
0 &=&v_{\xi }^{2}\left( \lambda _{2}+\lambda _{3}\right) +v_{\Phi
_{1}}v_{\Phi _{2}}\left\{ \lambda _{10}\cos {(\alpha -\theta )}+i\lambda
_{11}\sin {(\theta -\alpha )}\right\} ,  \notag \\
0 &=&v_{\Phi _{1}}\left[ \frac{\mu _{\Phi }^{2}}{2}+\lambda _{5}\left(
v_{\Phi _{1}}^{2}+v_{\Phi _{2}}^{2}\right) +2v_{\Phi _{2}}^{2}\left\{
\lambda _{6}\cos ^{2}{(\alpha -\theta )}-\lambda _{7}\sin ^{2}{(\theta
-\alpha )}\right\} -\lambda _{8}\left( v_{\Phi _{2}}^{2}-v_{\Phi
_{1}}^{2}\right) \right.  \notag \\
&&\left. +\frac{v_{\Phi _{2}}}{2v_{\Phi _{1}}}\left\{ \lambda _{10}\cos {%
(\alpha -\theta )}+i\lambda _{11}\sin {(\theta -\alpha )}\right\} v_{\xi
}^{2}\right] ,  \notag \\
0 &=&v_{\Phi _{2}}\left[ \frac{\mu _{\Phi }^{2}}{2}+\lambda _{5}\left(
v_{\Phi _{1}}^{2}+v_{\Phi _{2}}^{2}\right) +2v_{\Phi _{1}}^{2}\left\{
\lambda _{6}\cos ^{2}{(\alpha -\theta )}-\lambda _{7}\sin ^{2}{(\theta
-\alpha )}\right\} +\lambda _{8}\left( v_{\Phi _{2}}^{2}-v_{\Phi
_{1}}^{2}\right) \right.  \notag \\
&&\left. +\frac{v_{\Phi _{1}}}{2v_{\Phi _{2}}}\left\{ \lambda _{10}\cos {%
(\alpha -\theta )}+i\lambda _{11}\sin {(\theta -\alpha )}\right\} v_{\xi
}^{2}\right]  \label{ap3}
\end{eqnarray}%
As one can notice, in the last to expressions in Eq. (\ref{ap3}), there is a
symmetry of interchange $v_{\Phi _{1}}\leftrightarrow v_{\Phi _{2}}$. Along
with this, we demand that $v_{\Phi _{1}}\neq 0\neq v_{\Phi _{2}}$ therefore $%
v_{\Phi _{1}}=v_{\Phi _{2}}\equiv v_{\Phi }$ from the last two expressions.
Finally, we end up having 
\begin{eqnarray}
0 &=&v_{\xi }^{2}\left( \lambda _{2}+\lambda _{3}\right) +v_{\Phi
}^{2}\left\{ \lambda _{10}\cos {(\alpha -\theta )}+i\lambda _{11}\sin {%
(\theta -\alpha )}\right\} ,  \notag \\
0 &=&\frac{\mu _{\Phi }^{2}}{2}+2v_{\Phi }^{2}\left\{ \lambda _{5}+\lambda
_{6}\cos ^{2}{(\alpha -\theta )}-\lambda _{7}\sin ^{2}{(\theta -\alpha )}%
\right\} +\frac{v_{\xi }^{2}}{2}\left\{ \lambda _{10}\cos {(\alpha -\theta )}%
+i\lambda _{11}\sin {(\theta -\alpha )}\right\} .  \label{ap4}
\end{eqnarray}%
This shows that the VEV pattern of the two $Q_{4}$ doublets $\xi $ and $\Phi 
$ shown in Eq. (\ref{Q4VEVpattern}) is consistent with the minimization
conditions of the scalar potential.

\section{Stability of the scalar potential for two $Q_{4}$ doublets}

With the aim to determine the stability conditions of the scalar potential
for the two $Q_{4}$ doublets $\xi $ and $\Phi $, we proceed to analyze its
quartic terms because they will dominate the behavior of the scalar
potential in the region of very large values of the field components. To
this end, we introduce the following hermitian bilinear combination of the
scalar fields: 
\begin{eqnarray}
a &=&\Phi _{1}\Phi _{1}^{\dagger },\hspace{1.5cm}b=\Phi _{2}\Phi
_{2}^{\dagger },\hspace{1.5cm}c=\Phi _{1}\Phi _{2}^{\dagger }+\Phi _{2}\Phi
_{1}^{\dagger },\hspace{1.5cm}d=i\left( \Phi _{1}\Phi _{2}^{\dagger }-\Phi
_{2}\Phi _{1}^{\dagger }\right) ,  \notag \\
e &=&\xi _{1}^{2},\hspace{1.5cm}f=\xi _{2}^{2}
\end{eqnarray}
and rewrite the quartic terms of the scalar potential for the two $Q_{4}$
doublets $\xi $ and $\Phi $: 
\begin{eqnarray}
V_{4} &=&\lambda _{2}\left( \xi _{1}^{2}-\xi _{2}^{2}\right) ^{2}+\lambda
_{3}\left( \xi _{1}^{2}+\xi _{2}^{2}\right) ^{2}+4\lambda _{4}\xi
_{1}^{2}\xi _{2}^{2}+\lambda _{5}\left( \Phi _{1}\Phi _{2}^{\dagger }-\Phi
_{2}\Phi _{1}^{\dagger }\right) ^{2}+\lambda _{6}\left( \Phi _{1}\Phi
_{1}^{\dagger }-\Phi _{2}\Phi _{2}^{\dagger }\right) ^{2}  \notag \\
&&+\lambda _{7}\left( \Phi _{1}\Phi _{1}^{\dagger }+\Phi _{2}\Phi
_{2}^{\dagger }\right) ^{2}+\lambda _{8}\left( \Phi _{1}\Phi _{2}^{\dagger
}+\Phi _{2}\Phi _{1}^{\dagger }\right) ^{2}+\lambda _{10}\left( \xi
_{1}^{2}-\xi _{2}^{2}\right) \left( \Phi _{1}\Phi _{1}^{\dagger }-\Phi
_{2}\Phi _{2}^{\dagger }\right)  \notag \\
&&+\lambda _{11}\left( \xi _{1}^{2}+\xi _{2}^{2}\right) \left( \Phi _{1}\Phi
_{1}^{\dagger }+\Phi _{2}\Phi _{2}^{\dagger }\right) +2\lambda _{12}\xi
_{1}\xi _{2}\left( \Phi _{1}\Phi _{2}^{\dagger }+\Phi _{2}\Phi _{1}^{\dagger
}\right)
\end{eqnarray}
in the following form: 
\begin{eqnarray}
V_{4} &=&\left( \lambda _{2}+\lambda _{3}\right) \left( e^{2}+f^{2}\right)
+2\left( \lambda _{3}-\lambda _{2}+2\lambda _{4}\right) ef-\lambda
_{5}d^{2}+\left( \lambda _{6}+\lambda _{7}\right) \left( a^{2}+b^{2}\right)
+2\left( \lambda _{7}-\lambda _{6}\right) ab  \notag \\
&&+\lambda _{8}c^{2}+\lambda _{10}\left( e-f\right) \left( a-b\right)
+\lambda _{11}\left( e+f\right) \left( a+b\right) +2\lambda _{12}\sqrt{ef}c
\end{eqnarray}
Defining 
\begin{equation}
\kappa _{1}=\lambda _{2}+\lambda _{3},\hspace{1.5cm}\kappa _{2}=2\left(
\lambda _{3}-\lambda _{2}+2\lambda _{4}\right) ,\hspace{1.5cm}\kappa
_{3}=\lambda _{6}+\lambda _{7},\hspace{1.5cm}\kappa _{4}=2\left( \lambda
_{7}-\lambda _{6}\right) ,
\end{equation}
The above given quartic scalar interactions can be rewritten as follows: 
\begin{eqnarray}
V_{4} &=&\kappa _{1}\left( e^{2}+f^{2}\right) +\kappa _{2}ef-\lambda
_{5}d^{2}+\kappa _{3}\left( a^{2}+b^{2}\right) +\kappa _{4}ab+\lambda
_{8}c^{2}  \notag \\
&&+\lambda _{10}\left( e-f\right) \left( a-b\right) +\lambda _{11}\left(
e+f\right) \left( a+b\right) +2\lambda _{12}\sqrt{ef}c  \notag \\
&=&\frac{\kappa _{1}}{2}\left[ \left( e-f\right) ^{2}+\left( e+f\right) ^{2}%
\right] +\frac{\kappa _{3}}{2}\left[ \left( a-b\right) ^{2}+\left(
a+b\right) ^{2}\right]  \notag \\
&&+\kappa _{2}ef-\lambda _{5}d^{2}+\kappa _{4}ab  \notag \\
&&+\lambda _{8}c^{2}+\lambda _{10}\left( e-f\right) \left( a-b\right)
+\lambda _{11}\left( e+f\right) \left( a+b\right) +2\lambda _{12}\sqrt{ef}c 
\notag \\
&=&\left[ \sqrt{\frac{\kappa _{1}}{2}}\left( e-f\right) +\sqrt{\frac{\kappa
_{3}}{2}}\left( a-b\right) \right] ^{2}+\left[ \sqrt{\frac{\kappa _{1}}{2}}%
\left( e+f\right) +\sqrt{\frac{\kappa _{3}}{2}}\left( a+b\right) \right] ^{2}
\notag \\
&&+\left( \lambda _{10}-\sqrt{\kappa _{1}\kappa _{3}}\right) \left(
e-f\right) \left( a-b\right) +\left( \lambda _{11}-\sqrt{\kappa _{1}\kappa
_{3}}\right) \left( e+f\right) \left( a+b\right)  \notag \\
&&-\lambda _{5}d^{2}+\kappa _{4}ab+\left[ \sqrt{\kappa _{2}}\sqrt{ef}+\sqrt{%
\lambda _{8}}c\right] ^{2}+2\left( \lambda _{12}-\sqrt{\kappa _{2}\lambda
_{8}}\right) \sqrt{ef}c
\end{eqnarray}
Following the procedure used for analyzing the stability described in Refs. 
\cite{Maniatis:2006fs,Bhattacharyya:2015nca}, we find that our scalar
potential of two $Q_{4}$ doublets will be stable when the following
conditions are fulfilled: 
\begin{eqnarray}
\lambda _{2}+\lambda _{3} &\geq &0,\hspace{1.5cm}\lambda _{6}+\lambda
_{7}\geq 0,\hspace{0.7cm}\lambda _{10}-\sqrt{\left( \lambda _{2}+\lambda
_{3}\right) \left( \lambda _{6}+\lambda _{7}\right) }\geq 0,\hspace{0.7cm}\lambda_{11}-\sqrt{\left( \lambda _{2}+\lambda
_{3}\right) \left( \lambda _{6}+\lambda _{7}\right) }\geq 0,  \notag \\
\lambda _{5}&\leq &0,\hspace{0.7cm}\lambda _{7} \geq\lambda _{6},\hspace{0.7cm}\lambda _{8}\geq 0,\hspace{0.7cm}\lambda _{3}-\lambda _{2}+2\lambda _{4}\geq 0,\hspace{0.7cm}\lambda
_{12}\geq \sqrt{2\left( \lambda _{3}-\lambda _{2}+2\lambda _{4}\right)
\lambda _{8}}.
\end{eqnarray}

\section{Analytical expressions for the entries of the CKM matrix}
Explicitly, the CKM entries are given as
	\begin{eqnarray}
	(\mathbf{V}_{CKM})_{ud}&=&-\frac{\vert m_{u}\vert}{\vert a_{u}\vert}\sqrt{\frac{\vert m_{u}\vert^{2}		\mathcal{N}_{2}\mathcal{N}_{3} \mathcal{M}_{2}}{\mathcal{D}_{1}}}~\cos{\theta_{d}}+ \sqrt{\frac{\mathcal{N}_{1} \mathcal{M}_{1}\mathcal{M}_{3}}{\mathcal{D}_{1}}	}~\sin{\theta_{d}}~e^{-i\bar{\eta}_{c}},\nn\\
	(\mathbf{V}_{CKM})_{us}&=&-\left[\frac{\vert m_{u}\vert}{\vert a_{u}\vert}\sqrt{\frac{\vert m_{u}\vert^{2}		\mathcal{N}_{2}\mathcal{N}_{3} \mathcal{M}_{2}}{\mathcal{D}_{1}}}~\sin{\theta_{d}}+\sqrt{\frac{\mathcal{N}_{1} \mathcal{M}_{1}\mathcal{M}_{3}}{\mathcal{D}_{1}}	}~\cos{\theta_{d}}~e^{-i\bar{\eta}_{c}}\right],\nn\\
	(\mathbf{V}_{CKM})_{ub}&=&\frac{1}{\vert a_{u}\vert}\sqrt{\frac{\mathcal{N}_{1} \mathcal{M}_{2}		\mathcal{K}}{\mathcal{D}_{1}}}~e^{-i\eta _{t}},\nn\\
	(\mathbf{V}_{CKM})_{cd}&=&-\left[\frac{	\vert m_{c}\vert}{\vert a_{u}\vert}\sqrt{\frac{\vert m_{c}\vert^{2}\mathcal{N		}_{1}\mathcal{N}_{3} \mathcal{M}_{1}}{\mathcal{D}_{2}}}~\cos{\theta_{d}}+\sqrt{\frac{ \mathcal{N}_{2} \mathcal{M}_{2}\mathcal{M}_{3}}{\mathcal{D}	_{2}}}~\sin{\theta_{d}}~e^{-i\bar{\eta}_{c}}\right],\nn\\
	(\mathbf{V}_{CKM})_{cs}&=& \frac{	\vert m_{c}\vert}{\vert a_{u}\vert}\sqrt{\frac{\vert m_{c}\vert^{2}\mathcal{N		}_{1}\mathcal{N}_{3} \mathcal{M}_{1}}{\mathcal{D}_{2}}}~\sin{\theta_{d}}+\sqrt{\frac{ \mathcal{N}_{2} \mathcal{M}_{2}\mathcal{M}_{3}}{\mathcal{D}		_{2}}}~\cos{\theta_{d}}~e^{-i\bar{\eta}_{c}},\nn\\	
	(\mathbf{V}_{CKM})_{cb}&=&-\frac{1}{\vert a_{u}\vert}\sqrt{\frac{\mathcal{N}_{2} \mathcal{M}_{1}		\mathcal{K}}{\mathcal{D}_{2}}}~e^{-i\eta _{t}},\nn\\
	(\mathbf{V}_{CKM})_{td}&=& 	\frac{\vert
		m_{t}\vert}{\vert a_{u}\vert}\sqrt{\frac{\vert m_{t}\vert^{2}\mathcal{N}_{1}		\mathcal{N}_{2} \mathcal{M}_{3}}{\mathcal{D}_{3}}}~\cos{\theta_{d}}-\sqrt{\frac{ \mathcal{N}_{3} \mathcal{M}_{1}\mathcal{M}_{2}}{		\mathcal{D}_{3}}}~\sin{\theta_{d}}~e^{-i\bar{\eta}_{c}},\nn\\
	(\mathbf{V}_{CKM})_{ts}&=& \frac{\vert
		m_{t}\vert}{\vert a_{u}\vert}\sqrt{\frac{\vert m_{t}\vert^{2}\mathcal{N}_{1}		\mathcal{N}_{2} \mathcal{M}_{3}}{\mathcal{D}_{3}}}~\sin{\theta_{d}}-\sqrt{\frac{ \mathcal{N}_{3} \mathcal{M}_{1}\mathcal{M}_{2}}{		\mathcal{D}_{3}}}~\cos{\theta_{d}}~e^{-i\bar{\eta}_{c}},\nn\\ 
	(\mathbf{V}_{CKM})_{tb}&=&	\frac{1}{\vert a_{u}\vert}\sqrt{\frac{\mathcal{N}_{2} \mathcal{M}_{3}		\mathcal{K}}{\mathcal{D}_{3}}}~e^{-i\eta _{t}}.
	\end{eqnarray}

\bibliographystyle{utphys}
\bibliography{BiblioQ4}

\end{document}